\documentclass[aps,prd,superscriptaddress,groupedaddress,nofootinbib,10pt]{article}

\usepackage{dcolumn}
\usepackage{bm}
\usepackage{lscape}
\usepackage{amsmath,amssymb,graphicx}
\usepackage{epsfig}
\usepackage{pstricks}
\usepackage{multirow}
\usepackage[bookmarks,linktocpage]{hyperref}
\usepackage{booktabs}
\usepackage{graphicx}
\usepackage{axodraw4j}
\usepackage{subfigure}
\usepackage{cite}
\usepackage{rotating}
\usepackage{hyperref}
\usepackage{float}
\usepackage[a4paper, left=2cm, right=2cm, top=2.5cm, bottom=2.5cm]{geometry}

\usepackage{array}
\newcolumntype{L}[1]{>{\raggedright\let\newline\\\arraybackslash\hspace{0pt}}m{#1}}
\newcolumntype{C}[1]{>{\centering\let\newline\\\arraybackslash\hspace{0pt}}m{#1}}
\newcolumntype{R}[1]{>{\raggedleft\let\newline\\\arraybackslash\hspace{0pt}}m{#1}}

\RequirePackage{lineno}
\newcommand{\Lag}{\mathcal L}
\newcommand{\MET}{\diagup\hskip -8ptE_T}

\newcommand{\Z}{$\mathbb Z_2$ }

\begin{document}

\begin{flushright}
\hspace{3cm} 
LYCEN 2018-01\\
SI-HEP-2018-05
% SISSA-xx-2018-FISI\\
\end{flushright}

\setlength{\arraycolsep}{1.5pt}

\begin{center}
{\huge \textbf{Characterising Dark Matter \\ 
Interacting with Extra Charged Leptons}}

\vskip.1cm
\end{center}

\vskip0.2cm

\begin{center}
\textbf{{D. Barducci$^{1}$, A. Deandrea$^{2,3}$, S. Moretti$^{4,5}$, L. Panizzi$^{6,4}$, H. Prager$^{7,4,5}$} \vskip 8pt }

{\small
$^1$\textit{SISSA and INFN, Sezione di Trieste, via Bonomea 265, 34136 Trieste, Italy}\\[0pt]
\vspace*{0.1cm} $^2$\textit{Universit\'e de Lyon, F-69622 Lyon, France;
Universit\'e Lyon 1, Villeurbanne}\\[0pt]
\vspace*{0.1cm} $^3$\textit{CNRS/IN2P3, UMR5822, Institut de Physique
Nucl\'eaire de Lyon, F-69622 Villeurbanne Cedex, France}\\[0pt]
\vspace*{0.1cm} $^4$\textit{School of Physics and Astronomy, University of
Southampton, Highfield, Southampton SO17 1BJ, UK}\\[0pt]
\vspace*{0.1cm} $^5$\textit{Particle Physics Department, Rutherford Appleton
Laboratory, Chilton, Didcot, Oxon OX11 0QX, UK}\\[0pt]
\vspace*{0.1cm} $^6$\textit{Dipartimento di Fisica, Universit\`a di Genova and INFN, Sezione di Genova,
via Dodecaneso 33, 16146 Genova, Italy}\\[0pt]
\vspace*{0.1cm} $^7$\textit{Theoretische Physik 1, Universita\"at Siegen, Walter-Flex-Stra\ss e 3, D-57068 Siegen, Germany
}\\[0pt]
}
\end{center}

\begin{abstract} 
\noindent
In the context of a simplified leptophilic Dark Matter (DM) scenario where the mediator is a new charged fermion 
carrying leptonic quantum number and the DM candidate is either scalar or vector, the
complementarity of different bounds is analysed. In this framework, the extra lepton and  
DM are odd under a \Z symmetry, hence the leptonic mediator can only interact with the DM state and Standard Model (SM) 
leptons of various flavours. We show that there is the possibility to characterise the DM spin (scalar or vector), as well as the nature of the mediator,  through a combined analysis of 
cosmological, 
flavour  
and collider data. We present an explicit numerical analysis for a set of benchmarks
points of the viable parameter space of our scenario.  
\end{abstract}

% \newpage
\tableofcontents
\newpage

\section{Introduction}

Among the open problems in physics, understanding the origin of DM has one of the highest priorities. A large number of experiments from different sectors of cosmology, particle and astroparticle physics have been designed to detect, directly or indirectly, signatures originating from DM but, remarkably, nothing has been observed so far. 
This lack of observations requires a joint effort from the cosmology and particle physics communities in order try excluding, or at least constraining, DM scenarios and narrow down the viable hypotheses to a small subset with distinctive and testable properties. However, due to the wide range of possibilities for embedding a DM candidate in extensions of the SM, constraints from individual observations may not be able to effectively test the parameter space of different models, while correlations between different observables may result in complementary and mutually-incompatible constraints which can help in excluding classes of DM scenarios. In this context, if any signal compatible with DM is observed, finding ways to characterise the DM properties would act as a further selector of which Beyond the SM (BSM) scenario to pursue. 

Of course it is not possible, in practice, to design and undertake dedicated experimental analyses for each BSM construction predicting DM candidates. Hence, it is a common and well-established practice to consider simplified models, where the SM particle content is extended minimally to be able to reproduce, with some model-dependent degree of approximation, broad classes of scenarios. Analysing DM scenarios from a model-independent perspective through simplified models can therefore make much easier the identification of exclusion regions (and hence of complementary ones where detection could occur) in a minimal set of new physics parameters. The interpretation in terms of theoretically motivated scenarios is then reduced to mapping the simplified model parameters in terms of those of the specific theory. Simplified models for DM consist in minimal extensions of the SM with a DM candidate which interacts with the SM through a mediator. The latter acts therefore as a portal between the DM and the SM sectors and can be either a SM particle or belong itself to new physics. Simplified models can then be classified according to the spin of the mediator and  DM states. Usually the stability of the DM candidate is guaranteed  by imposing a discrete \Z symmetry, under which all SM states are even and the DM state is the lightest odd particle. If the mediator is not a SM particle, a further subdivision can then be done by considering models where the mediator is even or odd under the same discrete \Z symmetry.

This analysis focuses on how applying complementary constraints from different observables from cosmology, flavour and collider physics can help in the characterisation of the spin of DM  within a specific class of simplified scenarios. 
We will focus on a simplified model where the mediator is a new fermion, odd under the \Z symmetry and carrying lepton number, while the DM is a boson, either scalar or vector, which does not carry a lepton number. The only allowed interactions between the mediator and  DM will therefore involve also SM leptons, due to the conservation of lepton number. 
We will discuss the constraints on the new parameters of this scenario (masses and couplings) and combine them to identify exclusion/detection regions for some representative Benchmarks Points (BPs) characterised by how the eXtra Lepton (XL) interacts with the SM ones. Once this is done, we shall proceed to the aforementioned characterisation of the DM spin, by concentrating on the parameter space surviving both space and ground experiments, as we shall detail below. The former shall include relic density 
while the latter shall exploit constraints emerging from the anomalous magnetic moment $(g-2)$ of the electron and muon as well as Lepton Flavour and Number Violating (LFV and LNV) processes. We can anticipate that the scope offered in this respect by the Large Hadron Collider 
(LHC) is minimal, {\emph{i.e.}}, its sensitivity to the spin properties of DM is more modest in comparison. However, due to the potential to exclude a large region of  parameter space, collider bounds will be used as a baseline for the subsequent characterisation of the DM spin in the allowed regions of it.
Before proceeding to our phenomenological analysis, we should acknowledge here the debt owed to previous literature
dealing with 
various aspects of our scenario. In Refs.\cite{Toma:2013bka, Ibarra:2014qma, Giacchino:2014moa} the 
$\gamma$-ray emission from scalar DM annihilation, also mediated by an XL, is discussed. Furthermore, Ref.\cite{Giacchino:2013bta} focuses on 
constraints from $\gamma$-ray emission and relic density for a real scalar singlet DM and a charged singlet vector-like lepton. The  dipole 
moments of electron and muon are analysed in Refs. \cite{Fukushima:2013efa, Agrawal:2014ufa, Freitas:2014pua}. An overview of different observables is 
performed in \cite{Chang:2014tea} for a subset of scenarios and with specific assumptions about the couplings. Constraints from the process 
$e^+e^-\to \mu^+\mu^-$ at the Large Electron-Positron (LEP) collider and a projection for the International Linear Collider (ILC) are provided in \cite{Freitas:2014jla}. 

It is also important to notice that the scenario we are considering can describe theoretically motivated models of new physics. A class of these,
which predict fermionic \Z odd partners of SM leptons and a scalar or vector DM, is represented by, e.g., Universal Extra-Dimensions (UEDs) 
\cite{Servant:2002aq,Cheng:2002ej,Cacciapaglia:2009pa,Dobrescu:2007ec}.
In these models, each SM state is the zero mode of a Fourier expansion of the multi-dimensional state in the extra-dimensional coordinates while the 
other states of the tier can be even or odd under a discrete \Z  (Kaluza-Klein) symmetry, the lightest state of the first tier being usually the 
partner of the photon or a mixture containing it, which is either scalar or vector depending on the number of extra-dimensions. Therefore, the 
leptonic sector of the theory can be described in terms of simplified models in which the lightest odd-tier partner of SM leptons is the mediator 
and the lightest odd-tier partner of the photon is the DM candidate. 

In this analysis we will first provide the necessary formalism for the description of the simplified scenario, discussing the Lagrangian terms 
and the BPs we will consider.  Furthermore,  in the subsequent sections, the constraints from collider, relic density, dipole 
moments of electron and muon and flavour  observables will be dealt with.
Then we shall combine such constraints to find which parameter configurations of our simplified scenario are 
 excluded and which ones are still allowed, with the purpose of testing hypotheses on the spin properties of DM. We will then summarise and give our conclusions.

\section{Notation and Parametrisation}

The most general Lagrangian terms for a minimal SM extension with one  XL and one DM candidate (scalar or vector) depend on 
the representations of the XL and the DM states and SM lepton(s) involved in the interaction. Minimal extensions of the SM 
involve only singlet or doublet representations for both XL and DM.

If the DM candidate transforms as a \textit{singlet} under $SU(2)$, the most general interaction terms between XL, DM and SM leptons are:
\begin{eqnarray}
\Lag^S_1 &=& \sum_{f=e,\mu,\tau} \left[\lambda_{11}^f \bar{E} P_R e_f + \lambda_{21}^f \bar \Psi_{-1/2} P_L {\nu_f \choose e_f} \right] 
S^0_{\rm DM} + h.c. 
\label{eq:LagSingletDMS}
\\
\Lag^V_1 &=& \sum_{f=e,\mu,\tau} \left[g_{11}^f \bar{E} \gamma_\mu P_R e_f + g_{21}^f \bar \Psi_{-1/2} \gamma_\mu 
P_L {\nu_f \choose e_f} \right] V^{0\mu}_{\rm DM} + h.c. \label{eq:LagSingletDMV}
\end{eqnarray}
If the DM candidate transforms as a \textit{doublet} under $SU(2)$, the most general Lagrangian terms are:
\begin{eqnarray}
\Lag^S_2 &=& \sum_{f=e,\mu,\tau} \left[\lambda_{12}^{f} \bar E P_L {\nu_f \choose e_f} + \lambda_{22}^{f} \bar \Psi_{-1/2} P_R e_f\right] 
\Sigma_{\rm DM} + \left[(\lambda_{22}^{f})^{\prime} \bar \Psi_{-3/2} P_R e_f\right] \Sigma_{\rm DM}^{c} + h.c. \\
\Lag^V_2 &=& \sum_{f=e,\mu,\tau} \left[g_{12}^{f} \bar E \gamma^\mu P_L {\nu_f \choose e_f} + g_{22}^{f} \bar \Psi_{-1/2} \gamma_\mu 
P_R e_f\right] \mathcal{V}_{\rm DM}^\mu + \left[(g_{22}^{f})^{\prime} \bar \Psi_{-3/2} \gamma_\mu P_R e_f\right] \mathcal{V}_{\rm DM}^{c,\mu} + h.c. 
\label{eq:LagDoubletDM}
\end{eqnarray}
In the equations above, XLs are denoted as $E$ (or $E^\pm$) if charged and as $N$ (or $N^0$) if neutral. If XLs belong to 
non-trivial representations of $SU(2)$, they are denoted according to their weak hypercharge\footnote{We adopt the convention $Q=T_3+Y$.}: 
$\Psi_{-1/2} = {N^0 \choose E^-}$ and $\Psi_{-3/2} = {E^- \choose E^{--}}$. The DM candidate is denoted as $S^{(*)}_{\rm DM}$ (or 
$S^{0(*)}_{\rm DM}$) if scalar, real (or complex), and $V^{(*)}_{\rm DM}$ (or $V^{0(*)}_{\rm DM}$) if vector. If the DM candidate is part of a non-trivial 
$SU(2)$ representation, the full multiplet is denoted as $\Sigma_{\rm DM} = {S^+ \choose S^0_{\rm DM}}$ (with its charge-conjugate 
$\Sigma_{\rm DM}^c = {S^{0(*)}_{\rm DM} \choose - S^-}$) if scalar or as $\mathcal{V}_{\rm DM} = {V^+ \choose V^0_{\rm DM}}$ (with its 
charge-conjugate $\mathcal{V}_{\rm DM}^c = {V^{0(*)}_{\rm DM} \choose V^-}$) if vector.

The notation for the generic coupling between XL, DM and SM states shows explicitly the representation of the new particles. If the DM 
is scalar, Yukawa couplings are labelled as $\lambda_{ij}^f$, where $i$ and $j$ indicate the representation of the XL and DM respectively 
(1 for singlet, 2 for doublet and so on), and $f$ is a flavour index. If the DM is vectorial, the notation is analogous, but the coupling are 
labelled as $g_{ij}^f$. The flavour index has been explicitly written to show that the couplings between XL, DM and SM leptons of different flavours
are considered as independent parameters, which can be individually set to specific values, including zero, to allow for flavour-specific interactions. The 
effective Lagrangian parametrisation we use allows therefore to discuss quite different situations, including both flavour blind DM interactions as well 
as flavour specific DM interactions, which may arise for example in models with specific parities or in composite models. In the following we shall consider 
only the effective approach using benchmark points which are useful for the phenomenological study without entering in the details of specific models.

Scenarios with a DM doublet representation imply the presence of further new states, a charged scalar $S^\pm$ or vector $V^\pm$, 
and an exotic doubly-charged XL is also allowed. These non-minimal scenarios will not be considered in the following analysis. 

The interactions between XLs and the SM gauge bosons is parametrised as:
\begin{eqnarray}
\Lag_{AXL} &=& -e A^\mu \bar{E} \gamma_\mu E \\
\Lag_{ZXL} &=& Z^\mu \bar{E} \gamma_\mu \left(g^{ZEE}_L P_L + g^{ZEE}_R P_R \right) E + Z^\mu \bar{N} \gamma_\mu \left(g^{ZNN}_L 
P_L + g^{ZNN}_R P_R \right) N \\
\Lag_{WXL} &=& W^{+\mu} \bar{N} \gamma_\mu \left(g^{WLN}_L P_L + g^{WLN}_R P_R \right) E + h.c
\end{eqnarray}
where the coupling with the $W$ is present only in the case of doublet XLs. From a model-independent point of view the gauge couplings can be treated as free parameters, to allow for new physics in the gauge sector which may induce mixing patterns. However, since the simplified model considered in this analysis consists of only the XL and DM additional states, the gauge couplings will be completely determined by the XL representation under the SM gauge group and, therefore, they will be the same as for SM states belonging to analogous representations.

\subsection{Vector-like and Chiral Extra-Leptons}

From the terms in Eq.~\eqref{eq:LagSingletDMS} and Eq.~\eqref{eq:LagSingletDMV} it is possible to identify some interesting and representative scenarios 
which depend on the nature of the XL.\\

\noindent\textit{\textbf{Vector-like XL (VLL)}}. If the XL is vector-like, its left-handed and right-handed components belong to the same representation of $SU(2)$. 
This means that, if the XL is a singlet, only the interactions proportional to $\lambda_{11}^f$ or $g_{11}^f$ (depending on the spin of the DM) are 
allowed while, if it is a doublet, the only allowed interactions are proportional to $\lambda_{21}^f$ or $g_{21}^f$. Therefore, for vector-like XL, the 
couplings are either purely left-handed or purely right-handed. It is interesting to notice that, since XLs are odd under the \Z 
 parity of the 
DM, there is no mixing between XL and SM leptons, and therefore there are no suppressed couplings with opposite chirality projections, unlike in 
scenarios where the extra-fermions are even and mix with the SM fermions. In this scenario, the mass term for a single XL can be written as:
\begin{equation}
 \Lag_{VLL} = - M_{E_{VLL}} \bar E E
\end{equation}

\begin{figure}[ht!]
\centering\includegraphics[width=.5\textwidth]{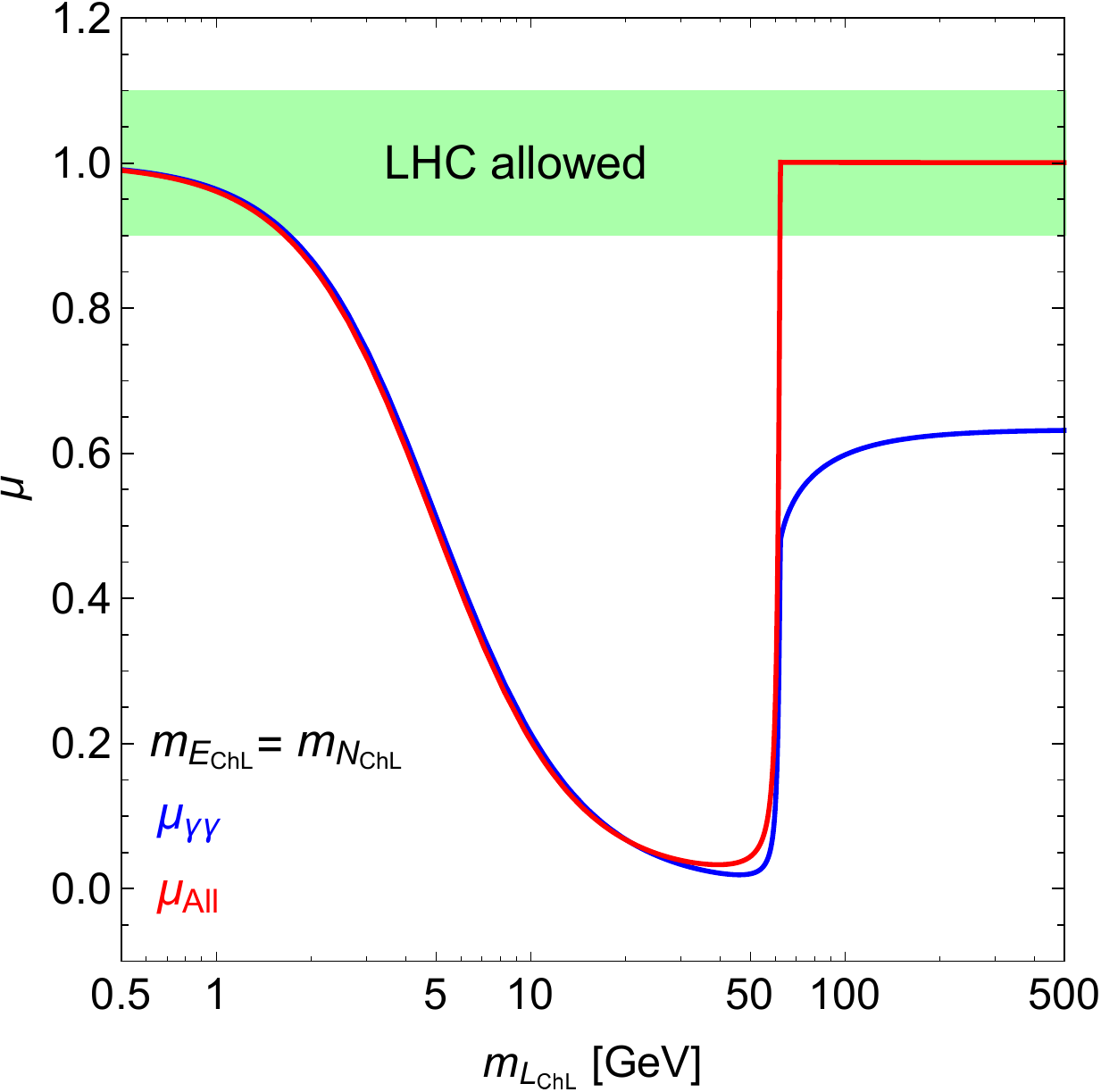}
\caption{\label{fig:ChLHiggsBounds}Signal strength for the SM Higgs boson into the $\gamma\gamma$ (red) and all the other (blue) final states in function of the common mass for the charged and neutral heavy lepton. The green shaded area corresponds to a 10\% deviation on these observables.}
\end{figure}

\noindent\textit{\textbf{Chiral XL (ChL)}}. If the left-handed and right-handed components of the XL belong to different representations of $SU(2)$, all the 
interactions of Eqs.(\ref{eq:LagSingletDMS}) or (\ref{eq:LagSingletDMV}) can be allowed at the same time. In particular, for scalar (vector) DM, if the left-handed component transforms as a singlet (doublet), the right-handed component transforms as a doublet (singlet) and the coupling constants are 
identical in absolute value, so  it is possible to have purely vector-like or purely axial-like interaction terms depending on the relative signs of the 
couplings. In this scenario, the XL can get its mass in the same way as SM leptons, through interactions with the Higgs boson,
\begin{equation}
\Lag_{ChL} = - y_{XL} \bar \Psi_{-1/2} H E + h.c. \quad\longrightarrow\quad - M_{E_{ChL}} \bar E E
\label{eq:ChLmass}
\end{equation}
where $M_{E_{ChL}} = {y_{XL} v \over \sqrt{2}}$ and $v$ is the Higgs Vacuum Expectation Value 
(VEV). Since ChLs acquire their mass through the Higgs mechanism, however, the presence of a heavy charged lepton can strongly modify the loop induced $H\gamma\gamma$ and $H\gamma Z$ interactions. Moreover, the Higgs boson will decay, if the process is kinematically allowed, in both a pair of charged and neutral heavy leptons, hence modifying the total Higgs decay width and altering in a universal way the Higgs decay rate in all possible final states, except those for the $\gamma\gamma$ and $\gamma Z$ final states which will be rescaled independently. These decay rates are measured at the LHC and the results are usually expressed in terms of signal strengths, {\emph{i.e.}} the ratio of the Higgs boson production cross sections times the Branching Ratio (BR) into a given final state over the SM expectation. 
We show in Fig.~\ref{fig:ChLHiggsBounds} the Higgs signal strengths in all the final state in function of the common mass of the charged and neutral heavy lepton, where the
green shaded area corresponds to a 10\% deviation in these observable which is compatible with the 7 and 8 TeV LHC ATLAS and CMS collaborations measurements on the Higgs boson couplings~\cite{ATLAS-CONF-2015-044}. As expected, due to the non-decoupling property of new chiral families, the constraints on these states are quite stringent, allowing only extremely light ChLs, roughly lighter than a few GeV. We have checked that the results do not drastically change assuming a mass for the heavy neutral lepton different from the one of the charged one.
For this reason, \textit{in the following we will not consider chiral leptons and focus only on vector-like lepton scenarios}. Therefore, in the rest of the paper, XLs will be referred to as VLL and their mass will be denoted generically as $M_E$ without ambiguities.

\subsection{Benchmark Points}

For the purposes of our phenomenological analysis, which aims at the characterisation of the spin of DM, and for the sake of simplicity, we will explore VLLs in the \textit{singlet representation}, {\it i.e.} interacting through right-handed couplings with SM leptons. It will be specified if any constraint does not depend on the VLL representation. 

Moreover, we have identified a set of BPs which represent various combinations of couplings between the VLL and the SM leptons, depending on their flavours. In particular, we have chosen to explore three BPs where the VLL couples only to one SM flavour and one BP where the VLL couples to all SM flavours with universal couplings. The BPs are summarised in Tab.~\ref{tab:BPandChir}. All couplings will be assumed to be real numbers.
We wish to stress that the benchmarks we have identified represent {\it extreme} representative scenarios, which are useful for a model-independent phenomenological analysis. It is beyond the scopes of this analysis to describe specific, theoretically motivated, scenarios of new physics. Furthermore such benchmarks are also justified by the fact that models with lepton-specific or lepton-non-universal couplings of the DM candidate have been studied in literature, see, {\it e.g.}, Refs.~\cite{vonderPahlen:2016cbw,Khoze:2017ixx}.

\begin{table}[H]
\begin{center}
\centering\begin{tabular}{c|ccc}
\toprule
    & $e$ + DM & $\mu$ + DM & $\tau$ + DM\\
\midrule
BP1 & $\lambda_{11}^e \neq 0$ or $g_{11}^e \neq 0$ & $\lambda_{11}^\mu = 0$ or $g_{11}^\mu = 0$ & $\lambda_{11}^\tau = 0$ or $g_{11}^\tau = 0$ \\
BP2 & $\lambda_{11}^e = 0$ or $g_{11}^e = 0$ & $\lambda_{11}^\mu \neq 0$ or $g_{11}^\mu \neq 0$ & $\lambda_{11}^\tau = 0$ or $g_{11}^\tau = 0$ \\
BP3 & $\lambda_{11}^e = 0$ or $g_{11}^e = 0$ & $\lambda_{11}^\mu = 0$ or $g_{11}^\mu = 0$ & $\lambda_{11}^\tau \neq 0$ or $g_{11}^\tau \neq 0$ \\
\midrule 
\multirow{2}{*}{BP4} & \multicolumn{3}{c}{universal couplings: same value for all flavours}\\
                     & $\lambda_{11}^e \neq 0$ or $g_{11}^e \neq 0$ & $\lambda_{11}^\mu \neq 0$ or $g_{11}^\mu \neq 0$ & $\lambda_{11}^\tau \neq 0$ or $g_{11}^\tau \neq 0$ \\
\bottomrule
\end{tabular}
\end{center}
\caption{\label{tab:BPandChir} Our BPs and the allowed interactions of the VLL with SM leptons and the DM candidate.}
\end{table}

\section{Width of the Extra-Lepton}

As we are considering scenarios where the VLL can only decay into DM and SM leptons, the only parameters which contribute to the width of the VLL 
are the VLL and DM masses and the couplings in Eq.~\eqref{eq:LagSingletDMS} or \eqref{eq:LagSingletDMV}, depending on the DM spin. It is thus important to determine 
which values of the couplings correspond to the Narrow and Large Width limits (NW and LW, respectively). More specifically, it is important to determine 
which values of the couplings can be accessed in any region of the mass  space without determining a too large width for the  VLL. (The interplay between 
the NW and LW regimes in heavy vector-like quark searches at the LHC has been studied in Refs.~\cite{Moretti:2016gkr,Moretti:2017qby,Prager:2017owg,Prager:2017hnt}.) 

The VLL width is given by the following expression (herein, $m_l$ is the SM lepton, $l_{\rm SM}$, mass):
\begin{equation}
\Gamma_E = \frac{ K_{\rm DM} }{32 \pi M_E^3} \sqrt{M_{\rm DM}^4-2 M_{\rm DM}^2 \left(M_E^2+m_l^2\right)+\left(M_E^2-m_l^2\right)^2 }
\end{equation}
where $K_{\rm DM}$ depends on the spin of DM
\begin{eqnarray}
 \begin{array}{ll}
  \text{Scalar DM: } & K_{S_{\rm DM}} = \left((\lambda_{11}^f)^2+(\lambda_{21}^f)^2\right) \left(M_E^2+m_l^2-M_{\rm DM}^2\right)+4 \lambda_{11}^f \lambda_{21}^f m_l M_E \\
  \text{Vector DM: } & K_{V_{\rm DM}} = \left((g_{21}^f)^2 + (g_{21}^f)^2\right) \left(2 M_{\rm DM}^4-M_{\rm DM}^2 \left(m_l^2+M_E^2\right)-\left(m_l^2-M_E^2\right)^2 \right)+12 g_{21}^f g_{11}^f M_{\rm DM}^2 m_l M_E
 \end{array}\nonumber\\
\end{eqnarray}

In the limit of $m_l\ll \{M_E,M_{\rm DM}\}$ and for singlet VLL ($\lambda_{21}^f=0$ or $g_{21}^f=0$, depending on the DM spin), the width expression simplifies to the following expressions:
\begin{equation}
  \text{\textit{\textbf{Scalar DM}:} } \Gamma_E \simeq \frac{ (\lambda_{11}^f)^2 \left(M_E^2-M_{\rm DM}^2\right)^2}{32 \pi M_E^3}  \quad\text{and}\quad
  \text{\textit{\textbf{Vector DM}:} } \Gamma_E \simeq \frac{ (g_{11}^f)^2 \left(M_E^6-3 M_E^2 M_{\rm DM}^4 +2 M_{\rm DM}^6 \right)}{32 \pi M_E^3 M_{\rm DM}^2} 
\end{equation}

\begin{figure}[htbp!]
\centering
\includegraphics[width=.48\textwidth]{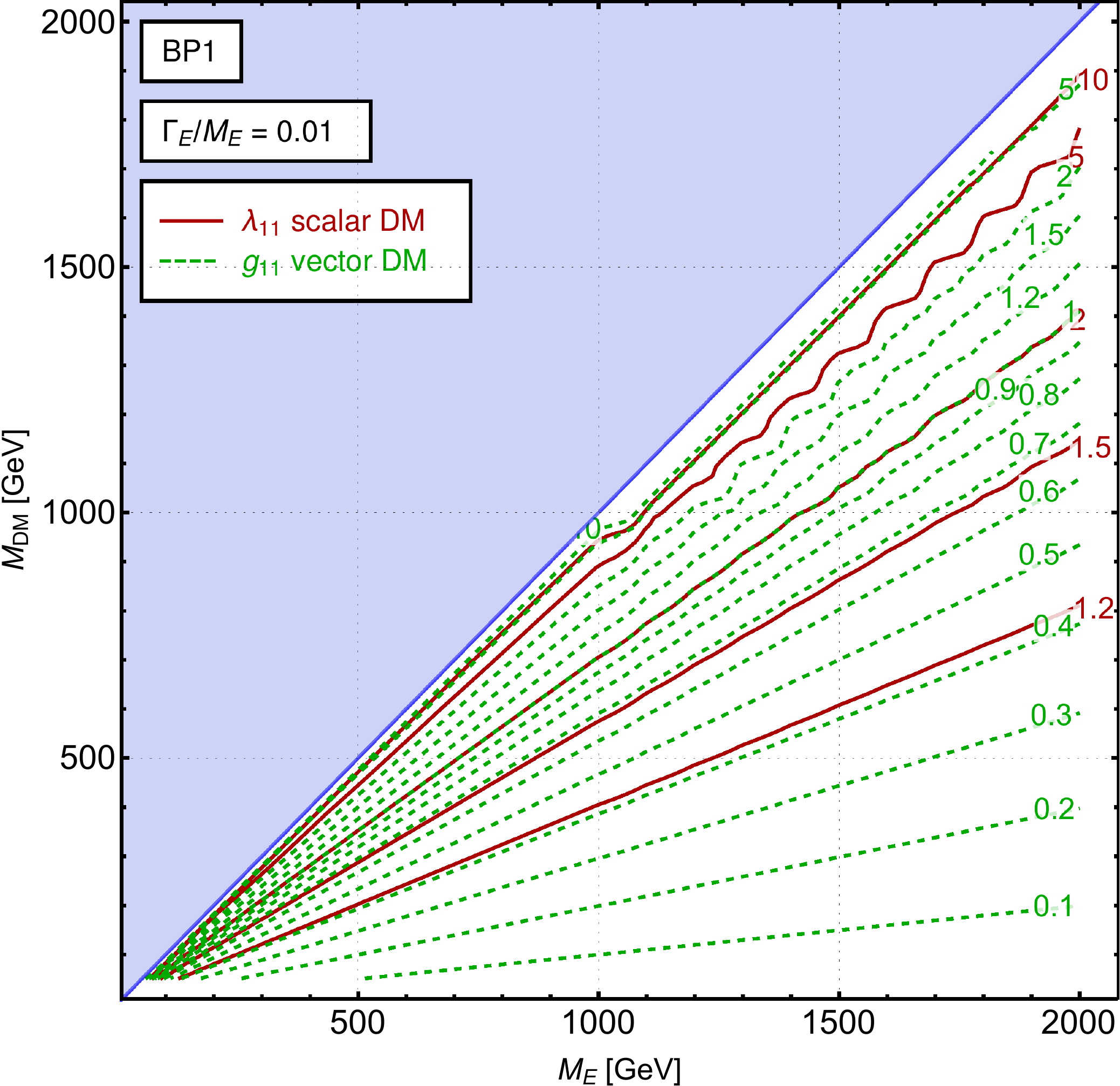}\hfill
\includegraphics[width=.48\textwidth]{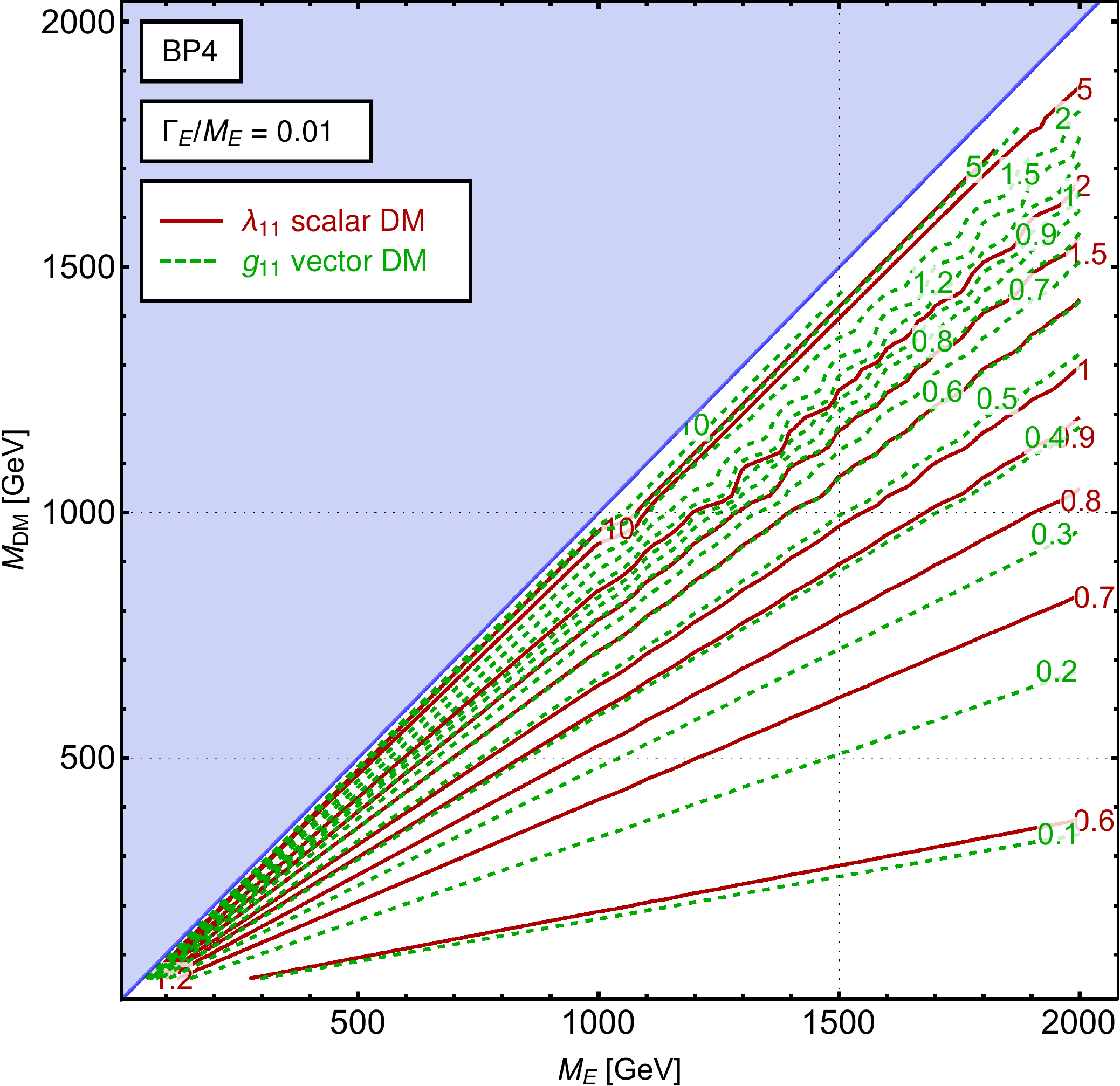}\\[20pt]
\includegraphics[width=.48\textwidth]{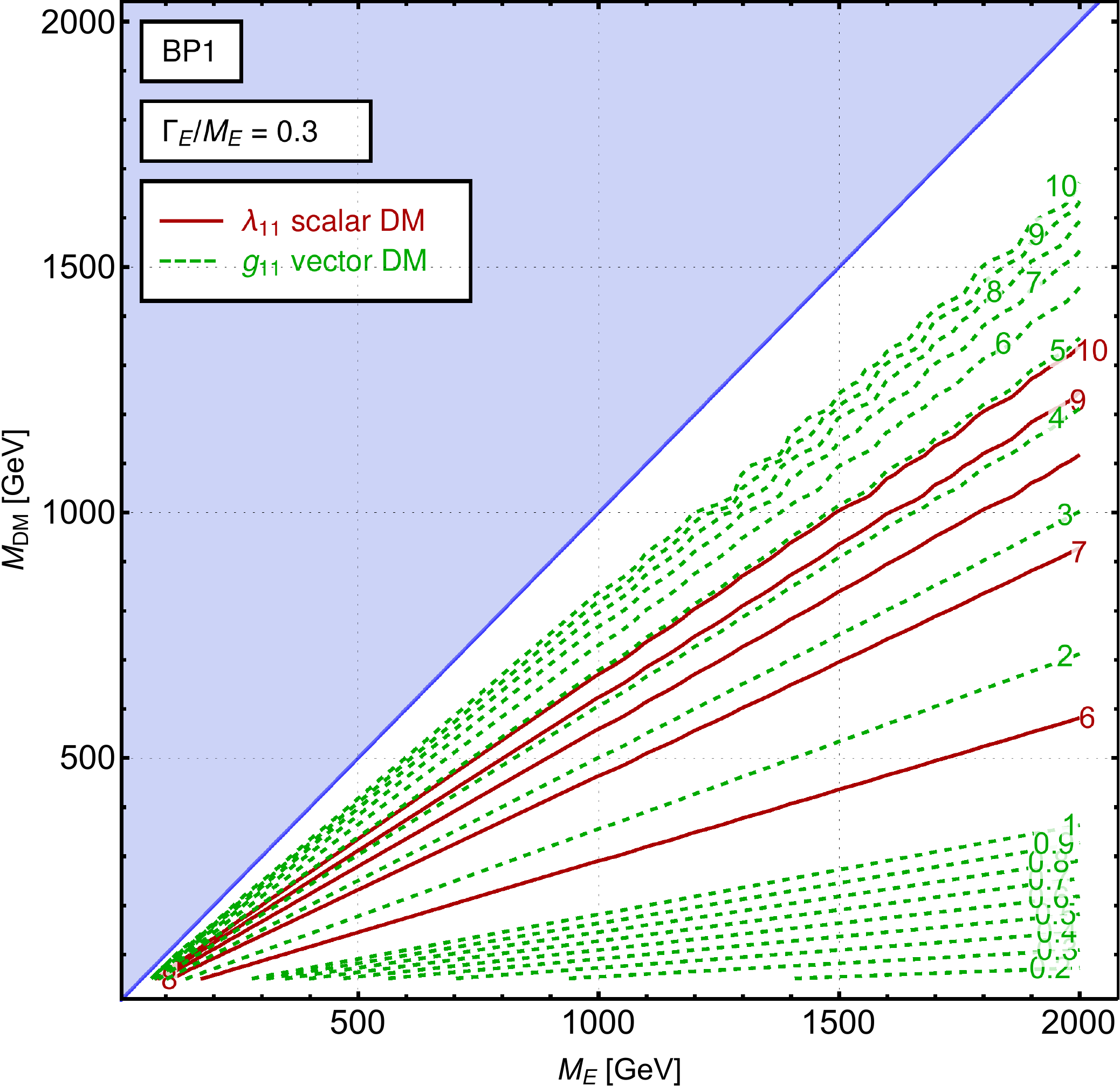}\hfill
\includegraphics[width=.48\textwidth]{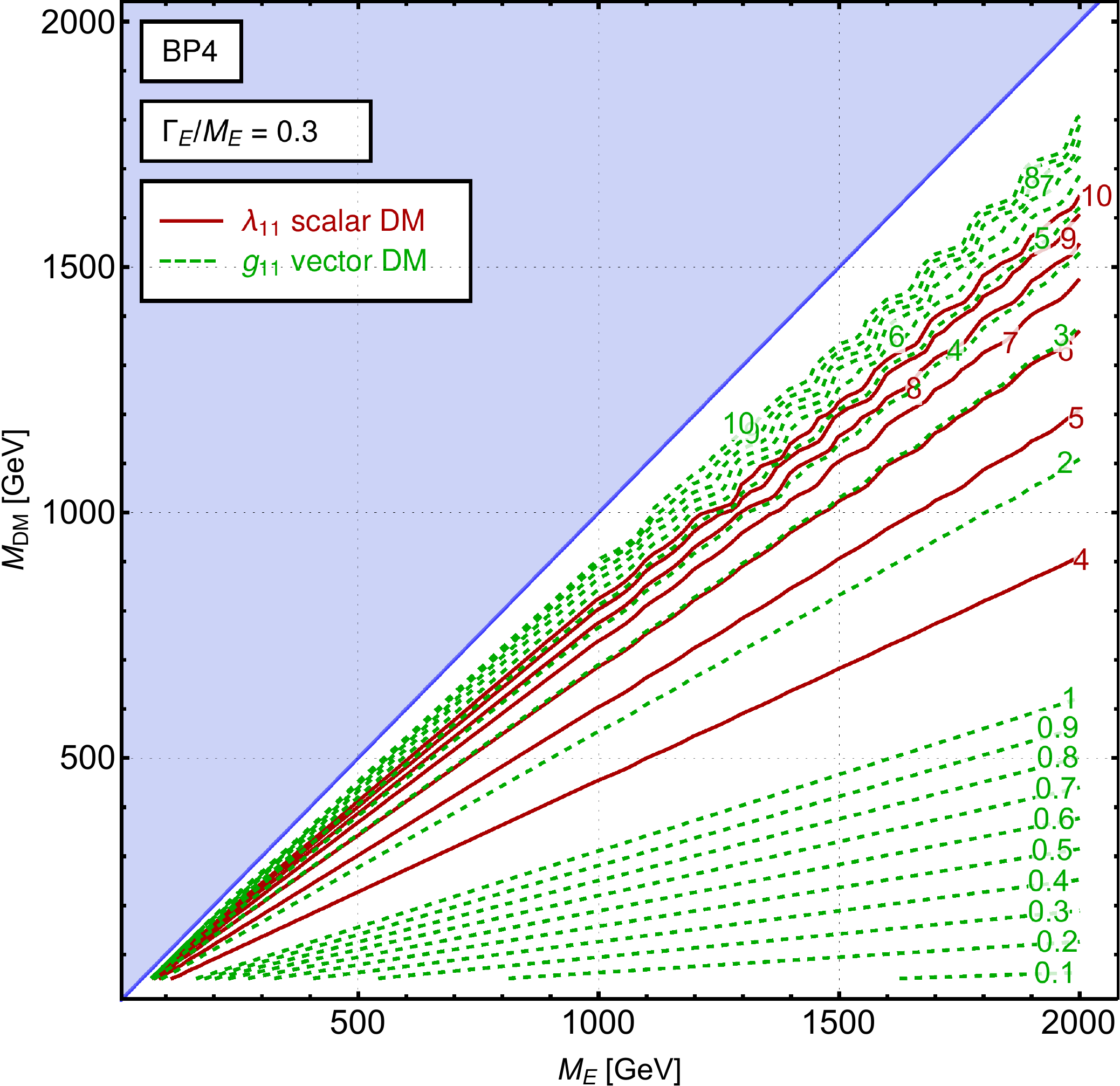}\\[20pt]
\caption{\label{fig:widths} Couplings between VLL, DM and SM lepton corresponding to a NW ratio $\Gamma_E/M_E=0.01$ (top row) or to a LW 
ratio $\Gamma_E/M_E=0.3$ (bottom row) for BP1 (left column) and BP4 (right column). BP2 and BP3 have the same qualitative behaviour as BP1 with small numerical differences.}
\end{figure}

The dependence of the width-to-mass ratios of the heavy leptons for all BPs and for different assumptions about their couplings with DM are 
summarised in Fig.~\ref{fig:widths}, where it is possible to infer the following.
\begin{itemize}
\item For \textit{\textbf{Scalar DM}}: couplings $\sim$1 always ensure that, in the whole mass range \{50-2000\} GeV for both VLL and DM, the width of the VLL is smaller or around of 1\% of its mass for BP1, 2 and 3, while for BP4 slightly smaller values $\sim 0.6$ are required. Of course larger coupling values would still generate small widths, but only in a limited VLL versus DM mass region: a coupling around 5, for example, would ensure a $\Gamma_E/M_E$ ratio smaller than 1\% only in the almost degenerate VLL and DM mass region. Still considering the whole VLL versus DM mass range, for BP1 to 3, a width smaller than 30\% of the mass can be obtained with couplings smaller than $\lambda_{ij}^f\sim6$ (LW limit), while for BP4 the limit is $\lambda_{ij}^f\sim4$.
\item For \textit{\textbf{Vector DM}}: the dependence of the width-to-mass ratio on the mass of the VLL is stronger than for the scalar DM case. Couplings smaller than $g_{ij}^f\sim0.1$ always ensure a $\Gamma_E/M_E$ ratio smaller than 30\% and, in most of the parameter space, also smaller than 1\%. 
\end{itemize}

\section{Constraints}
\label{sec:Constraints}

A crucial feature of heavy fermions which interact with DM particles is that they do not mix with their SM partners because the heavy fermions are odd under the same \Z parity of the DM while their SM partners are even. The absence of such a mixing implies that any contribution of heavy leptons to processes with SM particles in both initial and final states can only be at loop-level, as it must involve at least two new vertices, suppressing therefore mixings which can have an impact on measured quantities.

For any observable discussed in the following, constraints will be represented as excluded vs allowed regions in the ($M_E$, $M_{\rm DM}$) plane. The contours will depend on the free parameters of the theory, which are the couplings between VLL, DM and  SM leptons. 

The determination of exclusion regions for complementary observables will provide the first element for discrimination between scalar and vector  DM. An observation of a signal in regions which are excluded for one of the two scenarios and allowed for the other could only be interpreted univocally. If regions exist where both scalar and vector DM are allowed after the combination of constraints, only a detailed analysis of the signal properties could possibly distinguish between the two scenarios.

\subsection{Colliders} \label{sec:LHC}

\begin{figure}[ht!]
\begin{center}
\hskip -15pt
\includegraphics[width=.9\textwidth]{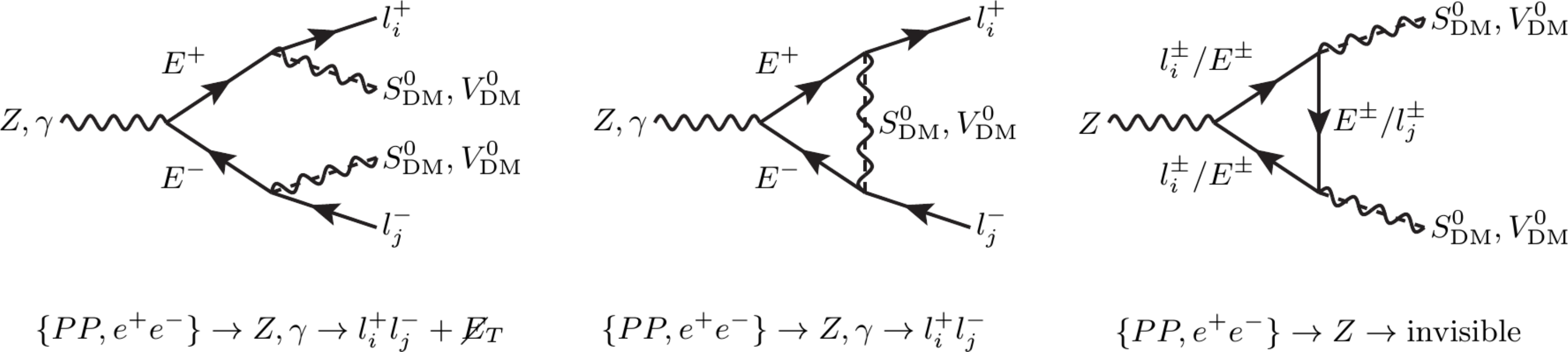}
\end{center}
\caption{\label{fig:ColliderFeynman} Collider signatures at tree- and loop-level.}
\end{figure}

At tree-level, the main mechanisms for the production of heavy leptons at colliders is through Drell-Yan (DY) channels via exchange of a $Z$ or photon, the heavy lepton then decays into SM leptons and DM. At loop-level, the heavy leptons contribute to the $Z$ to lepton and $Z$ to invisible decays. The Feynman diagrams corresponding to all these channels are shown in Fig.~\ref{fig:ColliderFeynman}.

\begin{figure}[ht!]
\includegraphics[width=.33\textwidth]{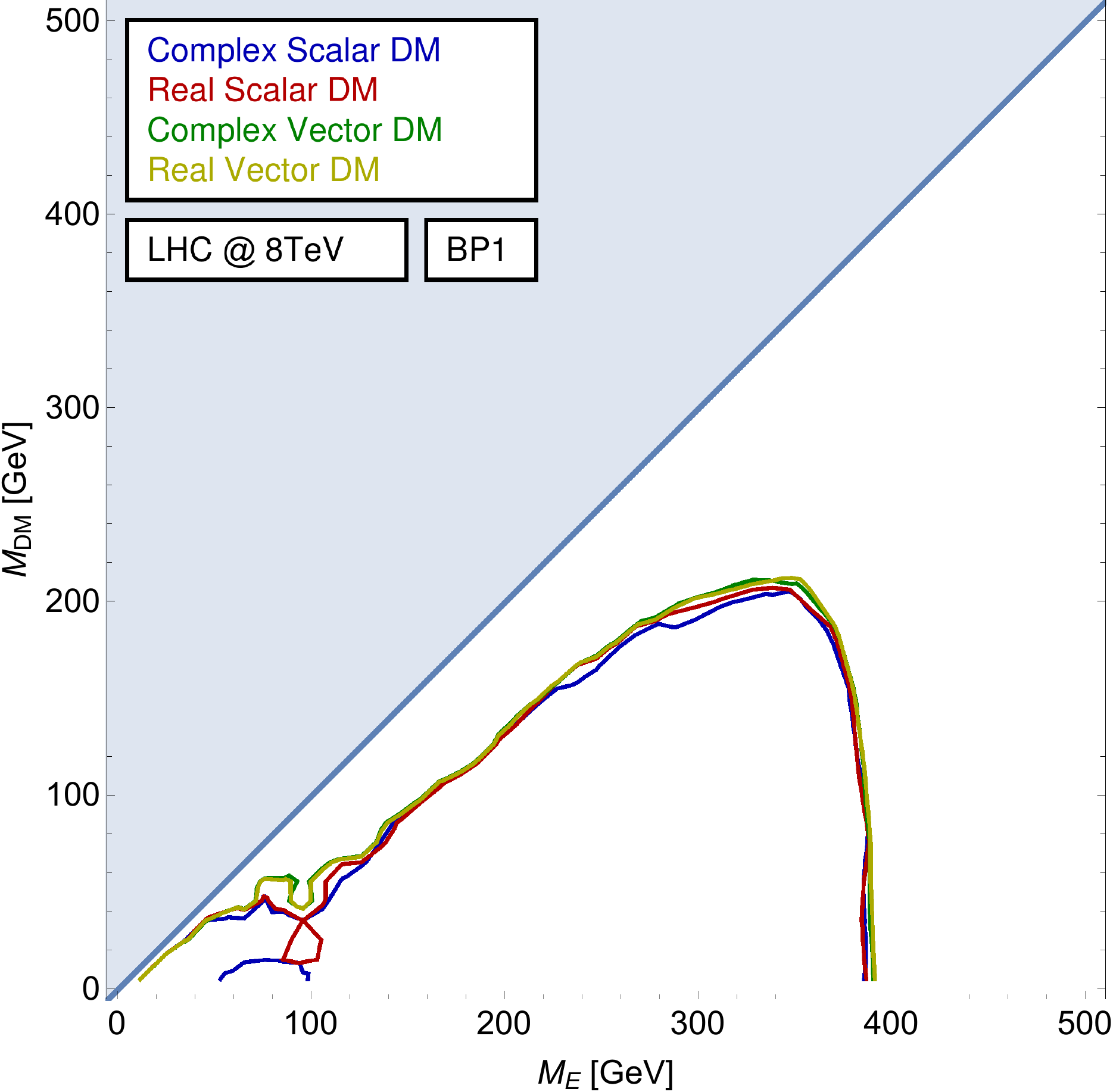}\hfill
\includegraphics[width=.33\textwidth]{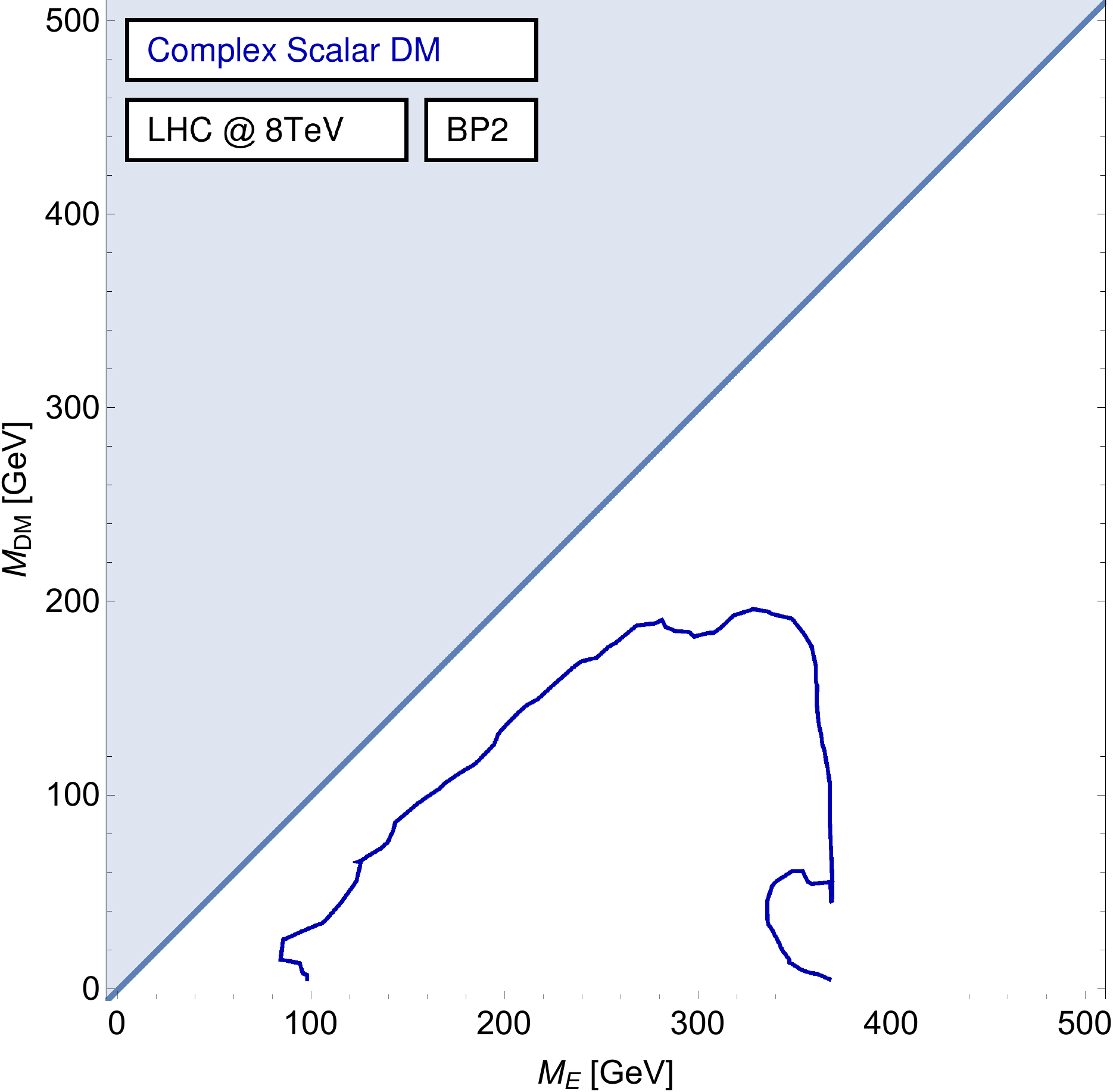}\hfill
\includegraphics[width=.33\textwidth]{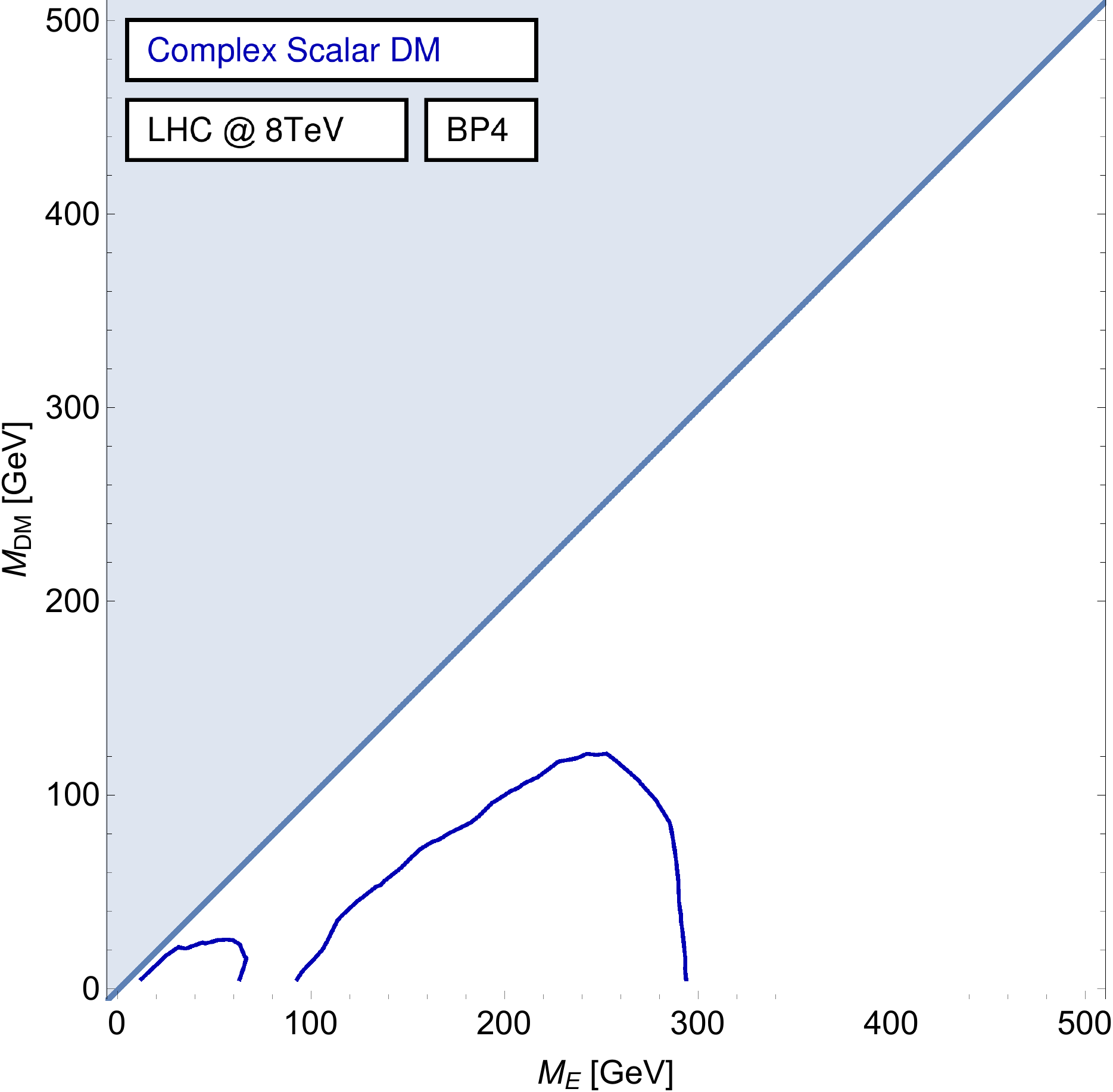}

\includegraphics[width=.33\textwidth]{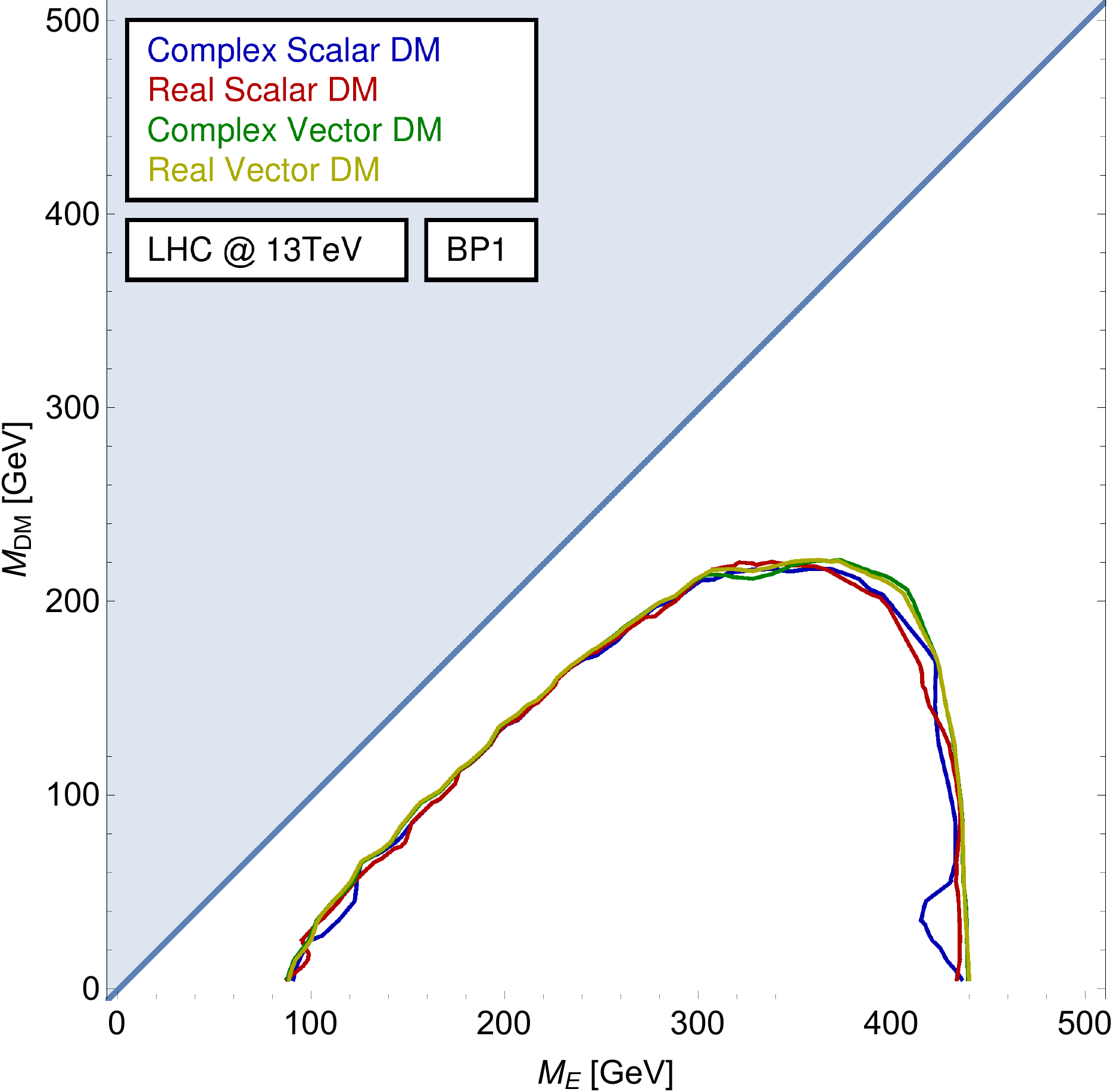}\hfill
\includegraphics[width=.33\textwidth]{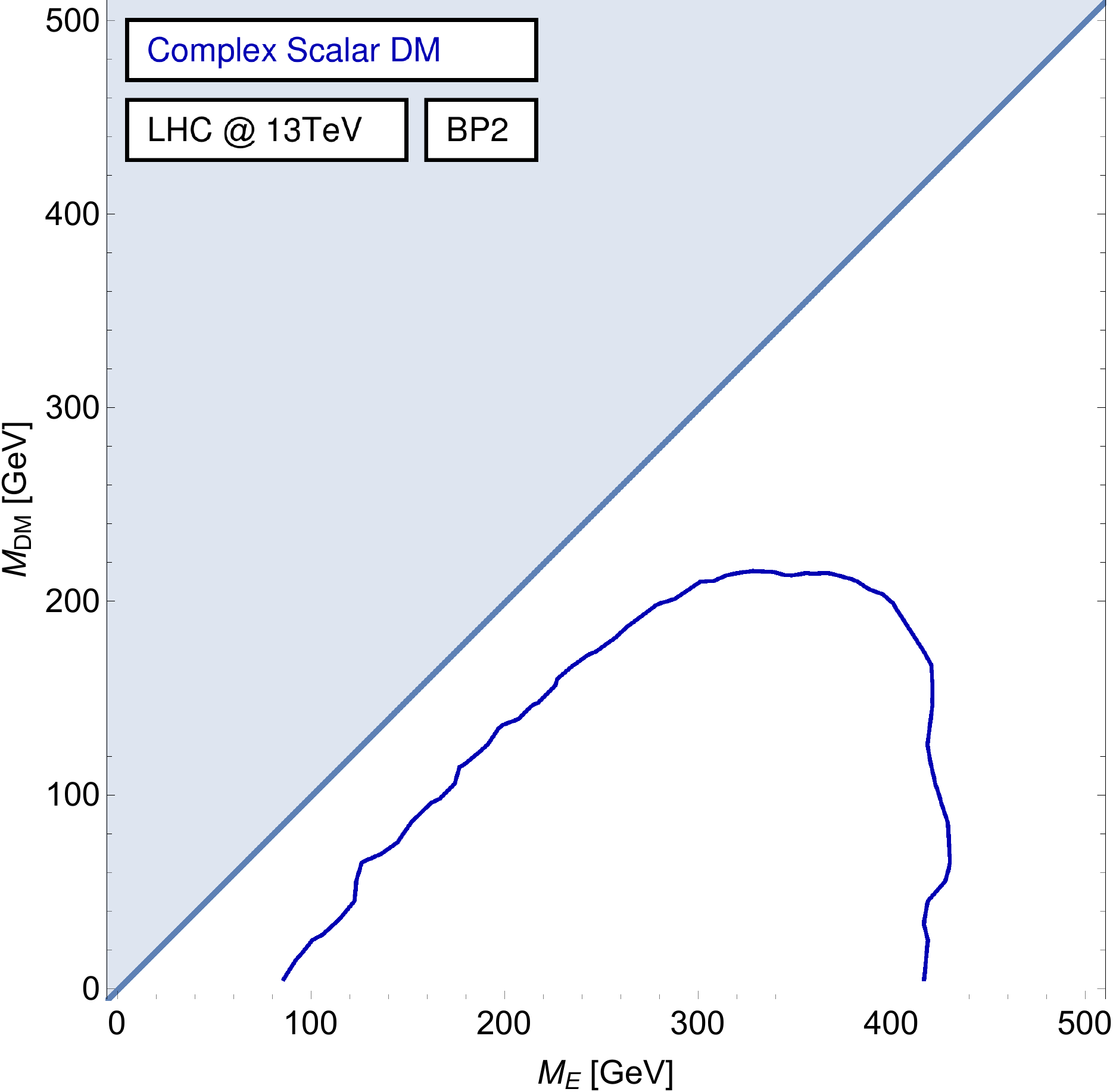}\hfill
\includegraphics[width=.33\textwidth]{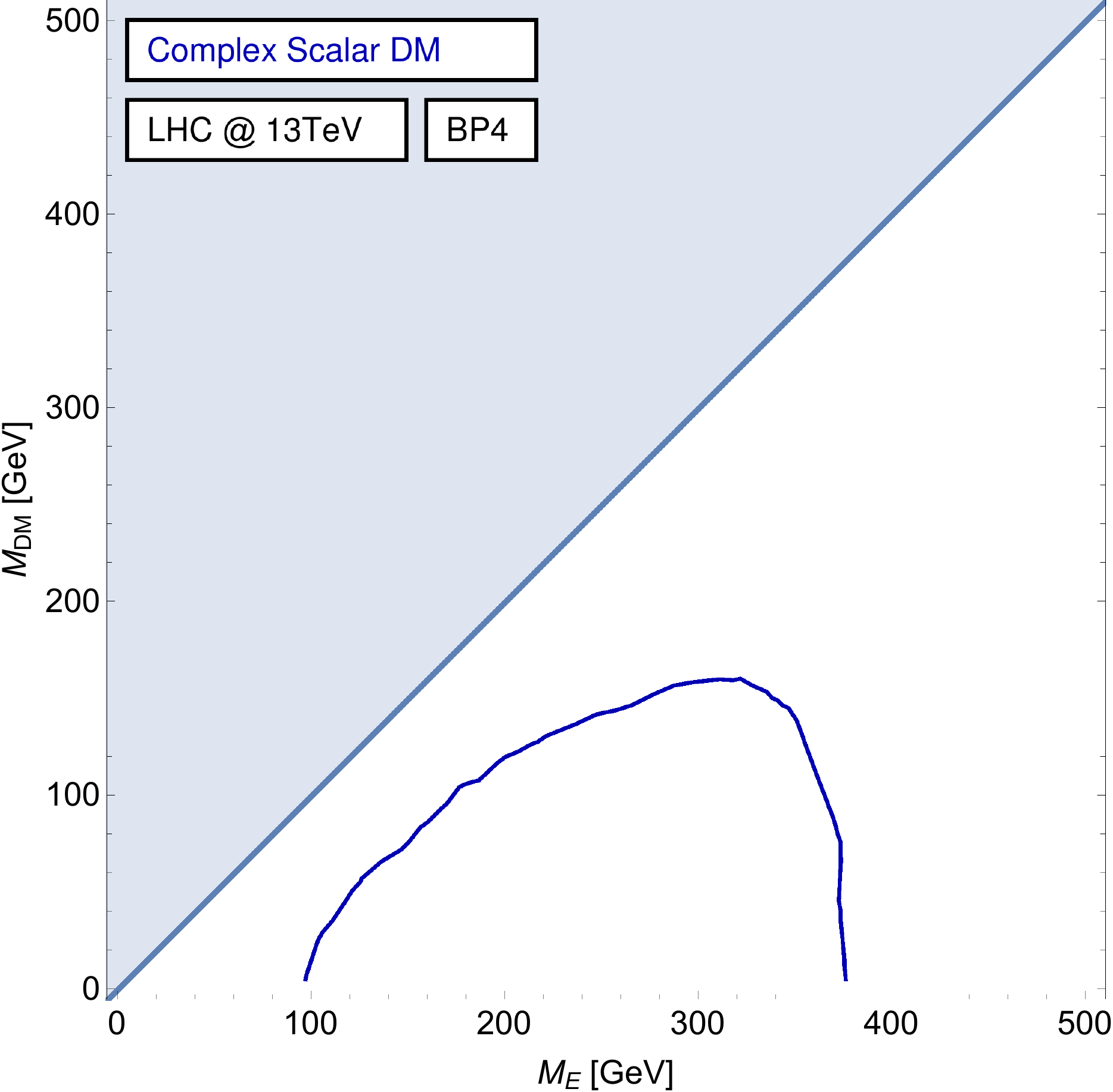}
\caption{\label{fig:LHCbounds} LHC bounds at 8 TeV (top row) and 13 TeV (bottom row). Contours for all DM scenarios are shown only for BP1, while for the other BPs the complex scalar scenario has been shown for the sake of simplicity. }
\end{figure}

In this exploratory analysis we will focus on the constraints from the DY processes at tree-level at the LHC and perform a scan over the VLL and DM masses. Simulations at parton level have been performed with {\sc MadGraph5} \cite{Alwall:2011uj,Alwall:2014hca} using a model implemented with {\sc FeynRules} \cite{Alloul:2013bka}: we have simulated the $2\to 4$ processes $PP\to {\rm DM}~{\rm DM}~l_i^+ l_j^-$ using the {\sc NNPDF2.3} PDF set~\cite{Ball:2012cx}. Hadronisation and parton shower have been performed through {\sc Pythia 8}\cite{Sjostrand:2014zea} and the events have been subsequently processed with {\sc CheckMATE 2}~\cite{Dercks:2016npn} to obtain the confidence level of the different points of our simulation grid analysis using a set of 8 and 13 TeV analyses from both ATLAS and CMS.
Due to the lack of sensitivity to the $2\tau+E_T^{\rm miss}$ final state from the searches implemented in {\sc CheckMATE 2}, the LHC limits for BP3 have been implemented in the following way. We have simulated the process $PP\to {E^+ E^-}$ and compared the cross section with the $2\sigma$ upper limit on the supersymmetric stau pair production cross section obtained by the ATLAS collaboration with 20.3 fb$^{-1}$ of data at 8 TeV~\cite{Aad:2014yka}. With this procedure we are neglecting the differences in the signal acceptances that can arise due to the different spin of the VLLs and the staus. This difference has however been shown to be negligible in the case of coloured scalar and fermionic top partners decaying to DM~\cite{Kraml:2016eti}. We are moreover assuming that the signal selection efficiency remains constant for scalar and vector DM (both real and complex), an assumption which is verified for the other considered BPs, see Fig.~\ref{fig:LHCbounds}.

\begin{figure}[ht!]
\begin{center}
\includegraphics[width=.33\textwidth]{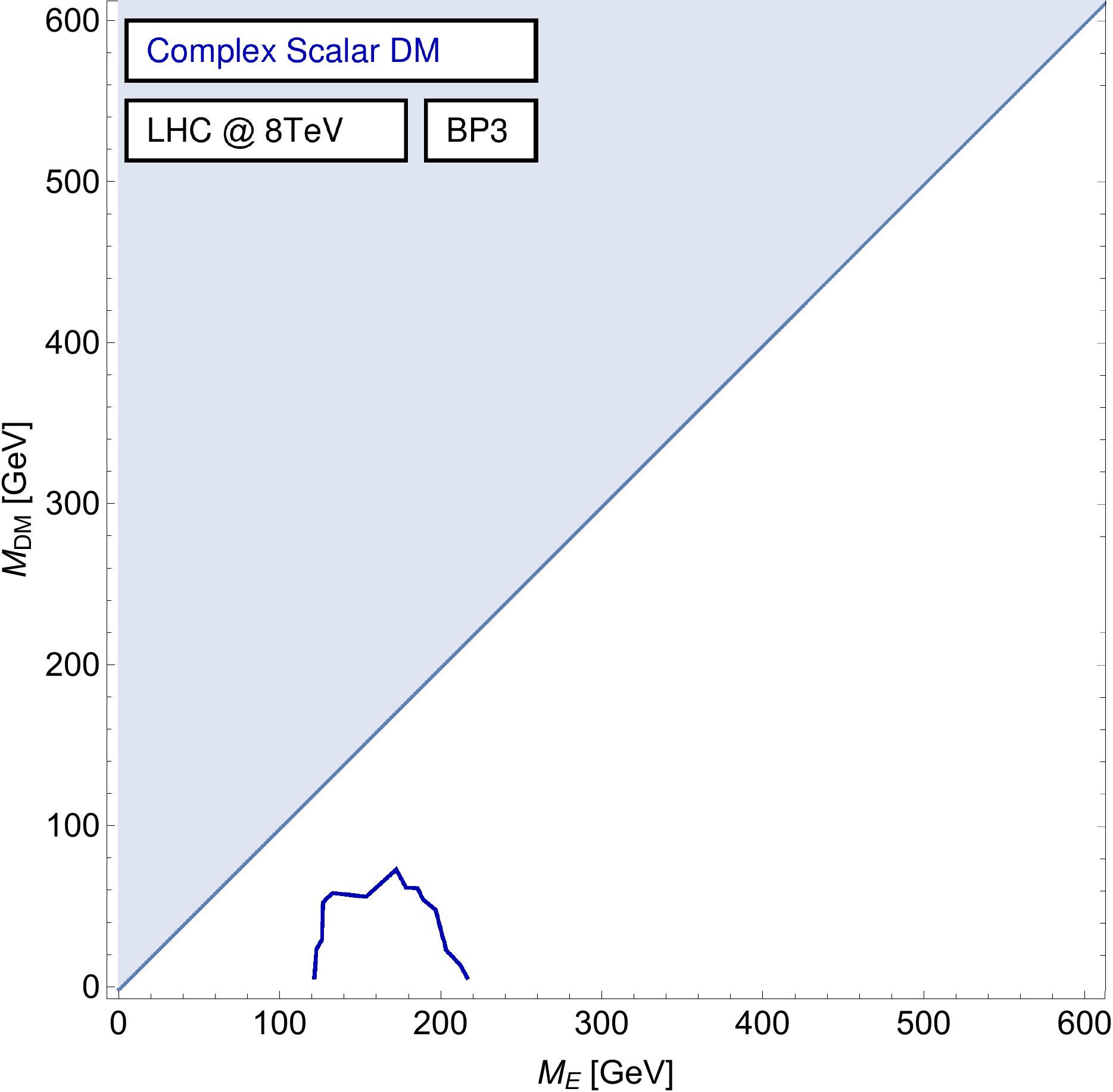}
\caption{\label{fig:LHCbounds2} LHC bounds at 8 TeV for BP3 implemented as described in the text. The bounds for the other DM scenario are assumed to be coincident  with the one of the complex scalar DM case. }
\end{center}
\end{figure}

We stress here that, as the purpose of this study is to discriminate between different DM candidates, our interest is twofold: 1) we are looking at the possibility of observing differences in the exclusion contours, and 2) we want to broadly identify the region in the ($M_E,M_{\rm DM}$) parameter space, which is allowed by collider data, in order to compare and correlate with observables from different areas (discussed in the next sections). Our results are shown as contours in the ($M_E,M_{\rm DM}$) plane in Fig.~\ref{fig:LHCbounds} and Fig.~\ref{fig:LHCbounds2}. Both the 8 and 13 TeV bounds are shown in in Fig.~\ref{fig:LHCbounds} due to the fact that the 8 TeV searches implemented in {\sc CheckMATE} appear to be more sensitive in the small mass region, while for the BP3 case the 13 TeV limits on the stau production cross section released by the CMS collaboration~\cite{CMS:2017rio} are consistent with the 8 TeV ones and thus not considered here.

\begin{figure}[ht!]
\begin{center}
\includegraphics[width=.33\textwidth]{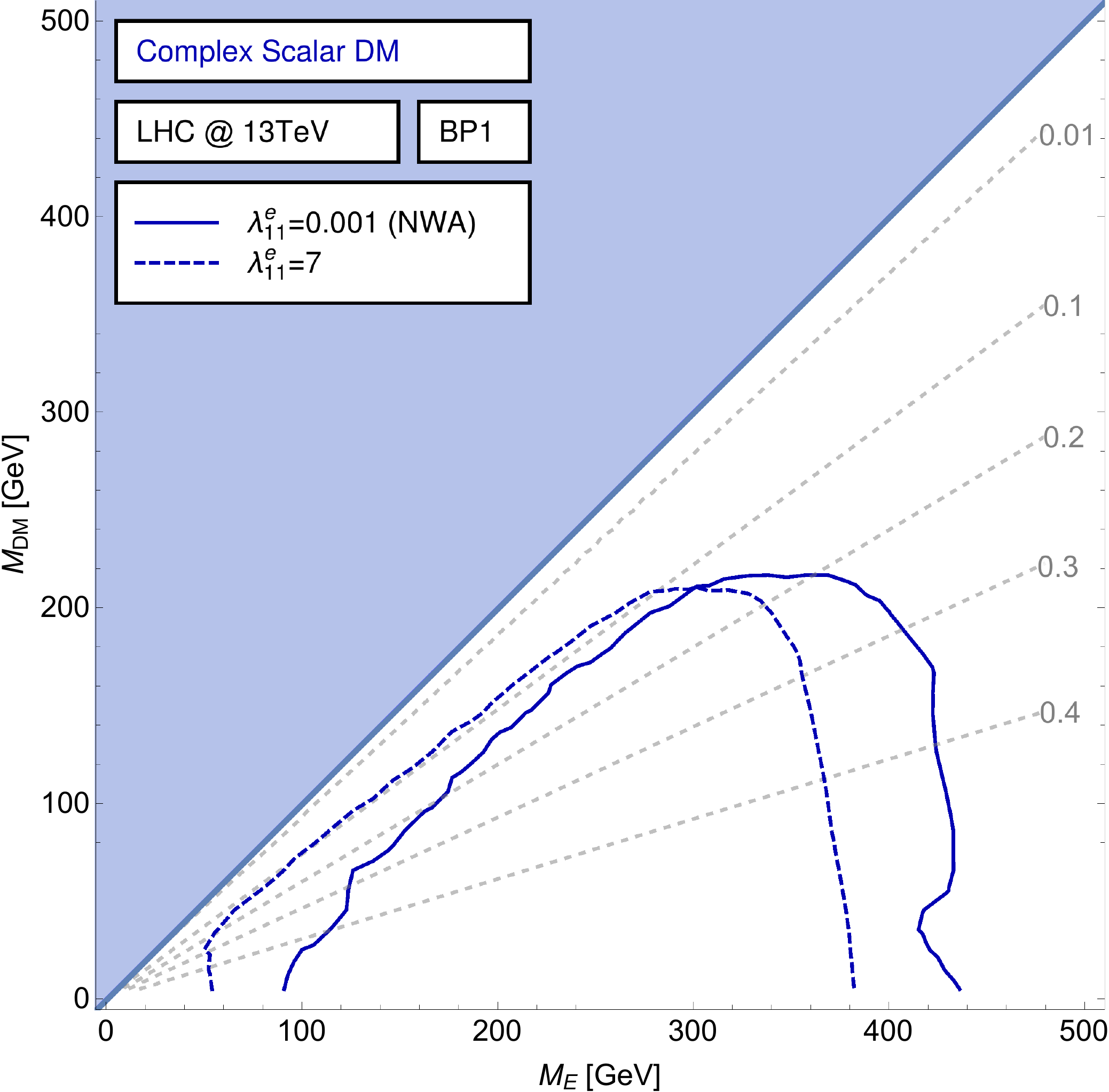}\hskip 20pt
\includegraphics[width=.33\textwidth]{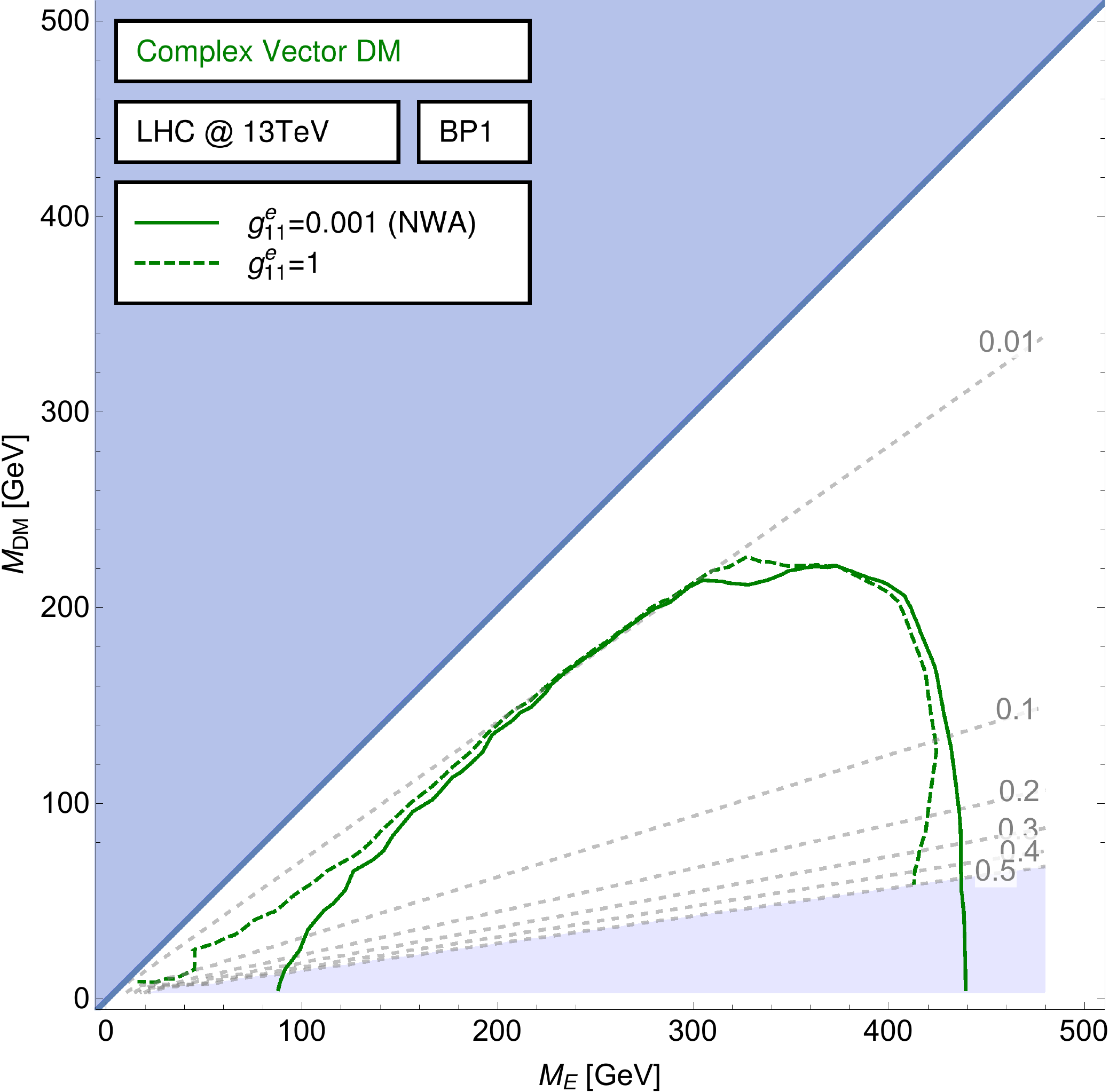}
\caption{\label{fig:LHCbounds_width} LHC bounds at 13 TeV for BP1 in the complex scalar (left) and complex vector (right) DM scenarios for different values of couplings, generating large VLL widths. Contours with constant width-to-mass ratio $\Gamma_E/M_E$ are represented by gray dashed lines for the largest coupling value. $\Gamma_E/M_E$ values larger than 50\% are not considered in this analysis.}
\end{center}
\end{figure}

We are not assuming here that the heavy lepton can decay in states other than DM and SM leptons: as we are in the context of a simplified model, the width of the heavy lepton is computed through the masses and couplings of the Lagrangians in Eq.~\eqref{eq:LagSingletDMS} or Eq.~\eqref{eq:LagSingletDMV} and we have considered a coupling value which is small enough for the VLL to be in the NW Approximation (NWA) regime whether it decays to scalar or vector DM. We have checked, however, that results for finite width are not sizably different. The exclusion region is slightly deformed and tend to exclude the small VLL and DM mass region, but the qualitative results are basically the same. When comparing the LHC results with other observables, the marginal role of colliders in the exclusion of such scenarios makes a detailed analysis of the large width regime not essential for our purposes, and we simply show how the excluded region changes when considering larger coupling values for BP1 in the complex scalar and vector DM scenarios, in Fig.~\ref{fig:LHCbounds_width}.

With the considered subset of searches, through DY pair production of VLL in the NW regime it is not possible to distinguish a scalar DM candidate from a vectorial one. This result is however expected since by factorising VLL production and decay in the NWA, the angular correlations between opposite charge VLLs are neglected. Not considering fluctuations due to Monte Carlo (MC) statistics, the limits on VLL masses are around 400 GeV in the light DM regime for all BPs, except BP3 which is in the 200 GeV region. The region in which the mass gap between VLL and DM is small is still allowed, except for BP1 in the small mass  region.

\subsection{Cosmological Data}

\subsubsection{Relic Density} \label{sec:relic}

The relic density of DM, $\Omega_{\rm DM}$, represents the relative quantity of DM in the Universe and in our scenario is determined by the annihilation cross section of the two \Z odd particles, VLL or DM candidate, into SM particles. 

\begin{figure}[ht!]
\centering
\includegraphics[width=.8\textwidth]{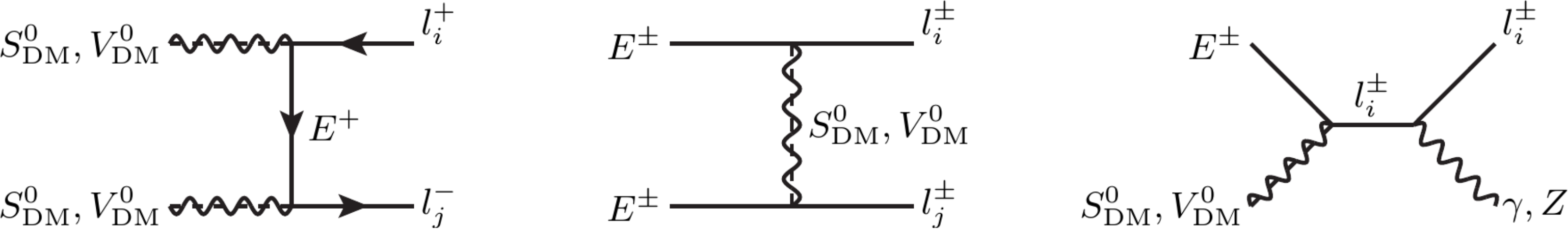}
% \vspace{-6mm}
\caption{\label{fig:RelicFeynman} Left: topology for the annihilation of two DM particle into two SM leptons. Centre: topology for the annihilation of two VLL into two SM leptons. Right: topology for the annihilation of VLL and DM into SM lepton and gauge boson. Notice the absence of neutrino + $W$ final state, due to the right-handed chirality of the coupling between singlet VLL and SM leptons.}
\end{figure}

If the mass gap between VLL and DM is not too small, the dominant topology is represented by a $t$-channel annihilation of two DM candidates into two SM leptons. When the masses of the VLL and  DM approach the degenerate region, topologies with the annihilation of two VLLs or annihilation of VLL and DM become dominant. The dominant topologies in the two parameter space regions are represented in Fig.~\ref{fig:RelicFeynman}.

\begin{figure}[ht!]
\centering

\includegraphics[width=.32\textwidth]{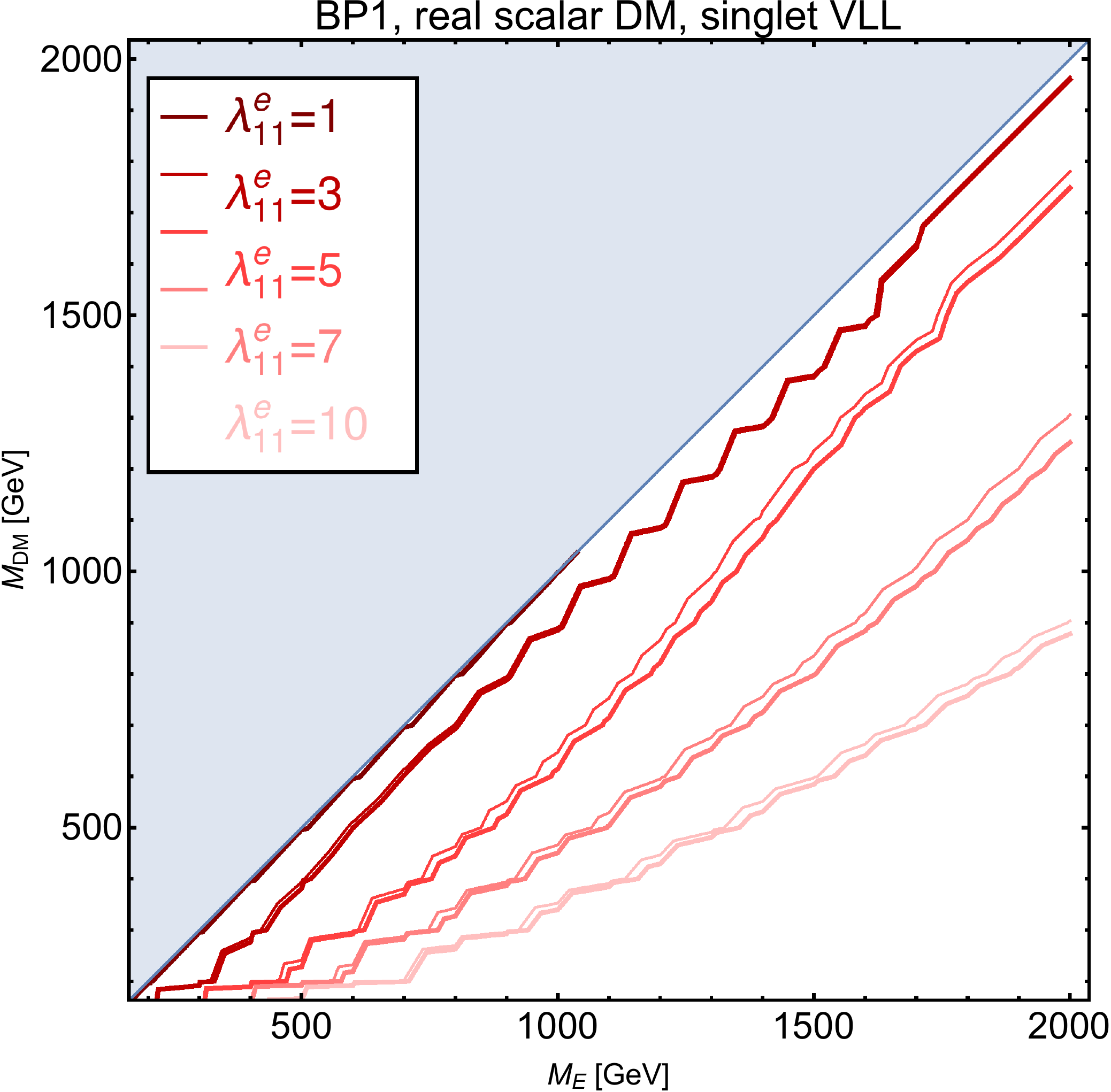}
\includegraphics[width=.32\textwidth]{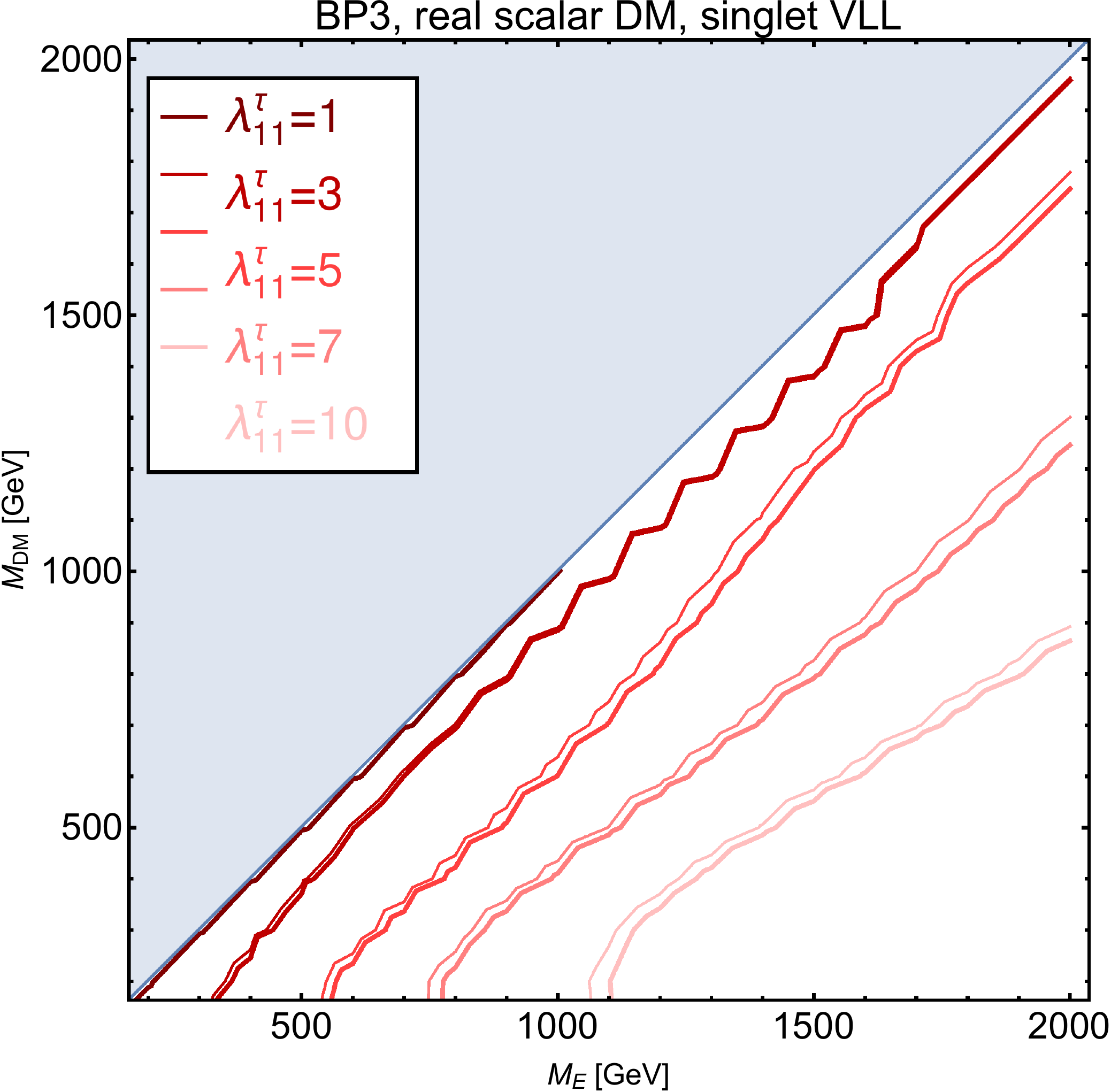}
\includegraphics[width=.32\textwidth]{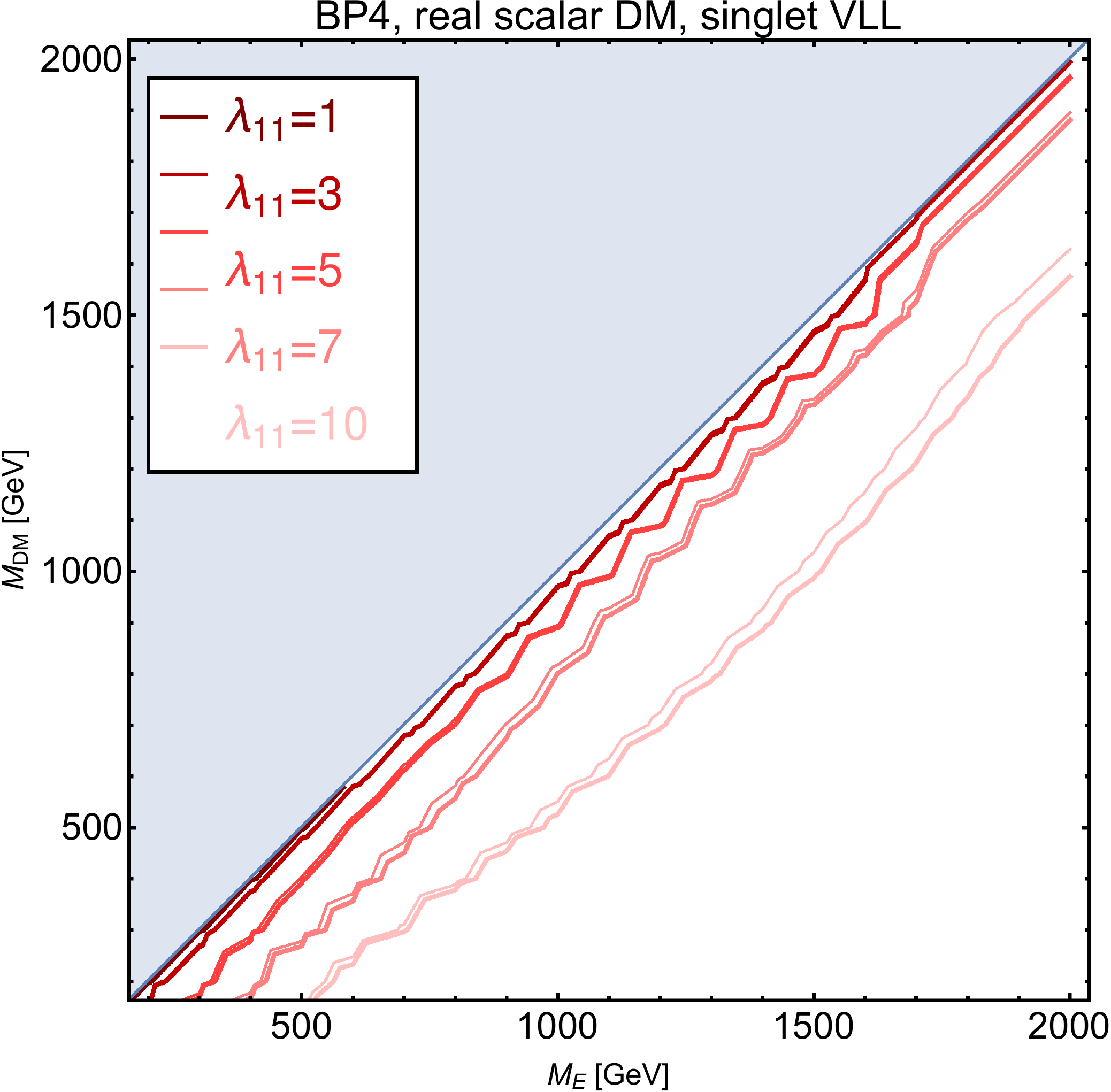} \\ \vspace{3mm}

\includegraphics[width=.32\textwidth]{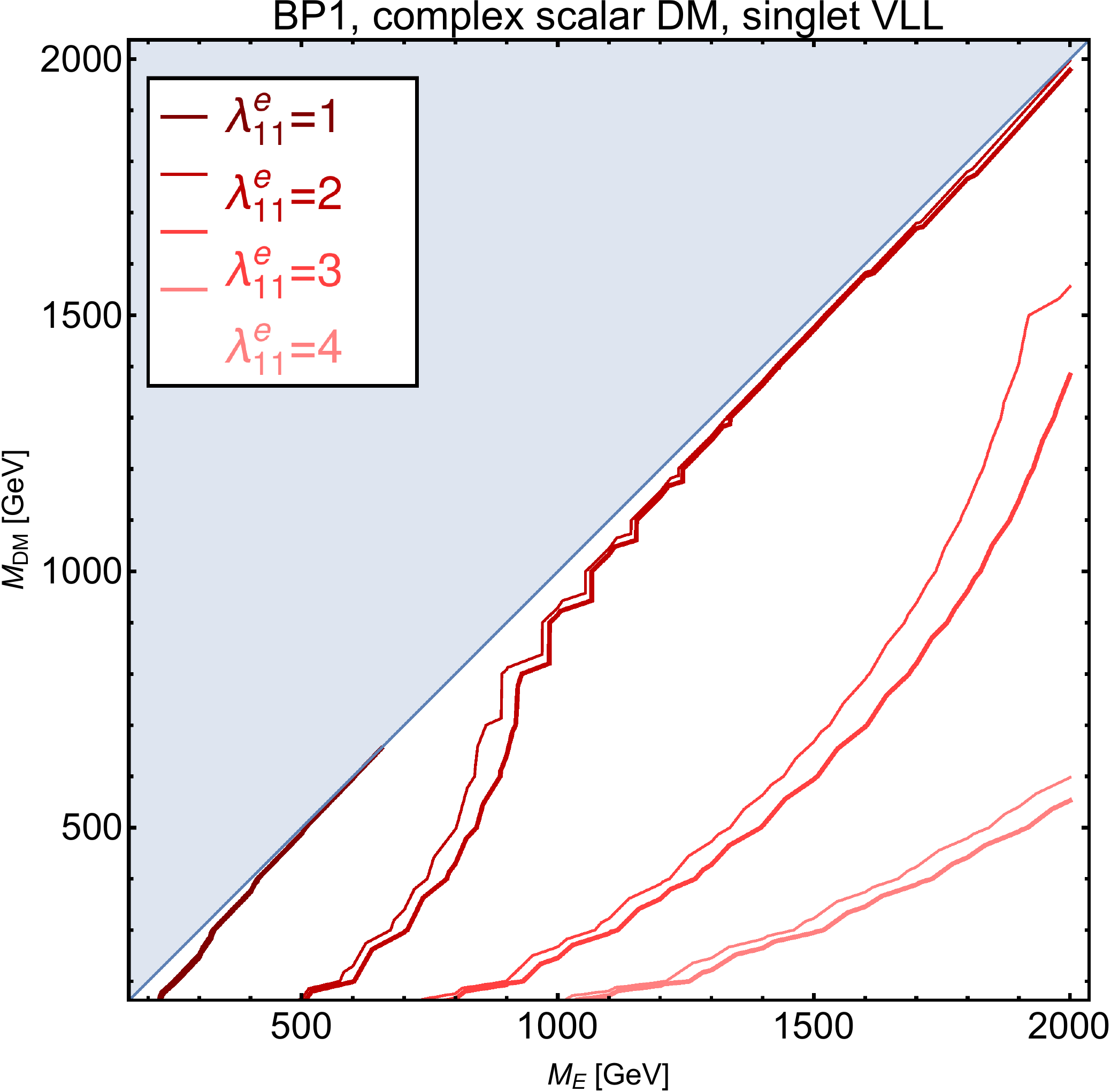}
\includegraphics[width=.32\textwidth]{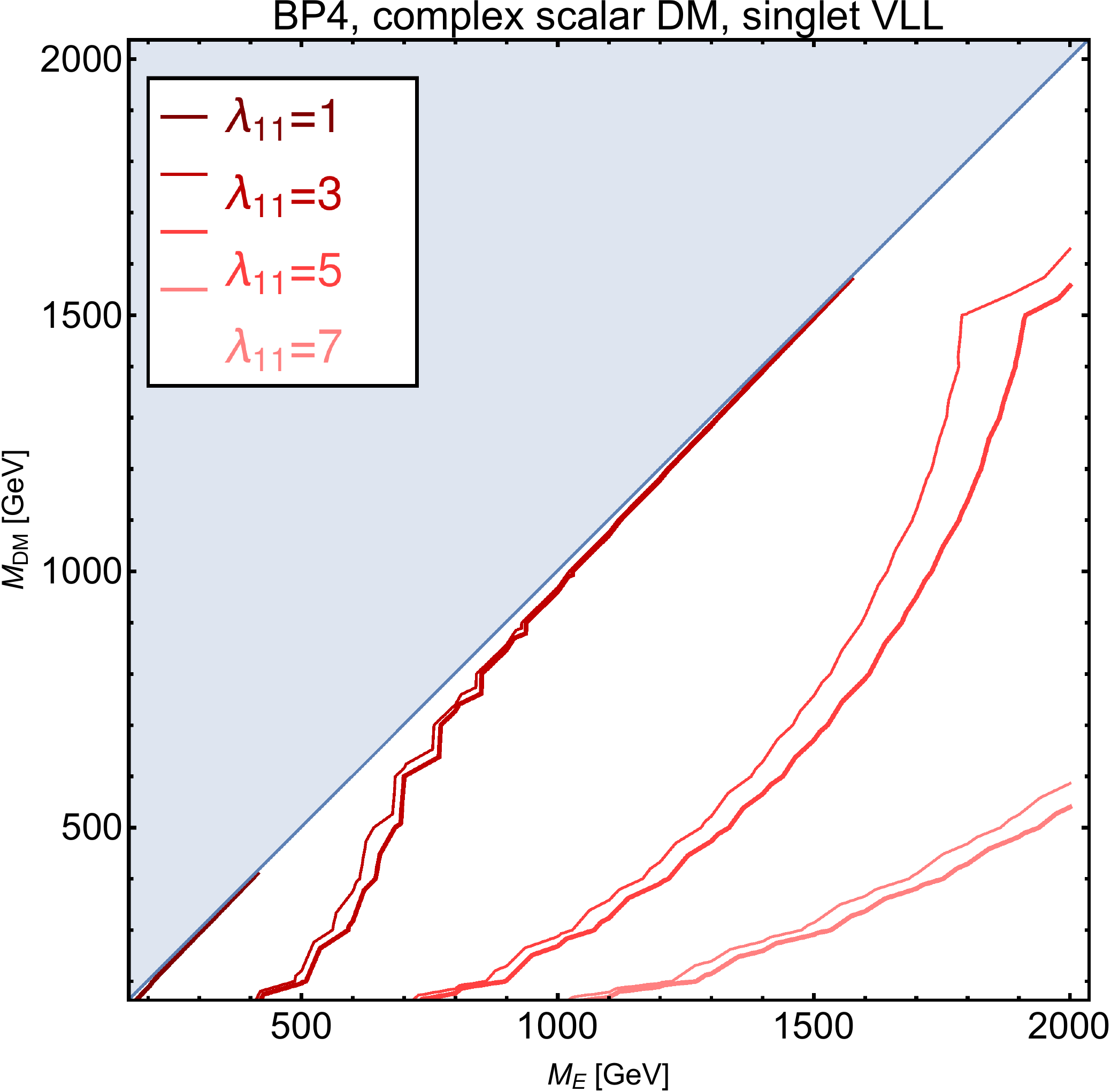} 
\caption{\label{fig:RelicScalar} Relic density constraints for scalar DM (real in the first row, complex in the second row) for the four BPs (when BP1, BP2 or BP3 have a qualitatively analogous behaviour only BP1 is shown). The different colour lines correspond to different values of the coupling. Thick lines represent the upper limit, thin lines the lower limit. The excluded region is for values of $\Omega_{\rm DM}$ above the upper limits, {\emph{i.e.}} on the right of the thick line.}
\end{figure}

\begin{figure}[ht!]
\centering

\includegraphics[width=.32\textwidth]{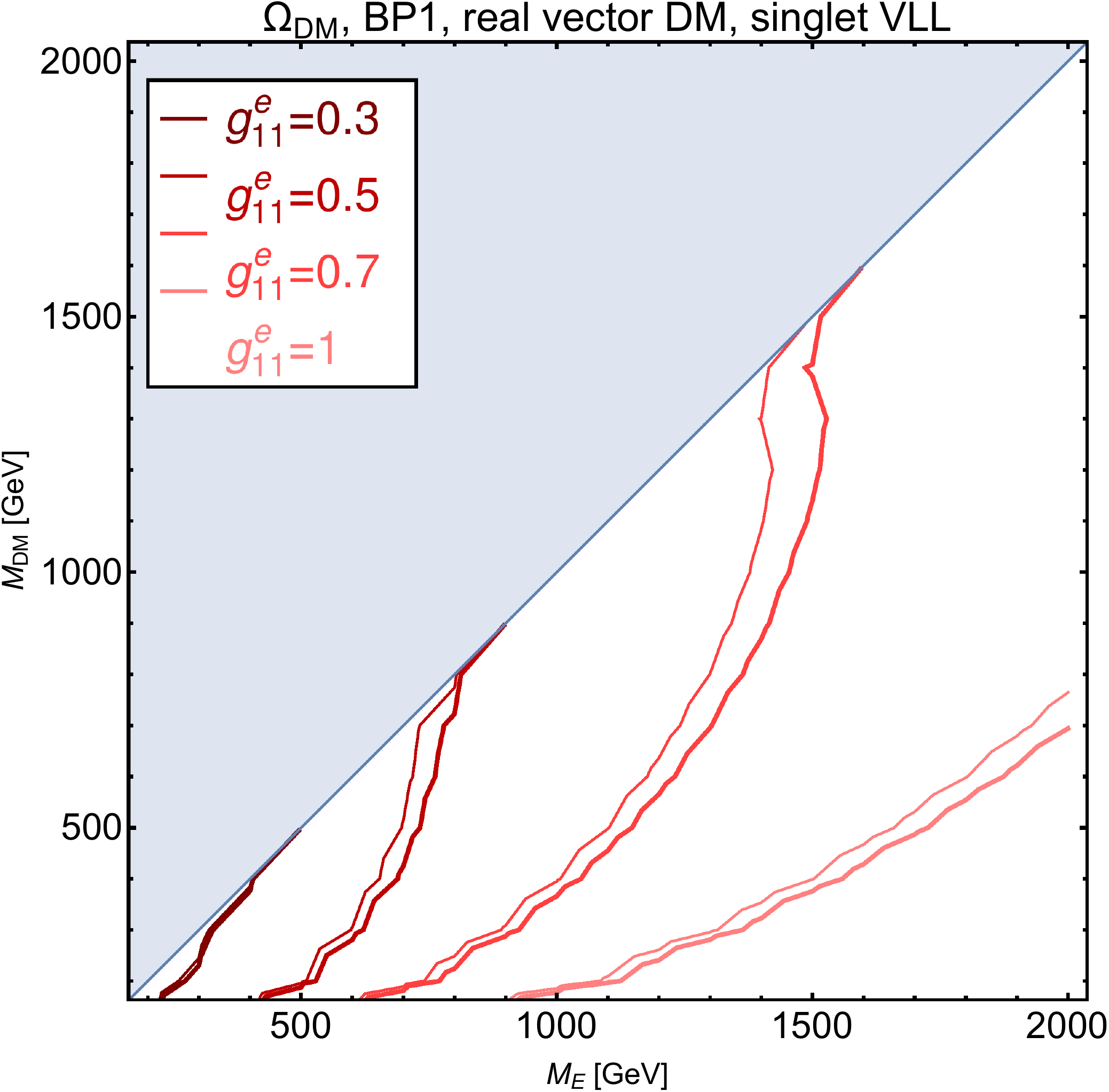}
\includegraphics[width=.32\textwidth]{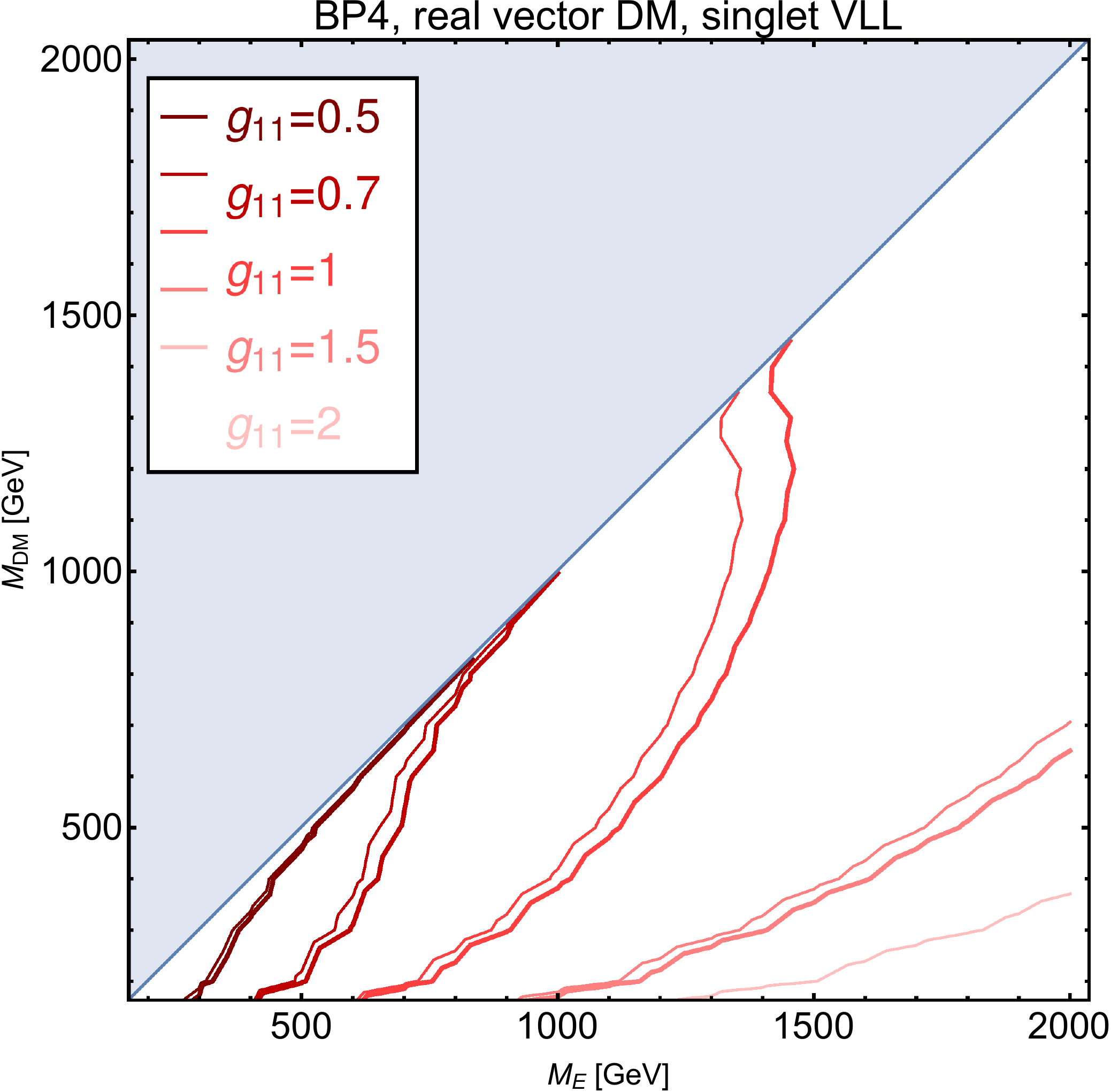}

\includegraphics[width=.32\textwidth]{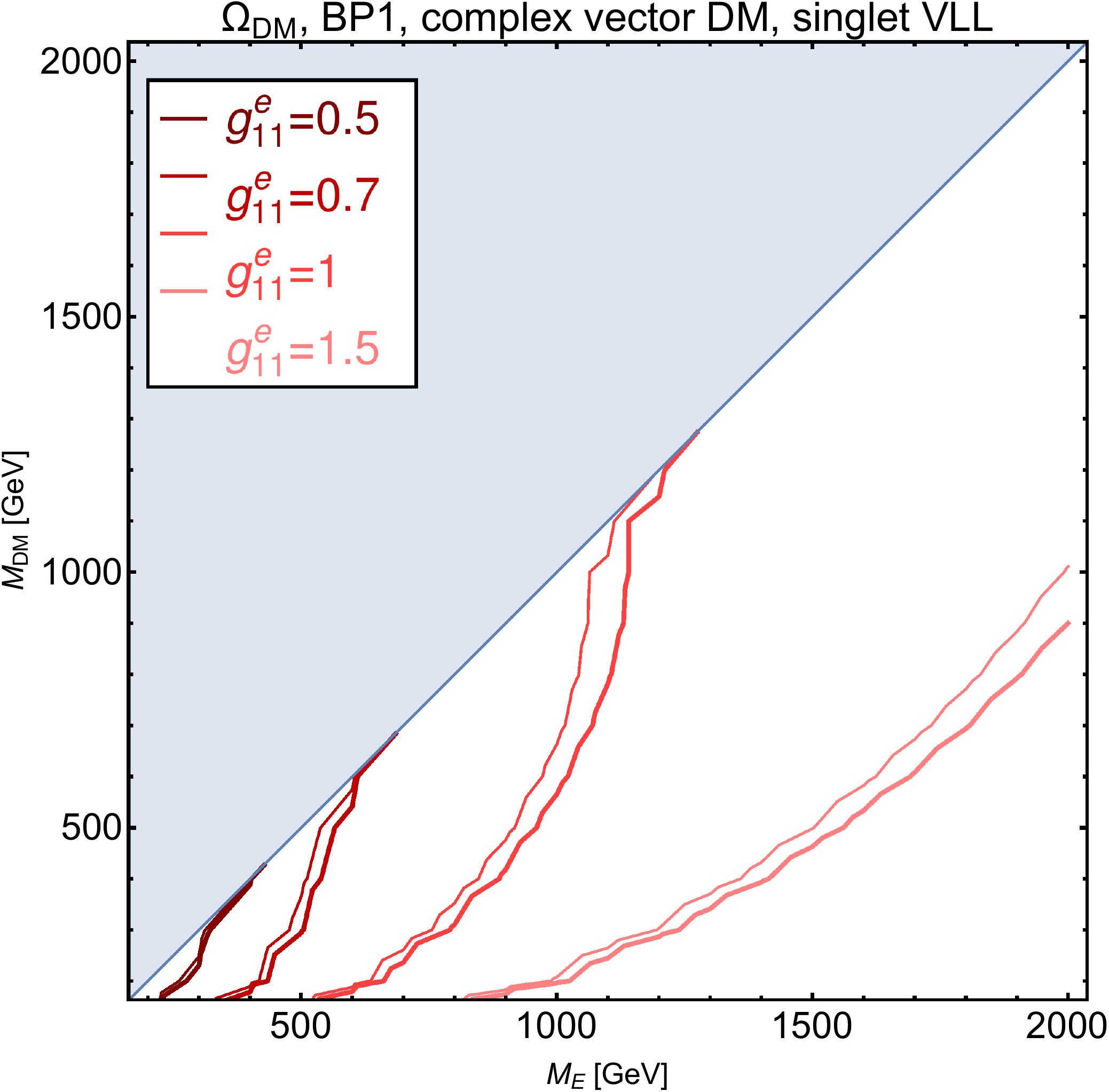}
\includegraphics[width=.32\textwidth]{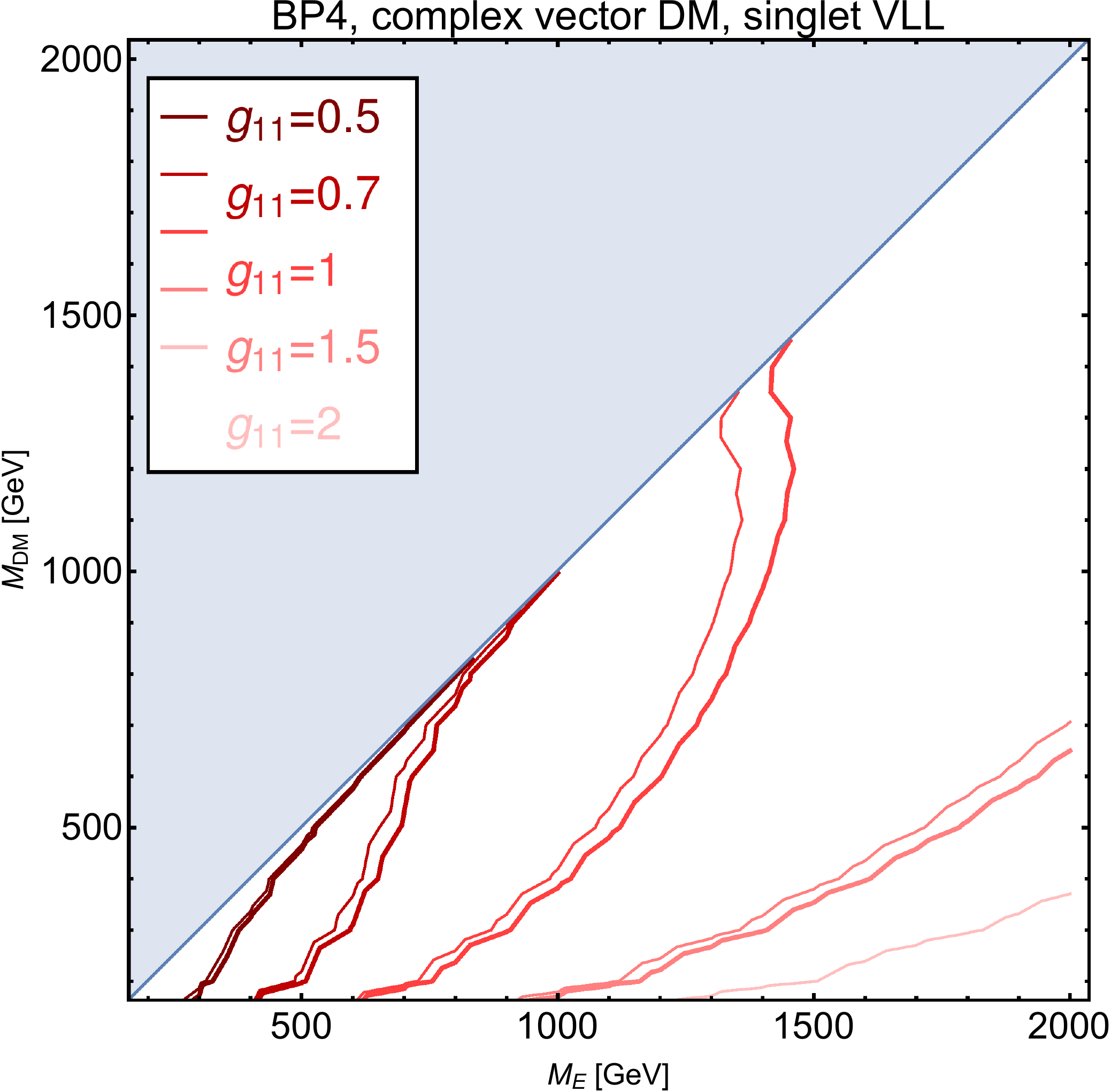} \\ \vspace{3mm}

\caption{\label{fig:RelicVector} Same as Fig.~\ref{fig:RelicScalar}, but for vector DM.}
\end{figure}

We have numerically computed the value of the relic density through the code {\sc micrOMEGAs} \cite{Belanger:2010pz,Barducci:2016pcb}. The obtained results depend on three parameters: the masses of the particles (VLL and DM) and the coupling strength. By doing a scan over these parameters for each BP and by comparing the results to the experimental value $\Omega_{\rm DM} = 0.1198 \pm 0.0026$ \cite{Agashe:2014kda} we can determine excluded regions in the parameter space, shown in Figs.~\ref{fig:RelicScalar} and \ref{fig:RelicVector}. 

We observe that the results for a separate coupling to electron (BP1), muon (BP2) and tau (BP3) are always qualitatively analogous. The different combinatorics associated with the real or complex DM scenarios, however, produces a factor 2  larger annihilation cross section in the complex DM case; for this reason the couplings needed to satisfy the relic density constraint are lower for complex DM, regardless of DM spin. In the light DM region the cross section becomes independent of the DM mass: the point at which this regime is achieved depends on the BP and on the DM spin and reality condition. Within the range of the plots, however, the region where the relic density bound becomes independent of the DM mass can only be seen in the real scalar DM scenarios of BP3 and 4.

The almost degenerate region is the hardest to exclude while the region of low DM mass and high VLL mass can only be allowed by increasing the value of the coupling; for scenarios with real scalar DM, models with a sufficiently large mass splitting are excluded even for couplings as large as 10. This is due to the fact that, if the VLL mass is much larger than the DM one, it will require a very energetic collision to annihilate two DM particles into two leptons with a VLL in $t$-channel, so the quantity of DM will easily become over-abundant.

It is interesting to notice that, for large parts of the parameter space, a relic density which determines at least an under-abundance of DM requires large values of the couplings, which in turn affect the width of the VLL. However, due to the fact that the VLL propagates in the $t$-channel and therefore has negative squared momentum, the imaginary part of the propagator is identically zero: for this reason the width of the VLL has no effect in the determination of the relic density bound. 

Looking at the influence of the DM spin we see that for a same value of the coupling the exclusion is much larger in the scalar case with respect to the vector one: for a VLL decaying to real scalar DM almost all the parameter space is excluded for any value of the coupling below 1, and a very high value of the coupling is needed to open a larger part of the parameter space. In comparison in the case of real vector DM a coupling of 0.5 already allows VLL with mass up to 800 GeV and a coupling $g_{11}\gtrsim$2 does not exclude anything in the mass region we consider in Figs.~\ref{fig:RelicVector}. This difference can be explained considering the amplitude of the $t$-channel process with propagation of the VLL, which is dominant for large mass splitting:
\begin{equation}
 \mathcal{A}_{S_{\rm DM}}^{t,E} = \bar u(l_j) (i \lambda_{11}^j P_R) \frac{\diagup\hskip -8pt k_E + M_E}{k_E^2 - M_E^2} (i \lambda_{11}^j P_L) u(l_i) \quad\text{and}\quad \mathcal{A}_{V_{\rm DM}}^{t,E} = \bar u(l_j) (i g_{11}^j \gamma_\mu P_R) \frac{\diagup\hskip -8pt k_E + M_E}{k_E^2 - M_E^2} (i g_{11}^j \gamma_\nu P_R) u(l_i) \epsilon^\mu_{V_{\rm DM}} \epsilon^\nu_{V_{\rm DM}}
\end{equation}
When squaring such amplitudes to obtain the cross section one obtains
\begin{equation}
 |\mathcal{A}_{S_{\rm DM}}^{t,E}|^2 \propto 2 M_E^2 m_l^2 \quad\text{and}\quad |\mathcal{A}_{V_{\rm DM}}^{t,E}|^2 \propto 32 M_E^2 m_l^2
\end{equation}
Such result, explains why larger couplings are needed to reach the observed bound for scalar DM.

\subsection{Flavour Data}

\subsubsection{$g-2$ of Electron and Muon} \label{sec:g-2}

\begin{figure}[ht!]
  \begin{center}
%     \begin{picture}(90,100)(0,0)
%       \SetWidth{1.0}
%       \SetColor{Black}
%       \Photon(40,80)(40,60){2}{4.5}
%       \Text(40,82)[cb]{\small{$\gamma$}}
%       \Line[arrow](20,40)(40,60)
%       \Text(23,55)[cc]{\small{$L$}}
%       \Line[arrow](0,20)(20,40)
%       \Text(-2,18)[rt]{\small{$e,\mu$}}
%       \Line[arrow](40,60)(60,40)
%       \Text(58,55)[cc]{\small{$L$}}
%       \Line[arrow](60,40)(80,20)
%       \Text(82,18)[lt]{\small{$e,\mu$}}
%       \Line[dash](20,40)(60,40)
%       \Photon[arrow](20,40)(60,40){2}{4}
%       \Text(40,36)[ct]{\small{$S^0_{\rm DM}, V^0_{\rm DM}$}}
%     \end{picture}
%     \epsfig{file=GM2topologies,width=.3\textwidth}
\includegraphics[width=.3\textwidth]{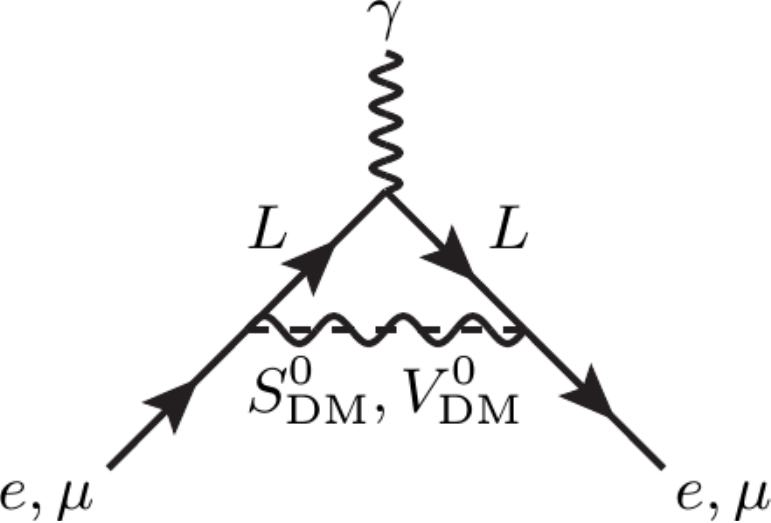}
  \end{center}
\caption{\label{fig:g2Feynman} Topology for $g-2$ of electron and muon with the contribution of the VLL and DM candidate states.}
\end{figure}

A stringent bound on the couplings of the DM particle and on the heavy vector-like fermions is given by the measurement of the 
electron and muon anomalous magnetic moments.  The diagrammatic contribution to the anomalous magnetic moment is given in 
Fig.~\ref{fig:g2Feynman}. In the following we shall use the present experimental values to estimate the bounds on the Scalar and Vector singlet DM
particle and on the heavy vector-like lepton mediator. These limits should be taken as an indication for the simplified models we are 
considering, but one should keep in mind that in a complete model extra contributions from other new particles can contribute too and even with 
opposite sign giving rise to cancellations. 
Note also that allowing at the same time couplings to the electron and the muon can induce
extra bounds, for example from charged LFV processes which is studied in Sec.~\ref{subsec-LFV}
In the following the limits coming from the electron and the muon 
anomalous magnetic moment are considered independently. We shall see that the limits are typically quite strong.

Considering firstly the case of a \textbf{\textit{scalar DM}} singlet from Eq. (\ref{eq:LagSingletDMS}) the extra contribution to $(g-2)/2$ can be obtained from
\cite{Leveille:1977rc} or \cite{Boehm:2003hm}:
\begin{equation}
\delta a_l = \frac{m_l^2}{32 \pi^2} \int_0^1 dx \frac{(\lambda^f_{11}+\lambda^f_{21})^2 (x^2(1+M_E/m_l)-x^3)+(\lambda^f_{11}
-\lambda^f_{21})^2 (x^2(1-M_E/m_l)-x^3)}{m_l^2 x^2 + (M_E^2 - m_l^2) x + M_{\rm DM}^2 (1 - x)}
\end{equation}
where $m_l$ is the mass of the light lepton (electron or muon), $M_E$ the mass of the VLL, and $M_{\rm DM}$ the one of the DM 
scalar particle. As the electron and muon are light, at first order in $m_l$ one obtains (we assume $M_E > M_{\rm DM} \gg m_l$):
\begin{equation}
\delta a_l \simeq\frac{m_l M_E \lambda^f_{11} \lambda^f_{21} \left(3 M_{\rm DM}^4 -4 M_{\rm DM}^2 M_E^2+M_E^4
-4 M_{\rm DM}^4 \log \left(\frac{M_{\rm DM}}{M_E}\right) \right)}{16 \pi ^2 \left(M_E^2-M_{\rm DM}^2 \right)^3}
\end{equation}   
In the limiting case in which $M_E \to M_{\rm DM}$ the previous formula reduces to
\begin{equation}
\delta a_l \simeq \frac{1}{24 \pi ^2} \frac{m_l}{M_{\rm DM}}\lambda^f_{11} \lambda^f_{21}
\label{eq:limchi} 
\end{equation} 
which shows more clearly that the suppression factor is ${m_l}/{M_{\rm DM}}$ in the small gap limit. 

In the VLL scenario, however, only one of the two couplings of Eq.(\ref{eq:LagSingletDMS}) can be allowed as the left and right handed components of the VLL belong to the same representation and can not simultaneously couple to the SM singlets and doublets. One has therefore to consider the next term in the expansion for small $m_l$:
\begin{equation}
\delta a_l \simeq\frac{m_l^2 ({\lambda^f_{11}}^2 + {\lambda^f_{21}}^2) \left(20 M_{\rm DM}^6 -39 M_{\rm DM}^4 M_E^2+ 24 M_{\rm DM}^2 M_E^4
-5 M_E^6 +12 M_{\rm DM}^4 (M_E^2-2 M_{\rm DM}^2)\log \left(\frac{M_{\rm DM}}{M_E}\right) \right)}{96 \pi ^2 \left(M_E^2-M_{\rm DM}^2 \right)^4}
\end{equation}   
which, in the limiting case $M_E \to M_{\rm DM}$, reduces to
\begin{equation}
\delta a_l \simeq \frac{7}{192 \pi^2} \left(\frac{m_l}{M_{\rm DM}}\right)^2 \left({\lambda^f_{11}}^2 + {\lambda^f_{21}}^2\right)
\label{eq:limvl} 
\end{equation}  
which shows that in the VLL scenario one extra power of $m_f/M_{\rm DM}$ suppresses $\delta a_l$ allowing for a larger $\lambda^f_{11}$ or $\lambda^f_{21}$ coupling (or smaller DM mass). \\

We next consider the case of a \textbf{\textit{vector DM}} singlet. The contribution to $(g-2)/2$ can be again extracted from
\cite{Leveille:1977rc}\footnote{Note that a small misprint is present in formula (3) of \cite{Leveille:1977rc} as a parenthesis is missing on the 
second line of that reference.}:
\begin{eqnarray}
\delta a_l &=& \frac{m_l^2}{16 \pi^2} \int_0^1 dx \left[ (g^f_{11}+g^f_{21})^2 \left((x-x^2) \left(x-2+\frac{2 M_E}{m_l}\right)
+\left(1-\frac{M_E}{m_l}\right)\frac{x^2(M_E^2-mf^2)}{2 M_{\rm DM}^2}-\frac{x^3 (M_E-m_l)^2}{2 M_{\rm DM}^2}\right)\right.\nonumber \\
&+&\left. (g^f_{11}-g^f_{21})^2 \left((x-x^2) \left(x-2-\frac{2 M_E}{m_l}\right)
-\left(1+\frac{M_E}{m_l}\right) \frac{x^2(M_E^2+mf^2)}{2 M_{\rm DM}^2}-\frac{x^3 (M_E+m_l)^2}{2 M_{\rm DM}^2}\right)\right] \times \nonumber \\
&\times& \left[ x^2 m_l^2 +(1-x) M_{\rm DM}^2+ x (M_E^2-m_l^2) \right]^{-1}
\end{eqnarray}
At the first order in the expansion for small $m_l$ we obtain:
\begin{eqnarray}
\delta a_l &\simeq& \frac{m_l M_E}{32 \pi^2 M_{\rm DM}^2 (M_E^2-M_{\rm DM}^2)^3} 
\left( 8 g^f_{11} g^f_{21} M_{\rm DM}^6 -3 ({g^f_{11}}^2+{g^f_{21}}^2) M_{\rm DM}^4 M_E^2
+4 (g^f_{11} -g^f_{21})^2 M_{\rm DM}^2 M_E^4 \right. \nonumber \\
&-& \left. ({g^f_{11}}^2+{g^f_{21}}^2) M_E^6 -4
({g^f_{11}}^2-8g^f_{11} g^f_{21} +{g^f_{21}}^2)M_{\rm DM}^4 M_E^2 \log{\frac{M_E}{M_{\rm DM}}} \right)
\end{eqnarray}
In the limit in which $M_E \to M_{\rm DM}$ the previous formula reduces to
\begin{equation}
\delta a_l \simeq \frac{1}{48 \pi^2} \frac{m_l}{M_{\rm DM}} \left( {g^f_{11}}^2+4g^f_{11} g^f_{21} +{g^f_{21}}^2\right)
\label{eq:limvecdm}
\end{equation}
In the VLL singlet scenario only $g_{11}^f$ is non-zero and therefore only the first term of the above expression survives. \\

\begin{figure}[htb!]
\centering
\begin{tabular}{C{.34\textwidth}|C{.68\textwidth}}
\toprule
$(g-2)_e$ & $(g-2)_\mu$ \\
\midrule
\midrule
\begin{minipage}{\textwidth}
\includegraphics[width=.32\textwidth]{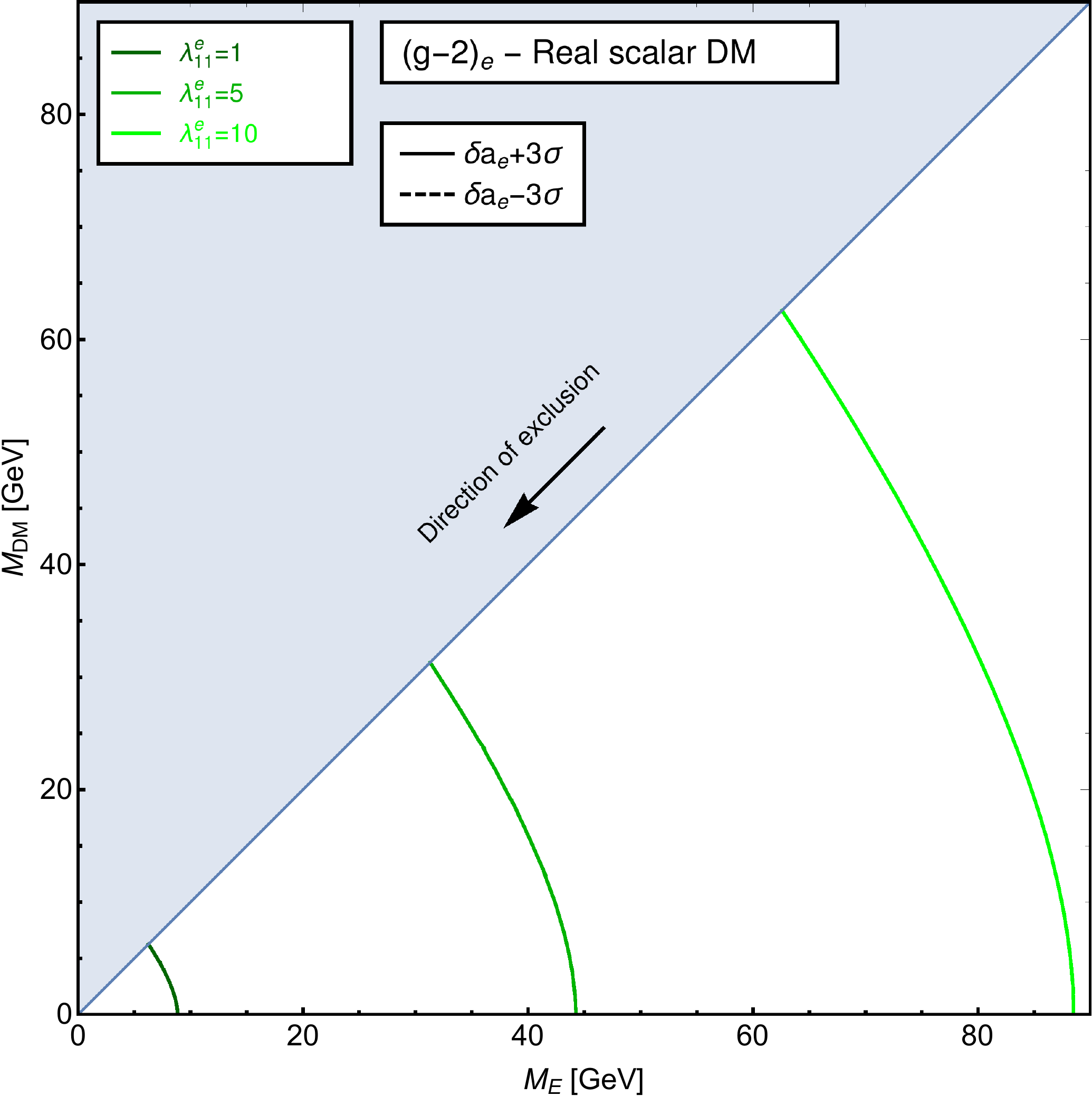} \\
\includegraphics[width=.32\textwidth]{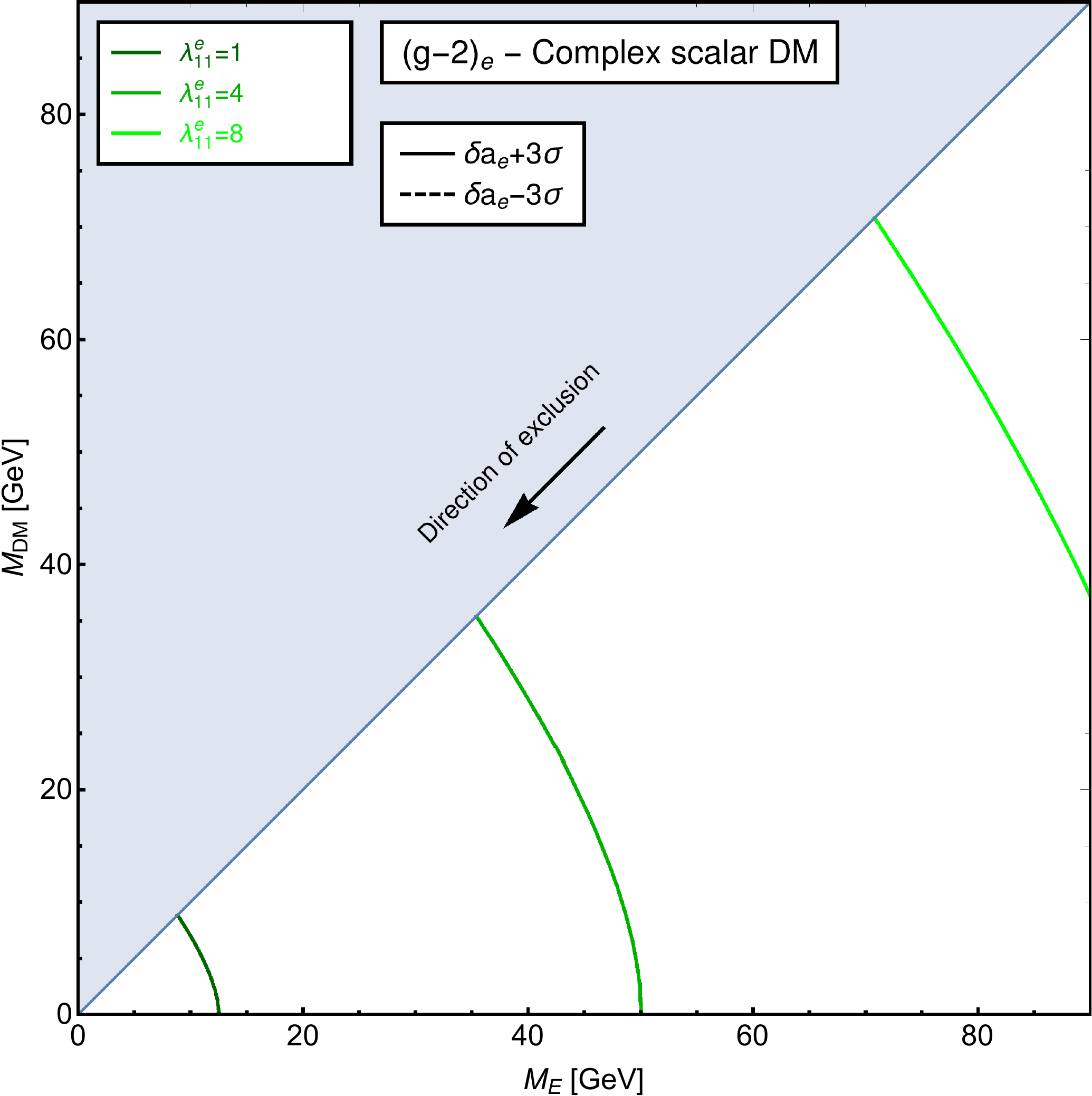}
\end{minipage}
{\small VLL and DM masses \textit{above} $\sim$100 GeV allowed for couplings $\lesssim\mathcal{O}(10)$} &
\begin{minipage}{\textwidth}
\includegraphics[width=.32\textwidth]{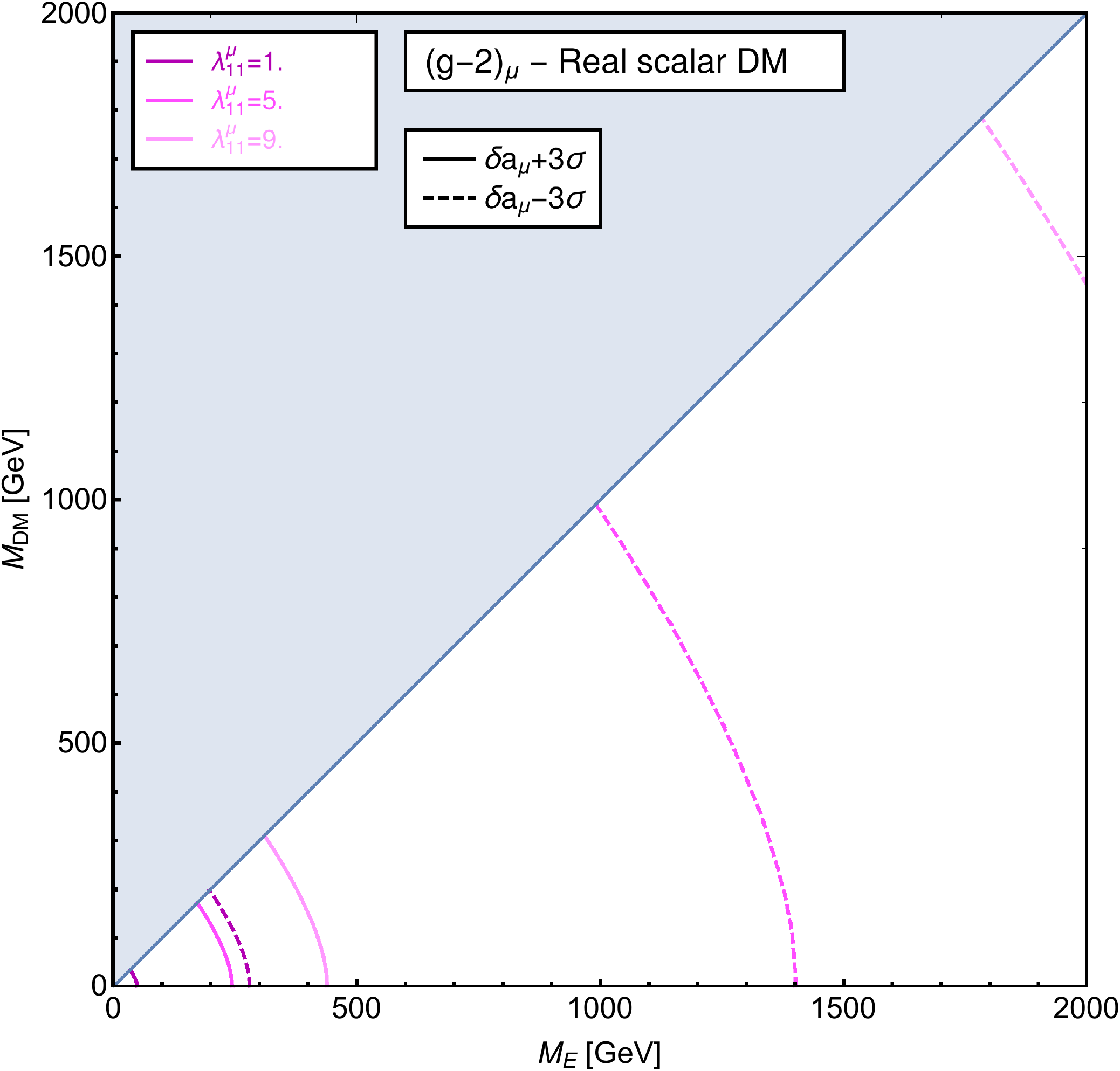}
\includegraphics[width=.32\textwidth]{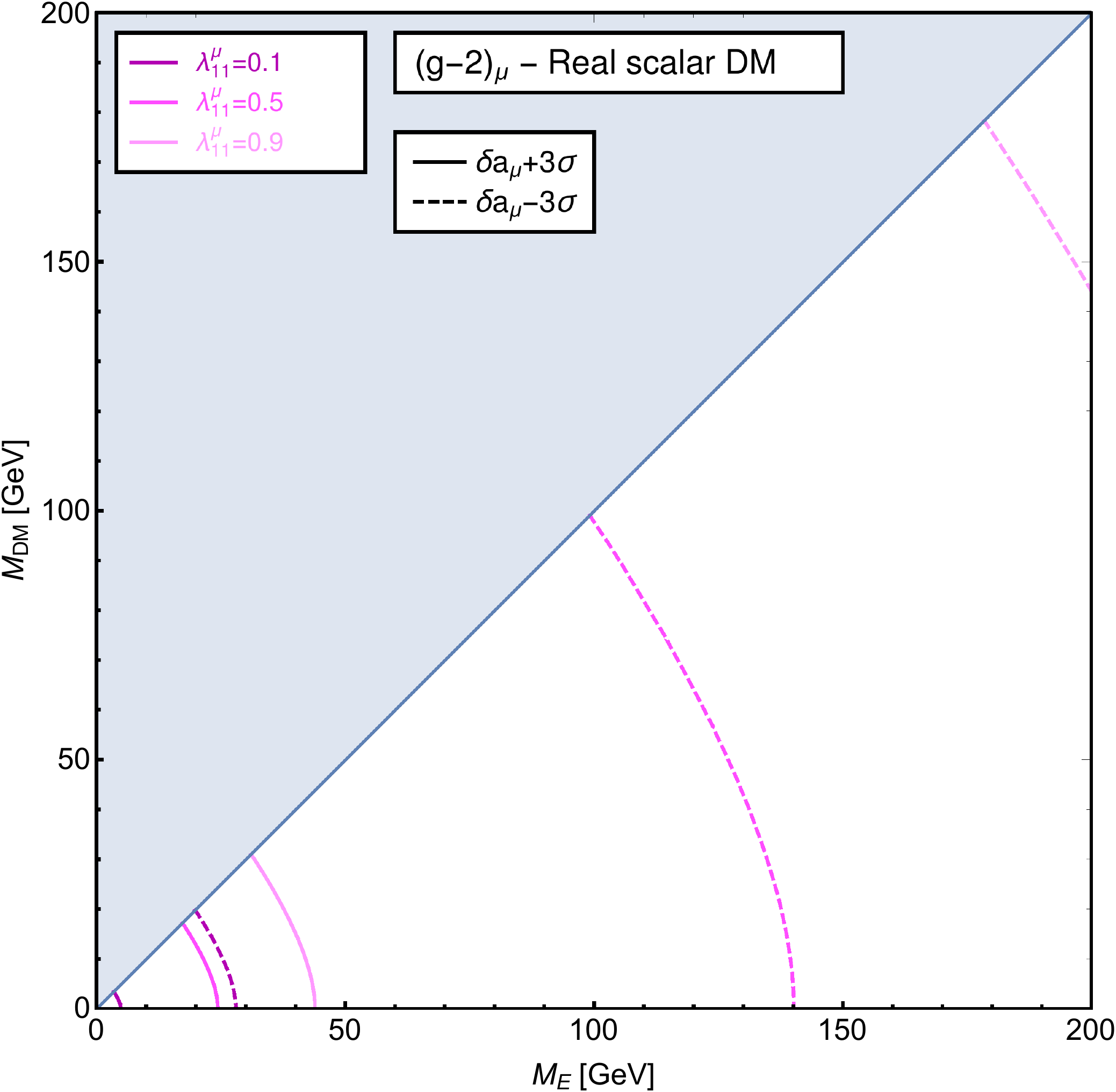} \\
\includegraphics[width=.32\textwidth]{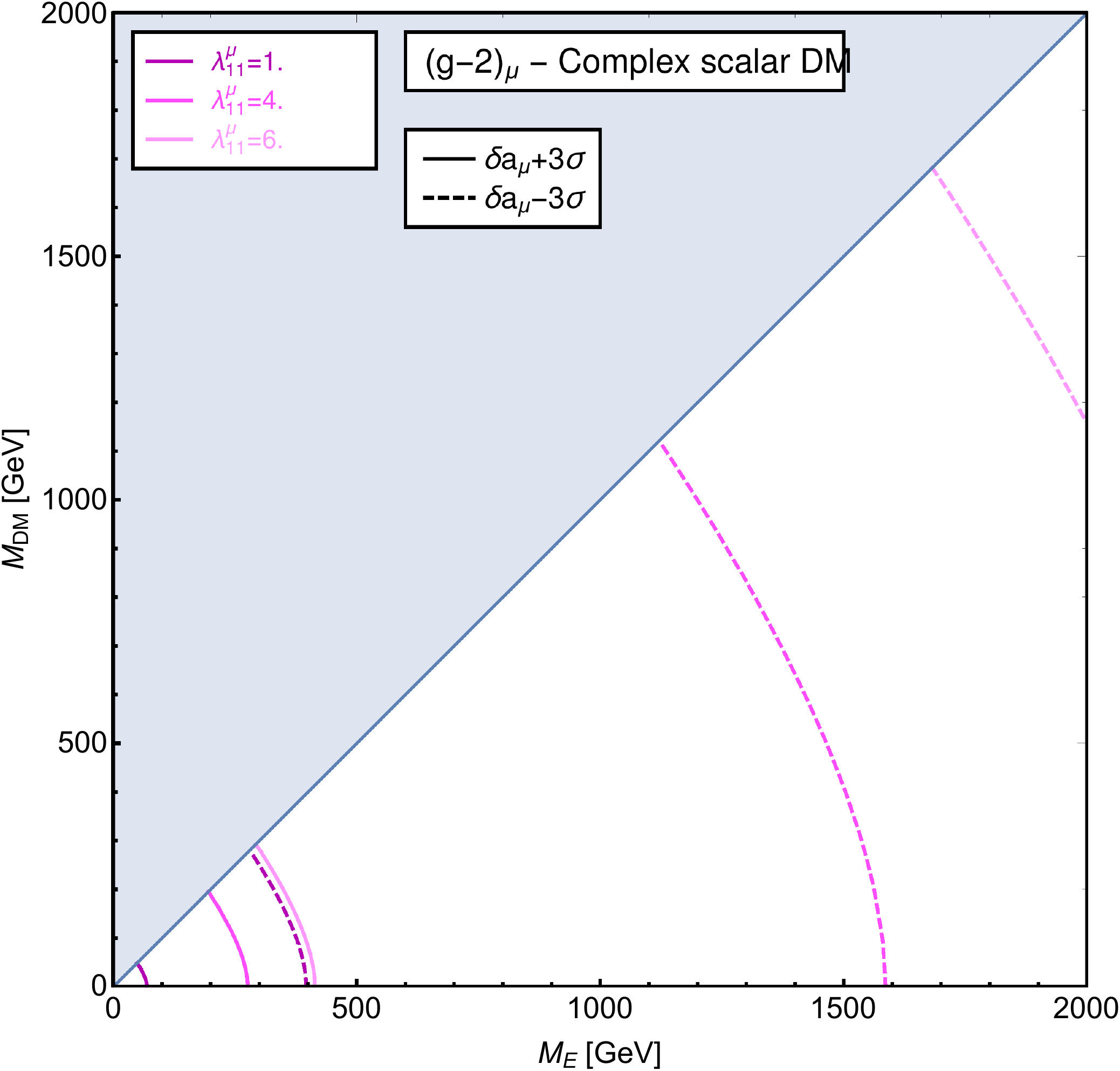}
\includegraphics[width=.32\textwidth]{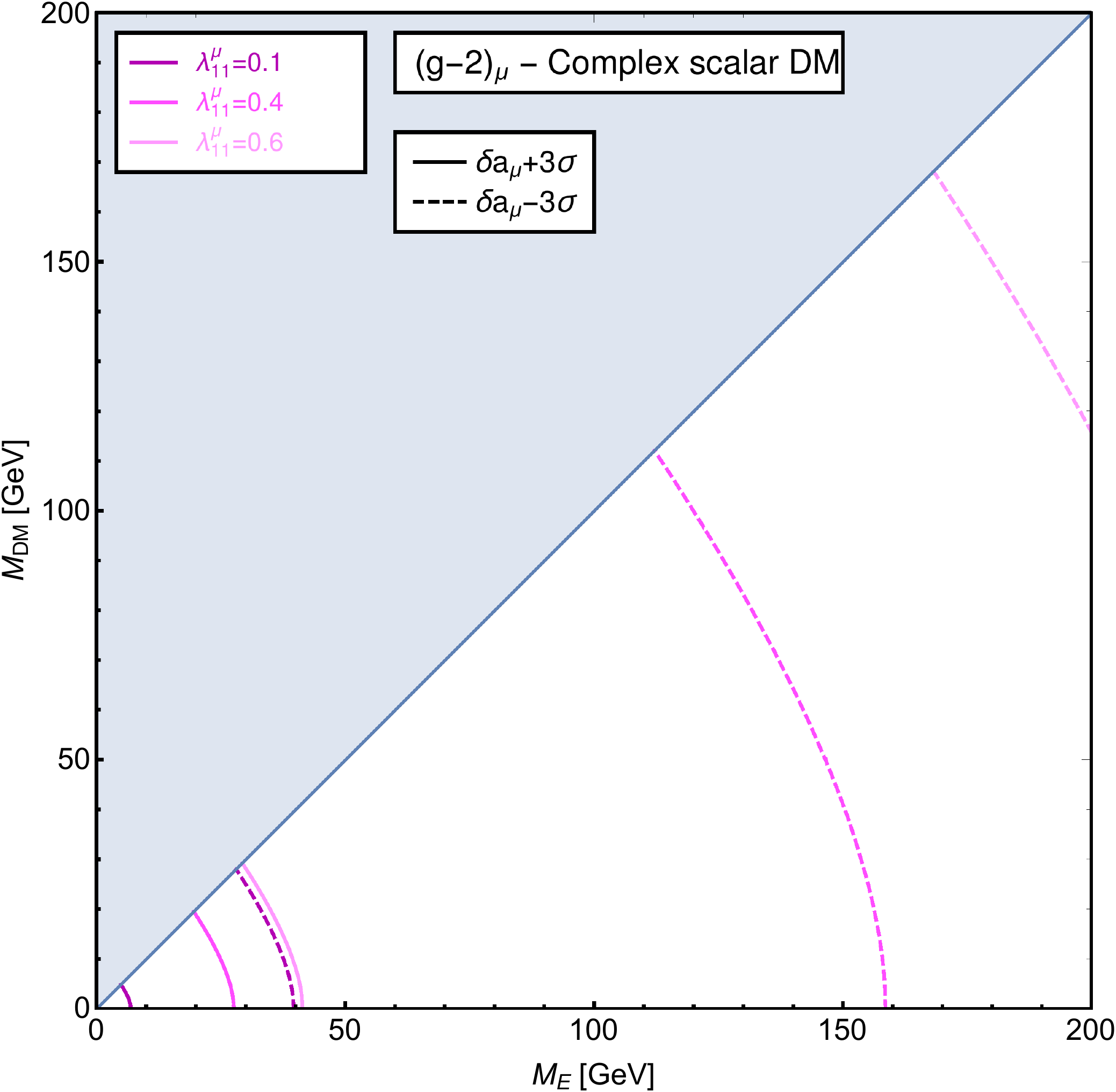}
\end{minipage}
{\small Bands of allowed VLL and DM masses are in the 100 GeV to TeV range for $\mathcal{O}(1)$ couplings and in the 1 GeV to 100 GeV range for $\mathcal{O}(0.1)$ couplings} \\
\bottomrule
\end{tabular}
\caption{\label{fig:g-2s} $(g-2)_f$ constraints for scalar DM. When only the $+3\sigma$ or the $-3\sigma$ limits apply, the direction of the excluded region is indicated by an arrow.}
\end{figure}

\begin{figure}[htb!]
\centering
\begin{tabular}{C{.68\textwidth}|C{.34\textwidth}}
\toprule
$(g-2)_e$ & $(g-2)_\mu$ \\
\midrule
\midrule
\begin{minipage}{\textwidth}
\includegraphics[width=.33\textwidth]{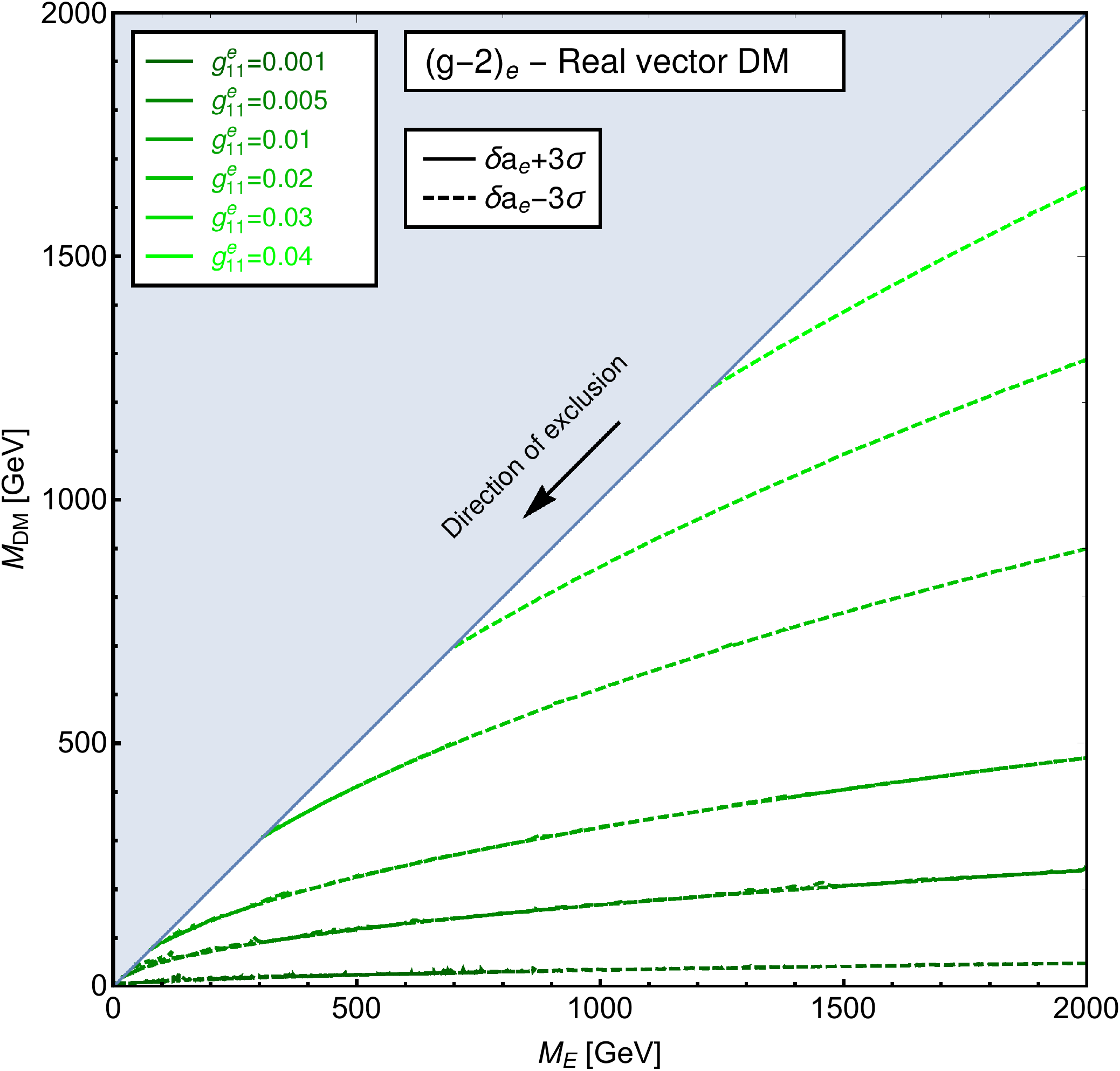}
\includegraphics[width=.33\textwidth]{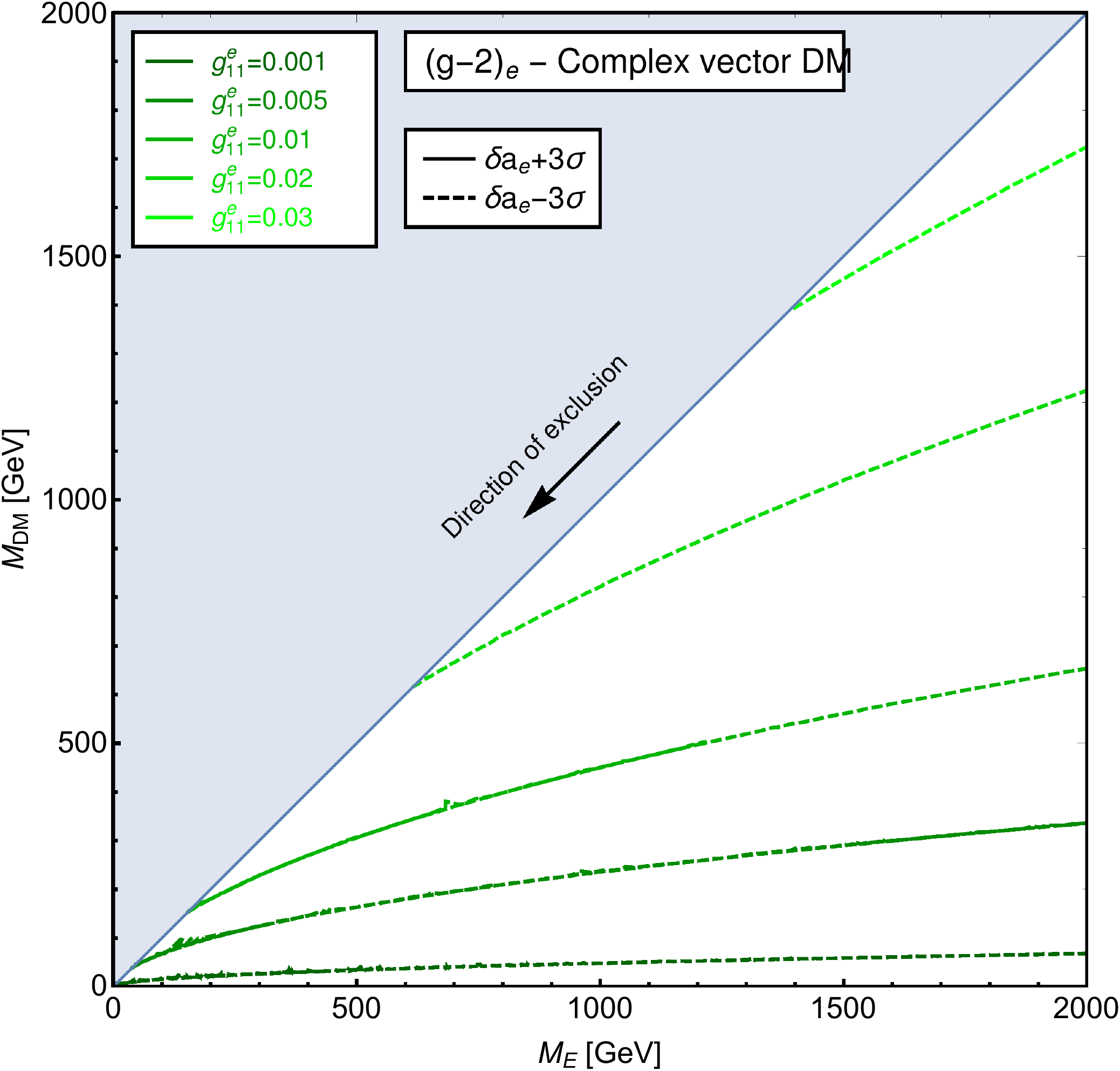}
\end{minipage}
{\small Couplings $\gtrsim\mathcal{O}(0.01)$ exclude the whole space in the considered range} &
\begin{picture}(0,0)(0,0)
\Text(0,10)[cc]{\small \begin{tabular}{c}
$|\delta a_\mu| > 3\sigma$ for all values \\ of couplings and masses
\end{tabular}}
\end{picture} \\
\bottomrule
\end{tabular}
\caption{\label{fig:g-2v} Same as Fig.~\ref{fig:g-2s}, but for vector DM.}
\end{figure}

The above results are valid for a real DM candidate. In case of complex DM, the contribution increases by a factor 2, due to fact that the VLL DM loop has to be counted twice, one for the DM particle and the other for the DM anti-particle.
To compare our results with the experimental values we will consider for the electron $\delta a_e = a_e({\rm exp}) - a_e({\rm SM}) =(1.06 \pm 0.82) \times 10^{-12}$ \cite{Aoyama:2012wj}. For the muon the anomalous magnetic moment measurement differs with the SM result by $\delta a_\mu = a_\mu({\rm exp}) - a_\mu({\rm SM}) = (25.5 \pm 8.0)\times 10^{-10} $ \cite{vonWeitershausen:2010zr}. The $(g-2)_f$ constraints can be quite stringent: the full dependence on the masses and couplings of the VLL and DM is shown in Fig.~\ref{fig:g-2s} for scalar DM and Fig.~\ref{fig:g-2v} for vector DM. 

The results from $(g-2)_f$ observables show sizeable differences in the allowed regions for given mass and coupling parameters depending on the DM spin. 

Considering $(g-2)_e$, for a scalar DM candidate (both real and complex) the bounds on the couplings always increase, if either the VLL or DM masses increase, and the dependence on the variation of the masses is analogous in all scenarios. For vector DM (real and complex), the dependence of the bound on the VLL and DM masses is largely different and with opposite sign: the bound on the coupling depends more strongly on the value of the DM mass and quite weakly on the value of the VLL mass plus it becomes stronger for increasing DM mass and decreasing VLL mass. It is important to notice that, for vector DM, couplings of order $\mathcal{O}(0.01)$ exclude the whole region below 2 TeV: this will be crucial when comparing the constraints with other observables such as relic density.

The scenario is different for $(g-2)_\mu$. In this case the $\sim 3\sigma$ tension between the theoretical and experimental values of $(g-2)_\mu$ affects the bounds on the VLL DM parameters by producing exclusion \textit{bands} instead of regions. For scalar DM, the allowed bands include larger values of VLL and DM masses as the coupling increases, and the functional dependence of the bound on the DM and VLL masses is analogous, as in the case of the electron $(g-2)_\mu$. For vector DM, in contrast, any value of the coupling for any combination of masses produces a $(g-2)_\mu$ values outside the allowed range, and more specifically, the $(g-2)_\mu$ parameter is always negative, while the experimental observation points towards a positive value within $3\sigma$: this result alone seems to indicate that scenarios with vector DM (either real or complex) and a singlet VLL coupling to muon are always excluded. Notice, however, that this is valid in the hypothesis of real couplings. If couplings are complex, the contribution to $(g-2)_\mu$ takes a different sign and allowed regions could be obtained.

\subsubsection{Lepton Flavour Violating Processes}\label{subsec-LFV}
\begin{figure}[ht!]
\begin{center}
\includegraphics[width=.7\textwidth]{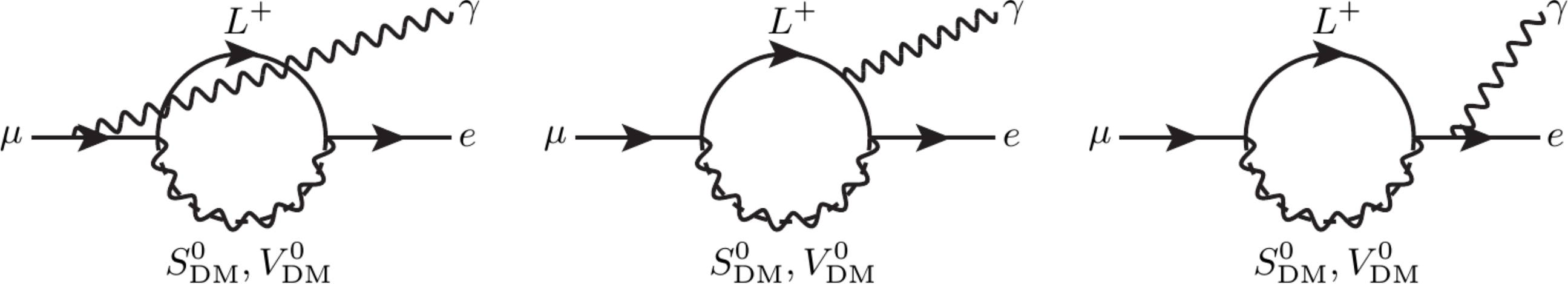}
\end{center}
\caption{\label{fig:LFV} Topologies for the LFV process $\mu\to e \gamma$ induced by the couplings between VLL, DM and the SM electron and muon.}
\end{figure}

The presence of a heavy lepton and of a DM boson interacting with the SM leptons may contribute significantly to yet unseen LFV processes, such as the process $\mu\to e \gamma$ represented in Fig.~\ref{fig:LFV}. Experimental limits on the rates of such processes pose constraints on the interactions between the new particles and SM leptons. It is important to notice that LFV limits apply only to scenarios where the VLL and the DM couple to more than one SM lepton, and therefore, considering our BPs, LFV results apply only to BP4. The analytical treatment of such processes can be found in Ref.\cite{Lavoura:2003xp} and we have exploited its results to obtain the LFV-induced bounds on the $E-{\rm DM}-l_{\rm SM}$ couplings, considering as numerical input the current bounds on the BRs from the Particle Data Group\cite{Olive:2016xmw}  reported in Tab.~\ref{tab:PDGBRLFV} for completeness.

\begin{table}[H]
\centering\begin{tabular}{cc}
\toprule
LFV process & Upper limit on BR \\
\midrule
$\mu\to e \gamma$    & $4.2 \times 10^{-13}$ \\
$\tau\to e \gamma$   & $3.3 \times 10^{-8}$  \\
$\tau\to \mu \gamma$ & $4.4 \times 10^{-18}$ \\
\bottomrule
\end{tabular}
\caption{\label{tab:PDGBRLFV} Upper limits on the BRs for $l_i\to l_j \gamma$ LFV processes from \cite{Olive:2016xmw}.}
\end{table}

\begin{figure}[ht!]
\centering
\begin{minipage}[t]{.33\textwidth}
\includegraphics[width=\textwidth]{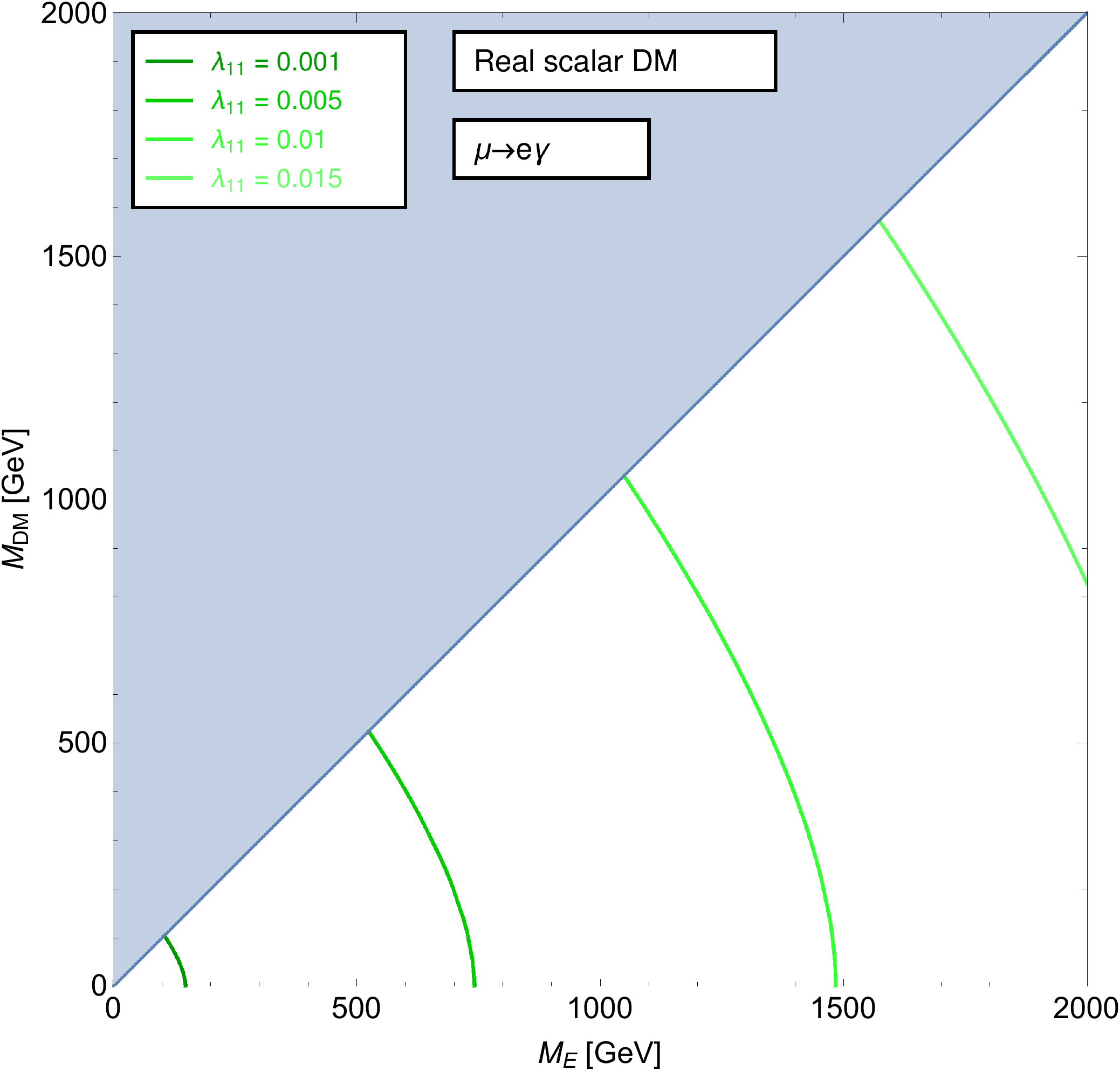}
\includegraphics[width=\textwidth]{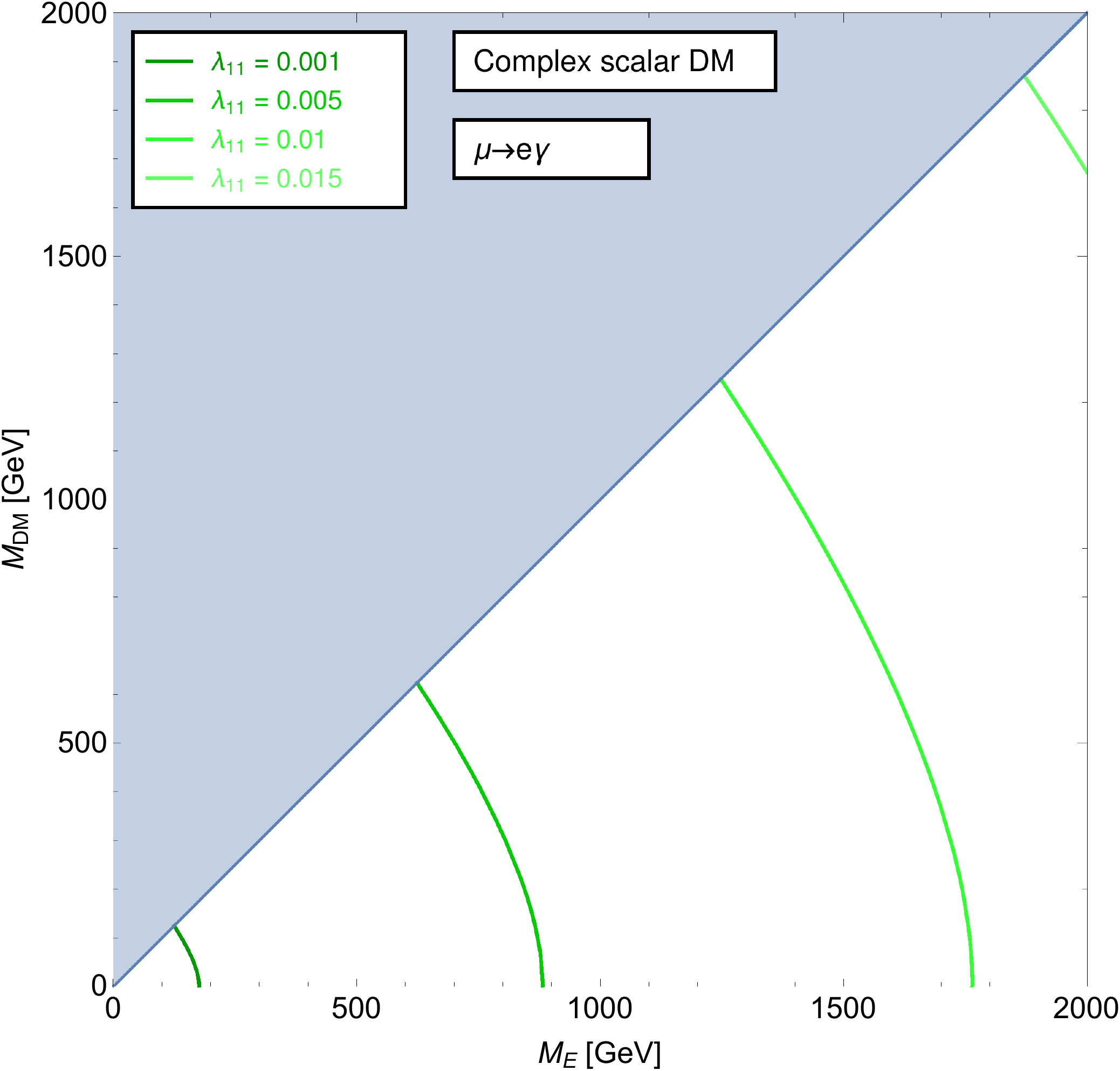}
\end{minipage}\hfill
\begin{minipage}[t]{.33\textwidth}
\includegraphics[width=\textwidth]{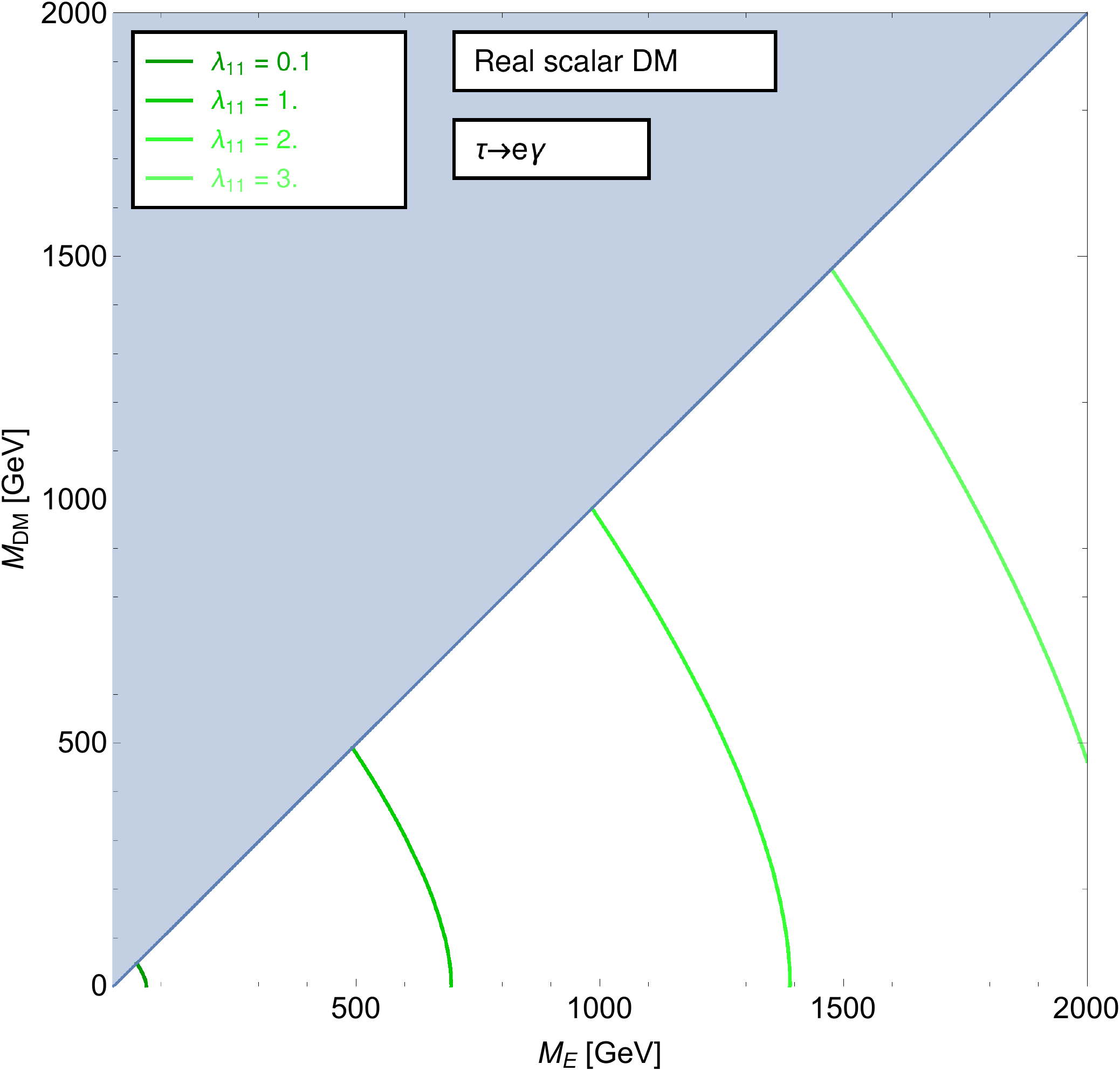}
\includegraphics[width=\textwidth]{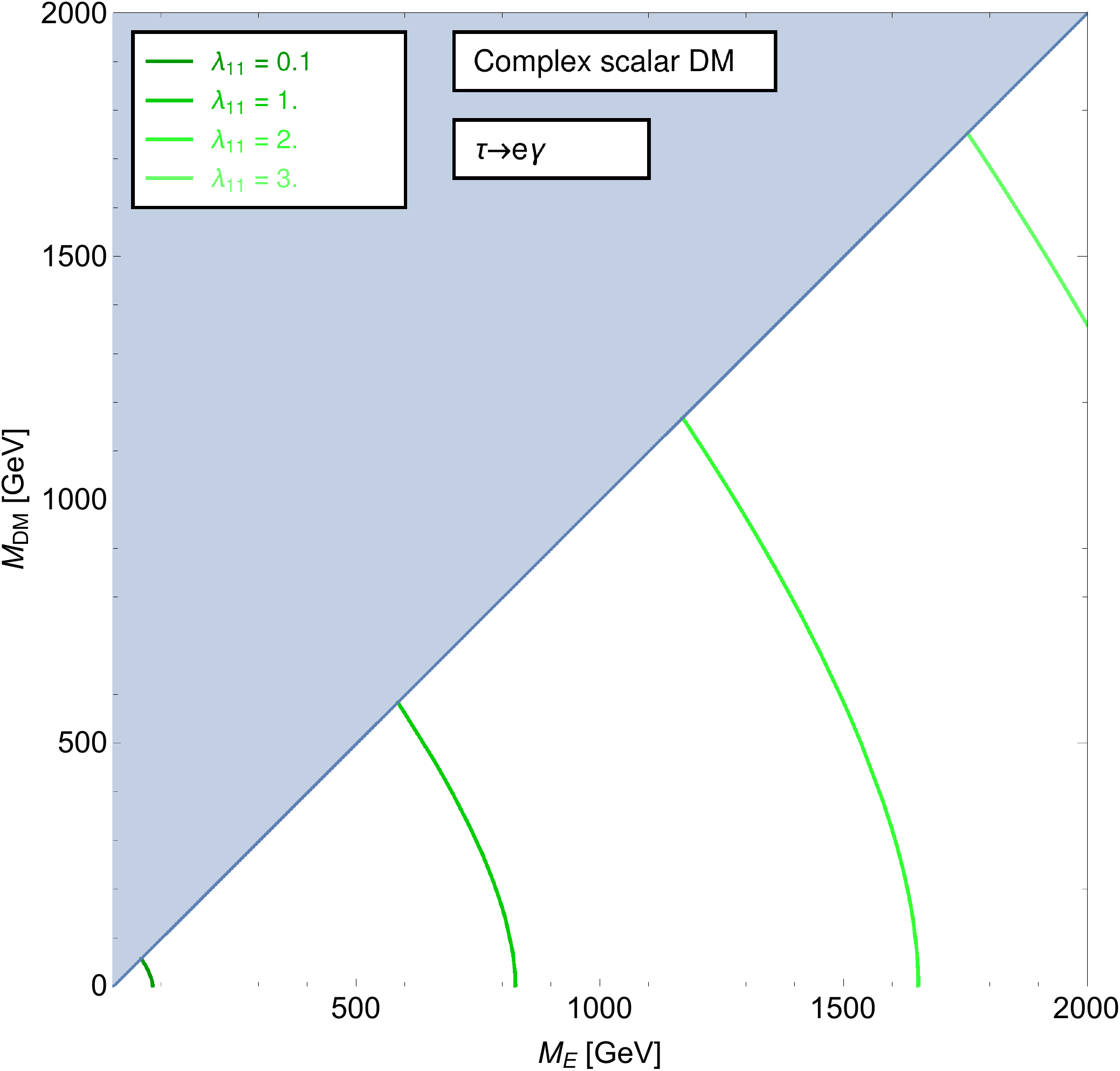}
\end{minipage}\hfill
\begin{minipage}[t]{.33\textwidth}
\includegraphics[width=\textwidth]{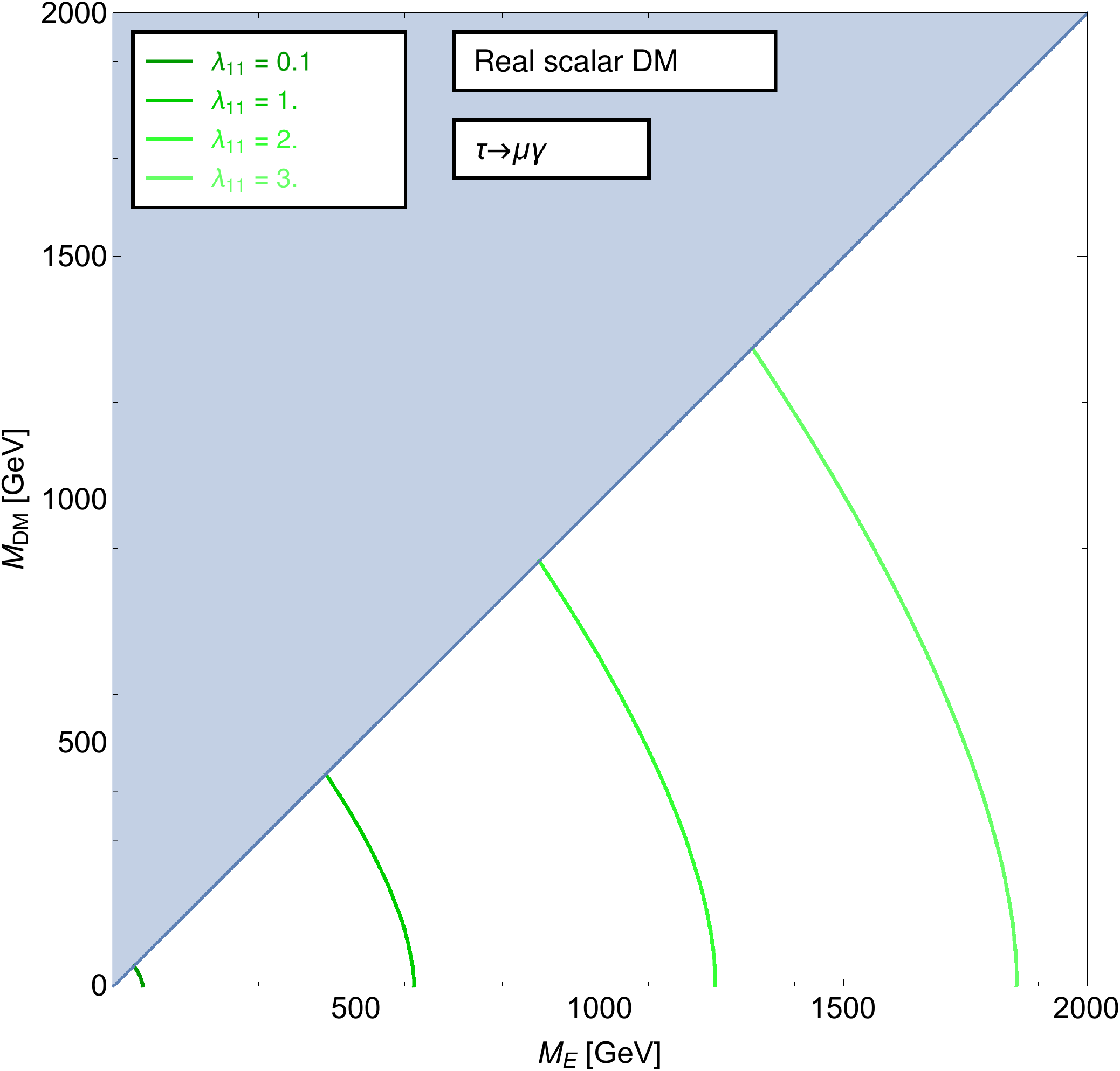}
\includegraphics[width=\textwidth]{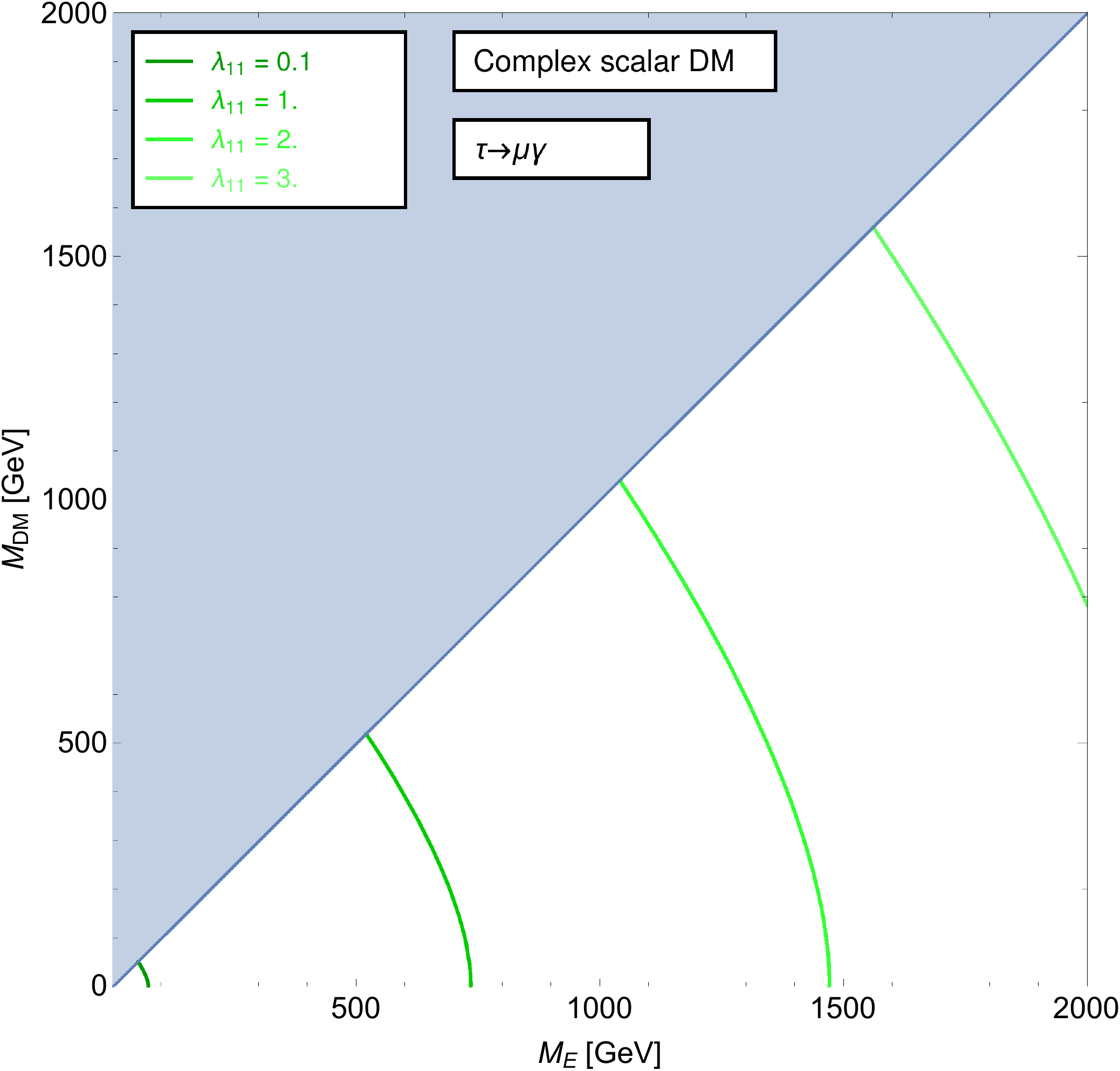}
\end{minipage}\hfill
\caption{\label{fig:LFV1s} LFV constraints for scalar DM. Column 1: constraints from $\mu\to e \gamma$; column 2: constraints from $\tau\to e \gamma$; column 3: constraints from $\tau\to \mu \gamma$. Top row: real scalar DM; bottom row: complex scalar DM. }
\end{figure}

\begin{figure}[ht!]
\centering
\begin{minipage}[t]{.33\textwidth}
\includegraphics[width=\textwidth]{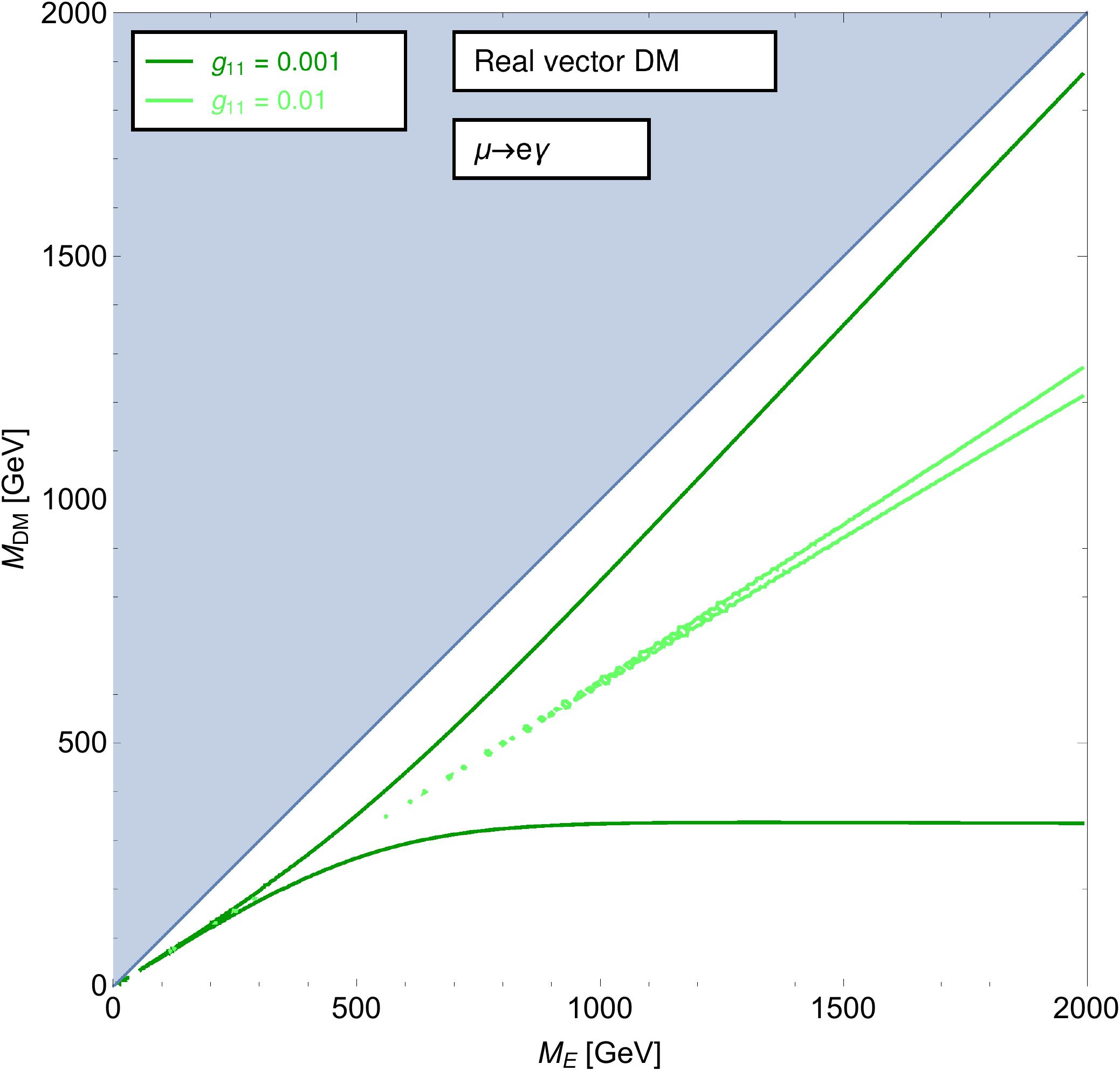}
\includegraphics[width=\textwidth]{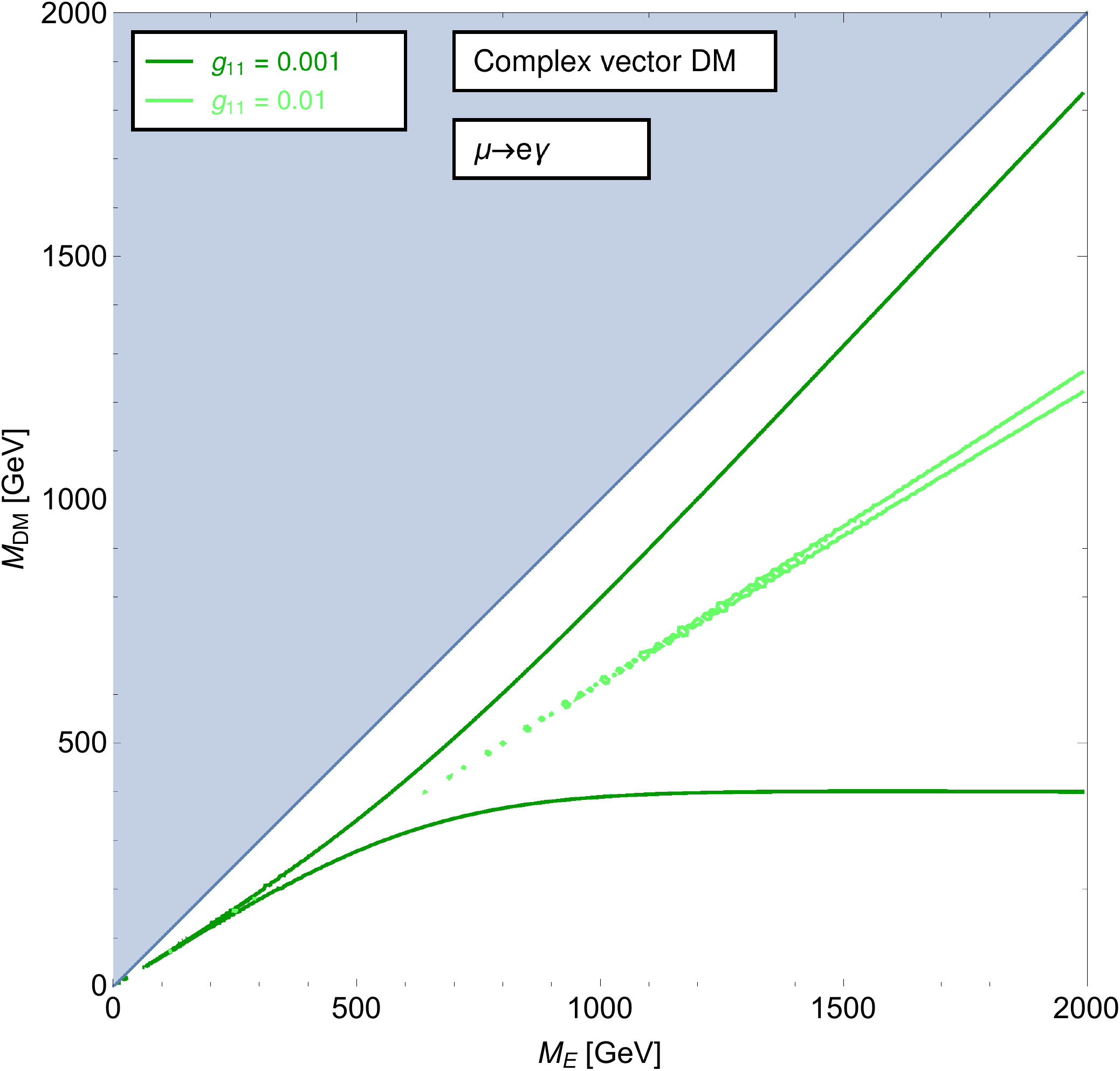}
\end{minipage}\hfill
\begin{minipage}[t]{.33\textwidth}
\includegraphics[width=\textwidth]{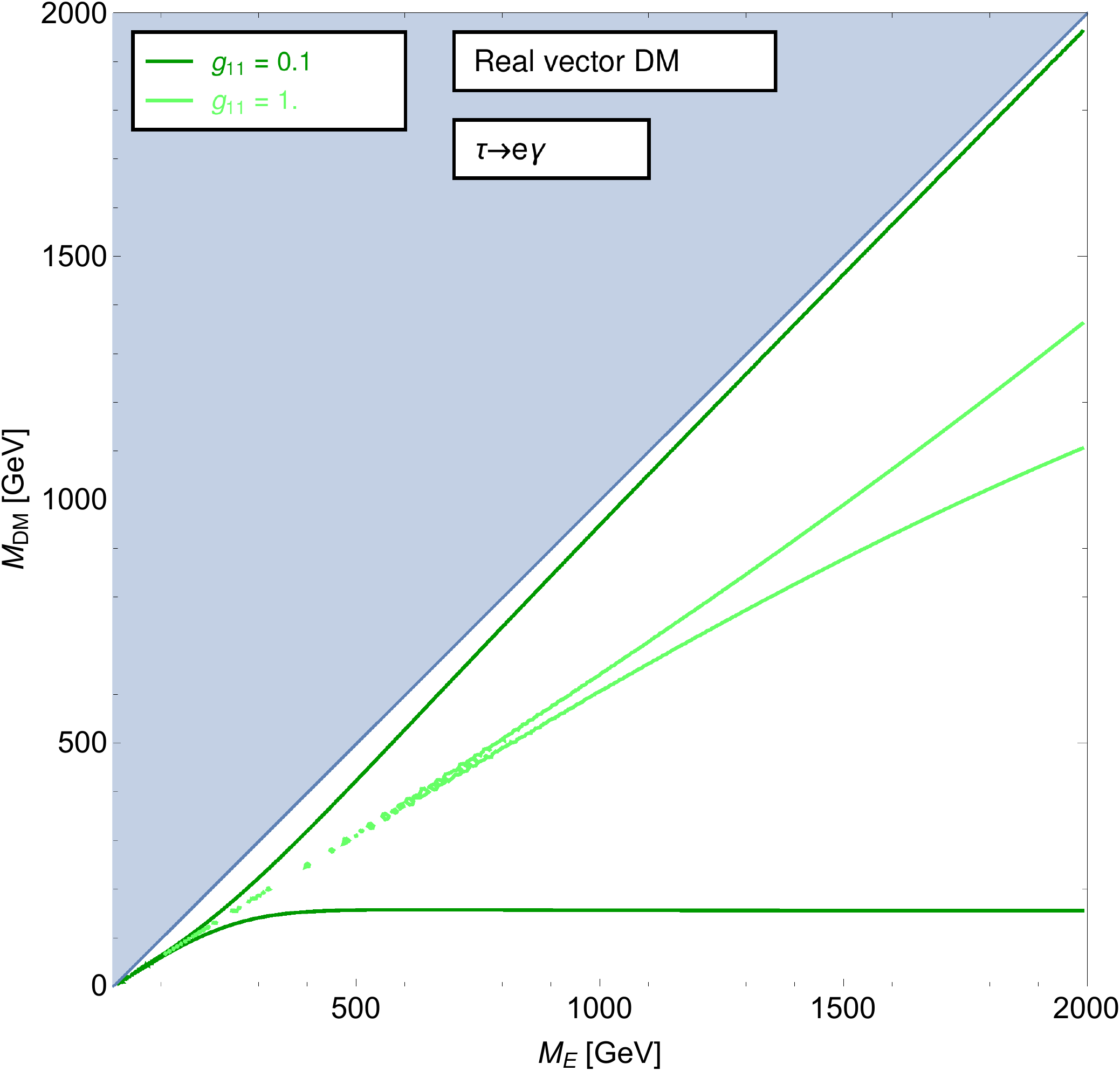}
\includegraphics[width=\textwidth]{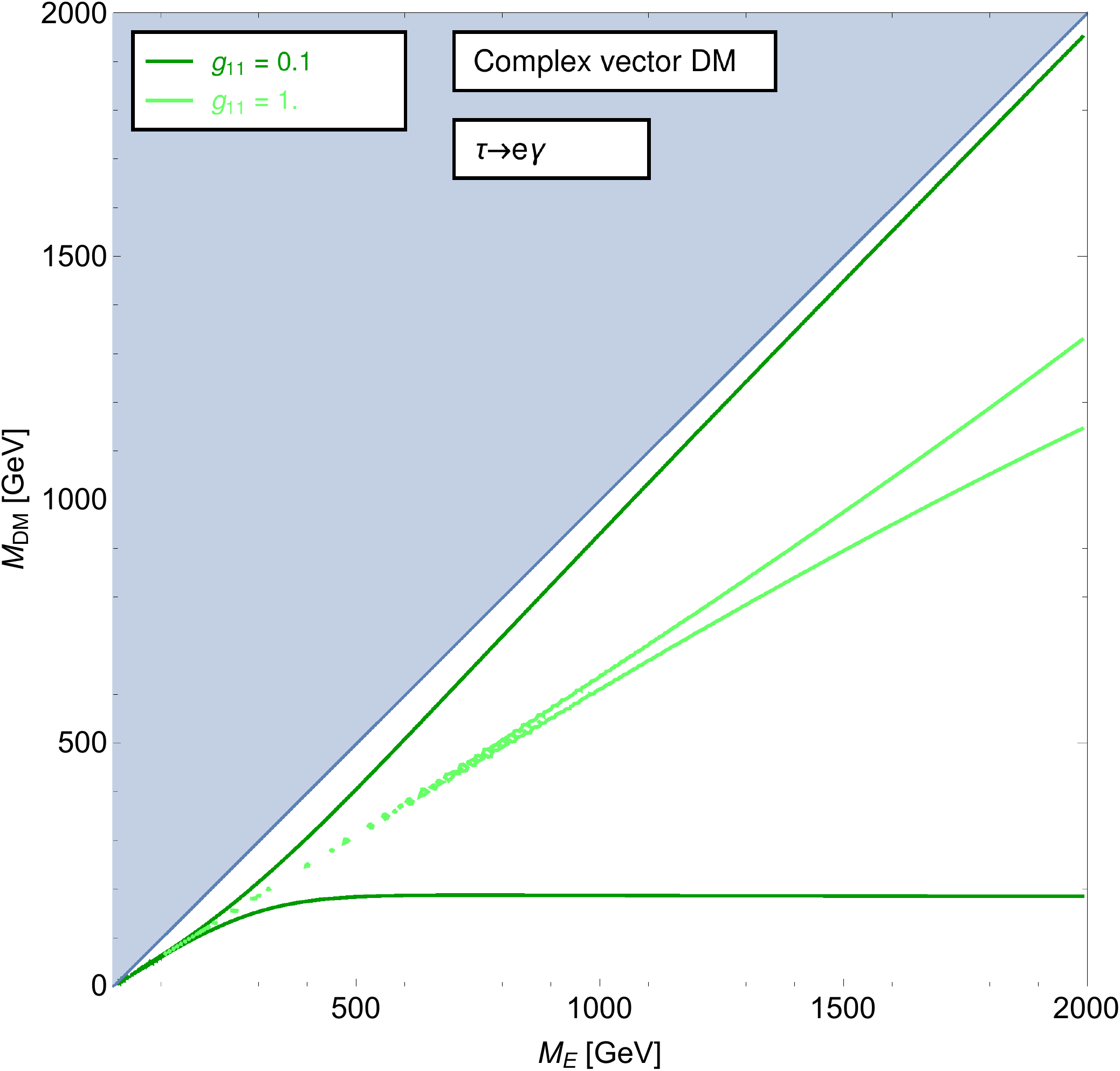}
\end{minipage}\hfill
\begin{minipage}[t]{.33\textwidth}
\includegraphics[width=\textwidth]{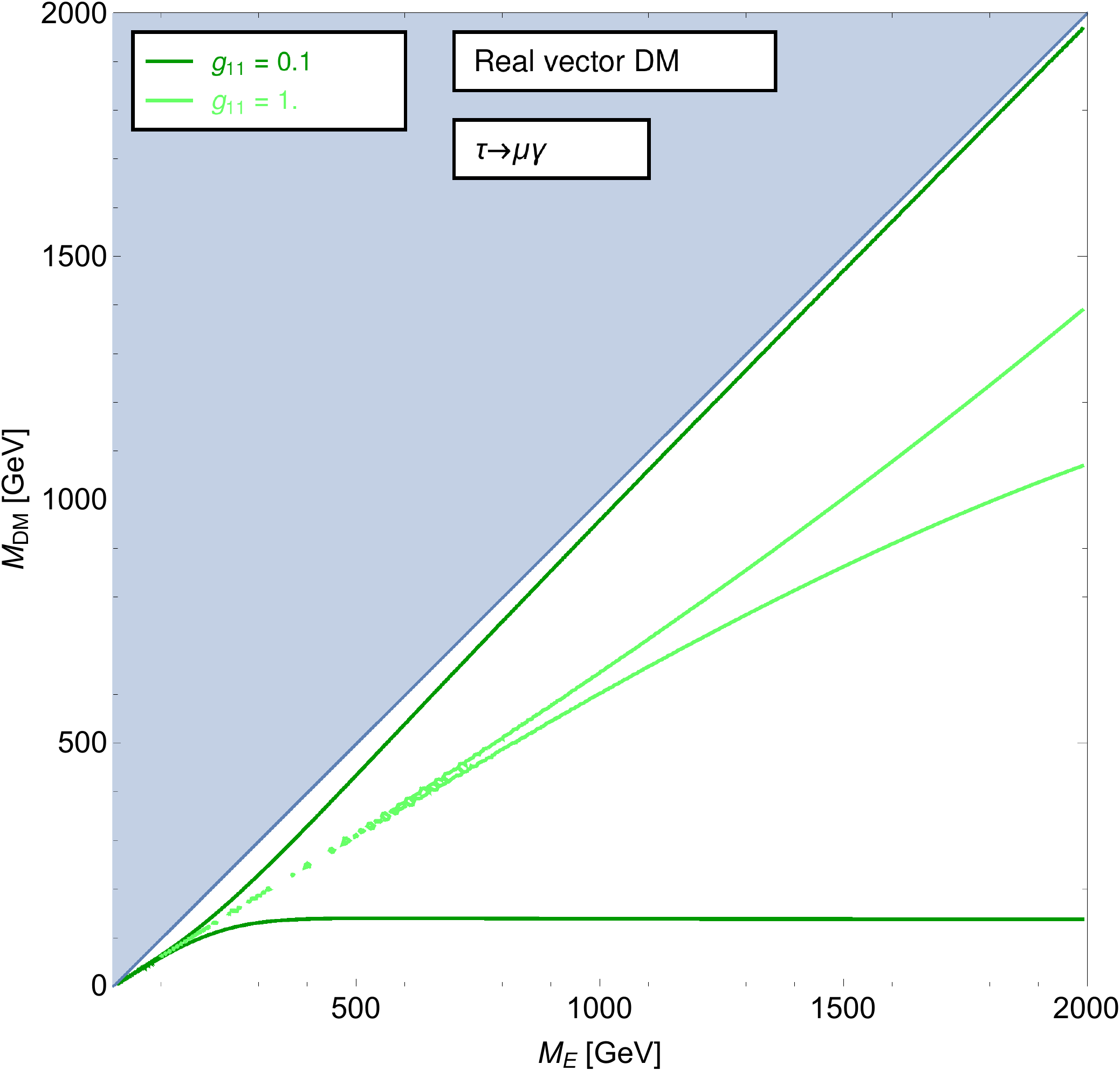}
\includegraphics[width=\textwidth]{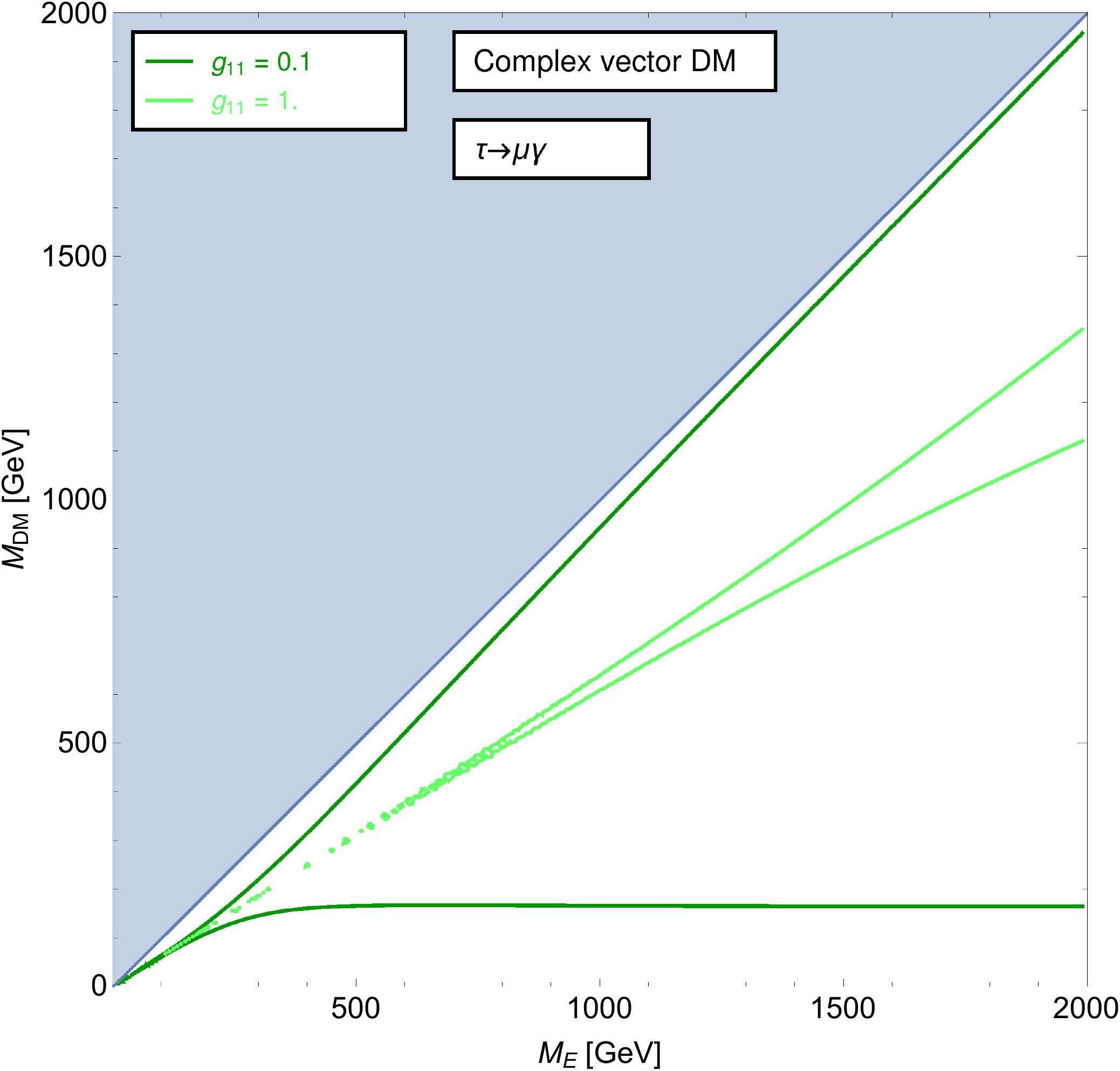}
\end{minipage}\hfill
\caption{\label{fig:LFV1v} Same as Fig.~\ref{fig:LFV1s}, but for vector DM.}
\end{figure}

Our numerical results are reported in Fig.~\ref{fig:LFV1s} for scalar DM and in Fig.~\ref{fig:LFV1v} for vector DM.  The plots show the upper bounds on the  relation $\sqrt{\lambda_{11}^i\lambda_{11}^j}$ for scalar DM or $\sqrt{g_{11}^ig_{11}^j}$ for vector DM in the ($M_E,M_{\rm DM}$) plane. However, due to the assumed universality of the couplings in BP4, such relations reduce to $\lambda_{11}$ and $g_{11}$ in the two DM spin scenarios. The dependence of the bound on the masses of BSM states is remarkably different whether the DM is scalar or vector. For scalar DM the bounds have analogous functional dependence on the VLL and DM masses, and an increase in the $E$ -- DM -- $l_{\rm SM}$ coupling excludes a larger range of both masses. For vector DM, the allowed range corresponds to a funnel region in the mass plane, which shrinks as the coupling increases. Analogously to the $(g-2)$ case, for complex DM, a factor of 2 has been included in the calculation of the constraints.

\section{Combination of all observables}

For the purpose of discriminating between different DM spins, the first step is to identify regions in parameter space which are allowed in the different scenarios. The possibility of excluding complementary regions for different DM spins would be essential for identifying the DM spin in case of signal discovery in one of such regions.
Therefore, in this section all the observables described so far will be compared to identify which scenarios are excluded and which are allowed. 

The constraints from relic density (Sec.~\ref{sec:relic}) impose a minimum value on the coupling between the VLL and the DM in order for a meaningful region in the VLL versus DM mass parameter space to be allowed. Such minimum value of the coupling is about $\lambda_{11}^f \simeq 1$ for a coupling to scalar DM, and $g_{11}^f \simeq 0.2-0.5$ for a coupling to vector DM. Large values of the couplings, in contrast, determine a large VLL width: we have not considered values of the width-to-mass ratio larger than 50\% for our analysis.
The results from the anomalous magnetic moment of electron and muon $(g-2)$ (Sec.~\ref{sec:g-2}) and LFV processes (Sec.~\ref{subsec-LFV}), on the other hand, show that with such values of the coupling the allowed region of parameter space often shrinks to tiny regions. 
Finally, the LHC constraints (Sec.~\ref{sec:LHC}) only exclude VLL masses between $\sim 100$ GeV and $\sim 400$ GeV for BP1 and BP2, and between $\sim 100$ GeV and $\sim 200$ GeV for BP3; however, they are almost independent of the interaction coupling between VLL, DM and SM leptons in the limit of small width. 

\begin{figure}[ht!]
\centering
\begin{minipage}{.32\textwidth}
\includegraphics[width=\textwidth]{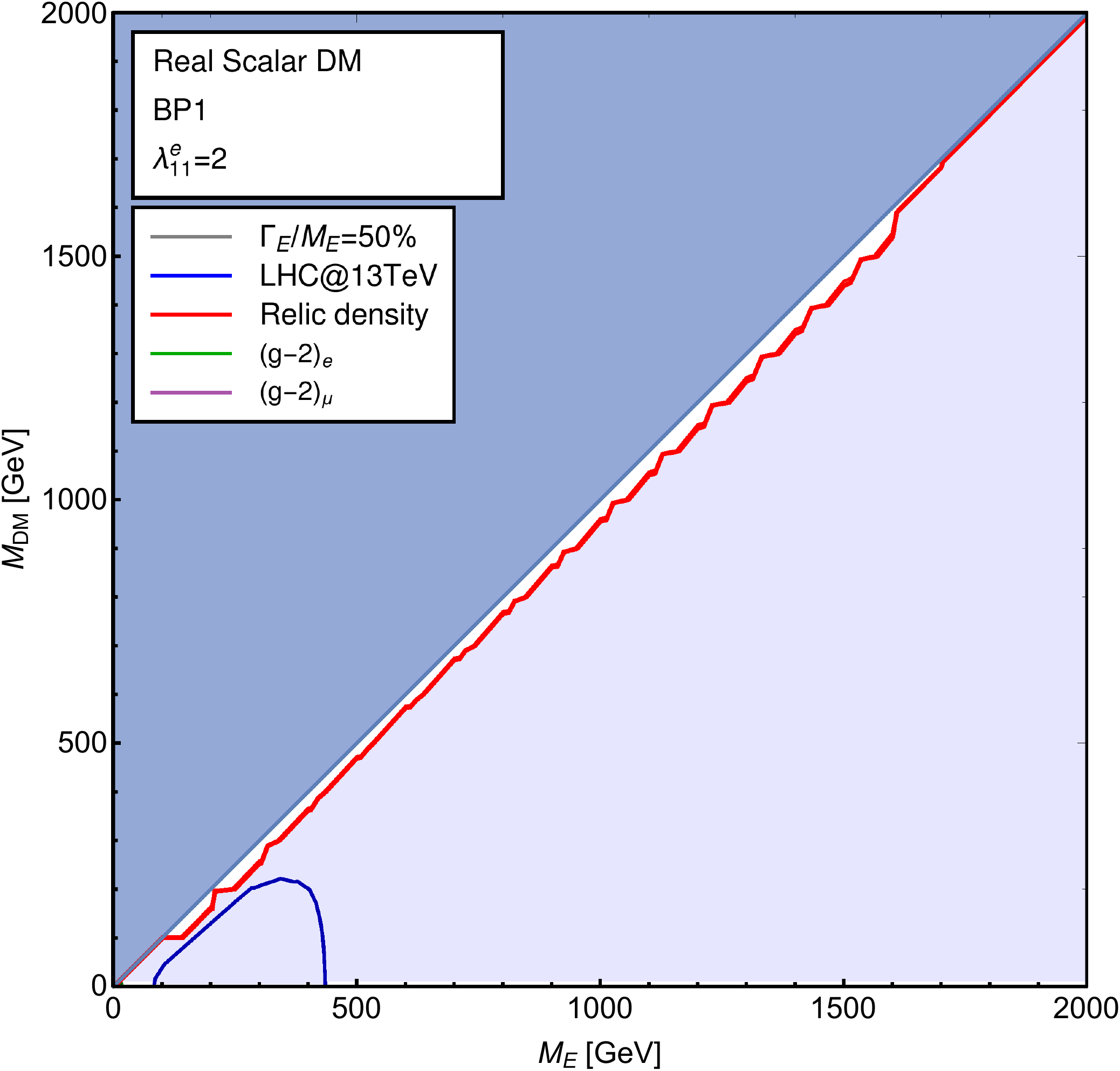}
\end{minipage}
\begin{minipage}{.32\textwidth}
\includegraphics[width=\textwidth]{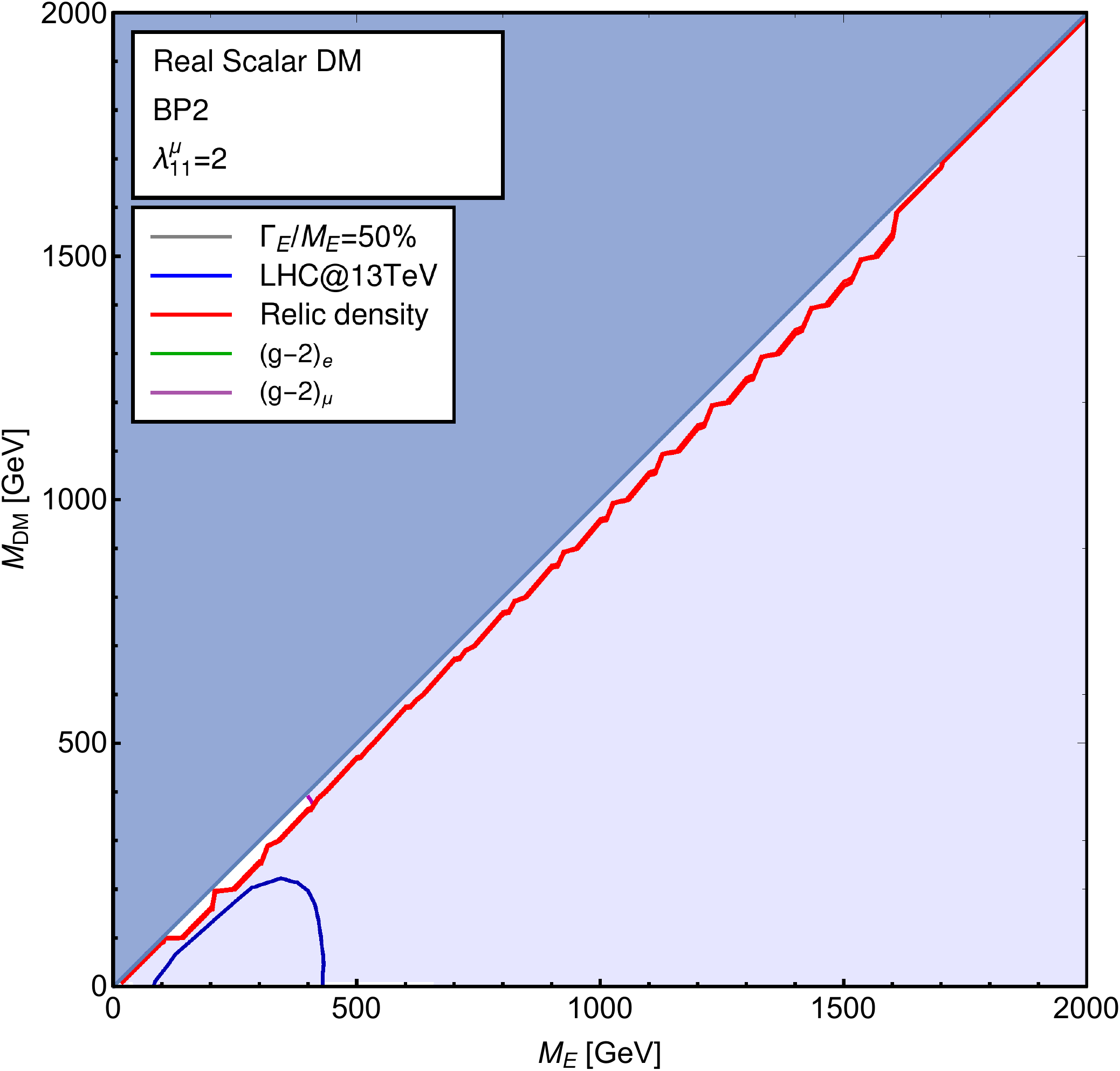}
\end{minipage}
\begin{minipage}{.32\textwidth}
\includegraphics[width=\textwidth]{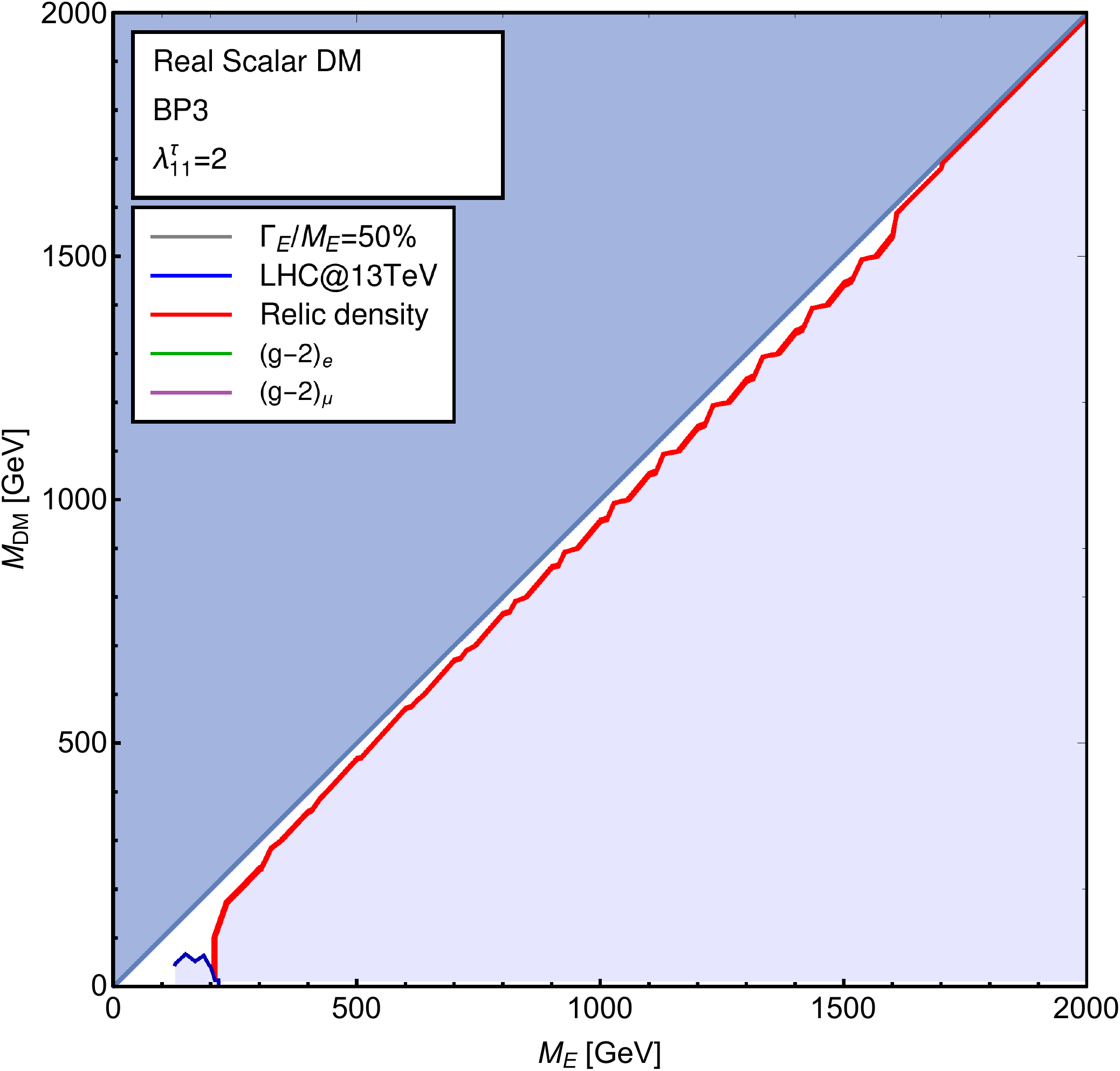}
\end{minipage}\\
\begin{minipage}{.32\textwidth}
\includegraphics[width=\textwidth]{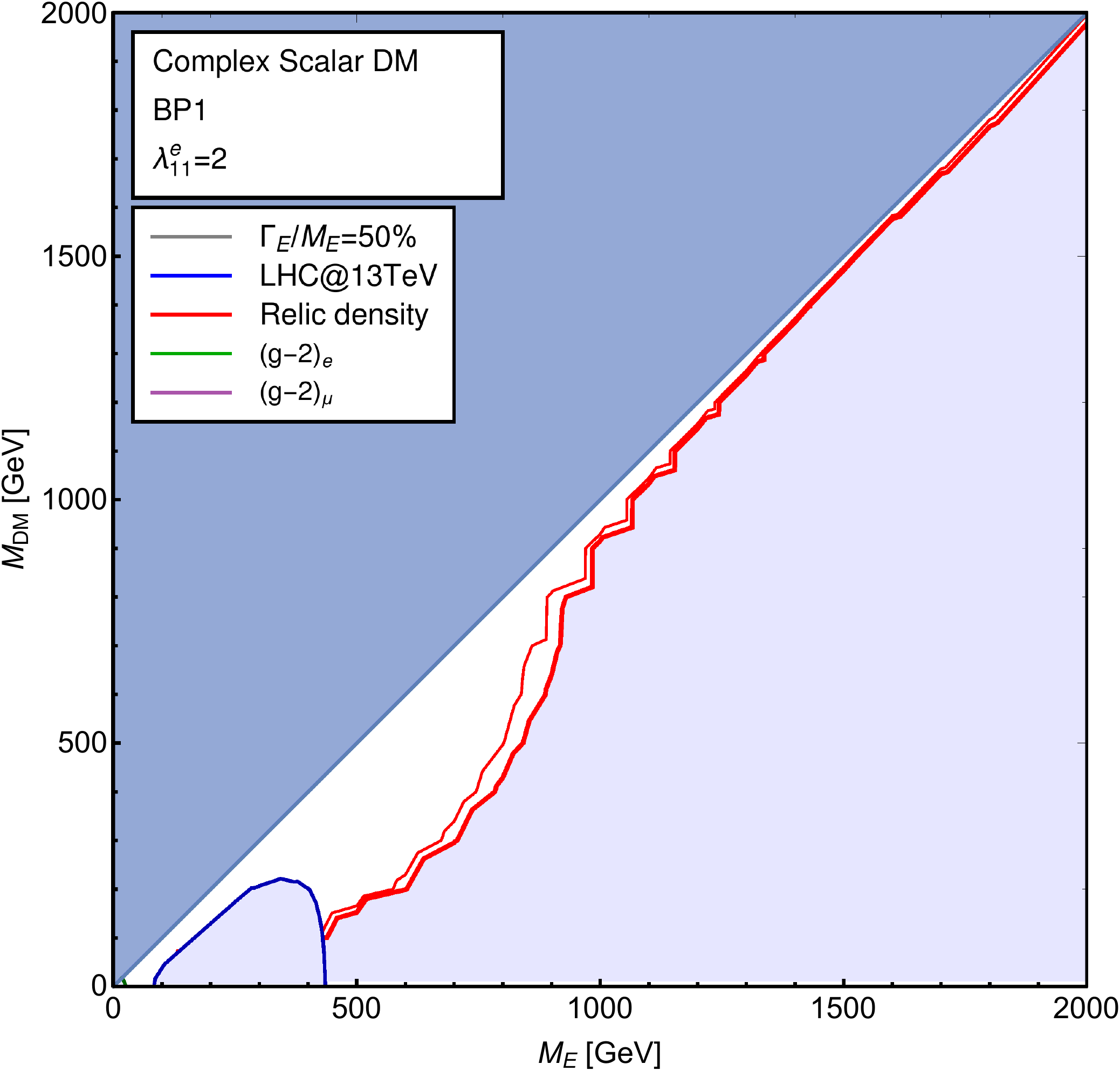}
\end{minipage}
\begin{minipage}{.32\textwidth}
\includegraphics[width=\textwidth]{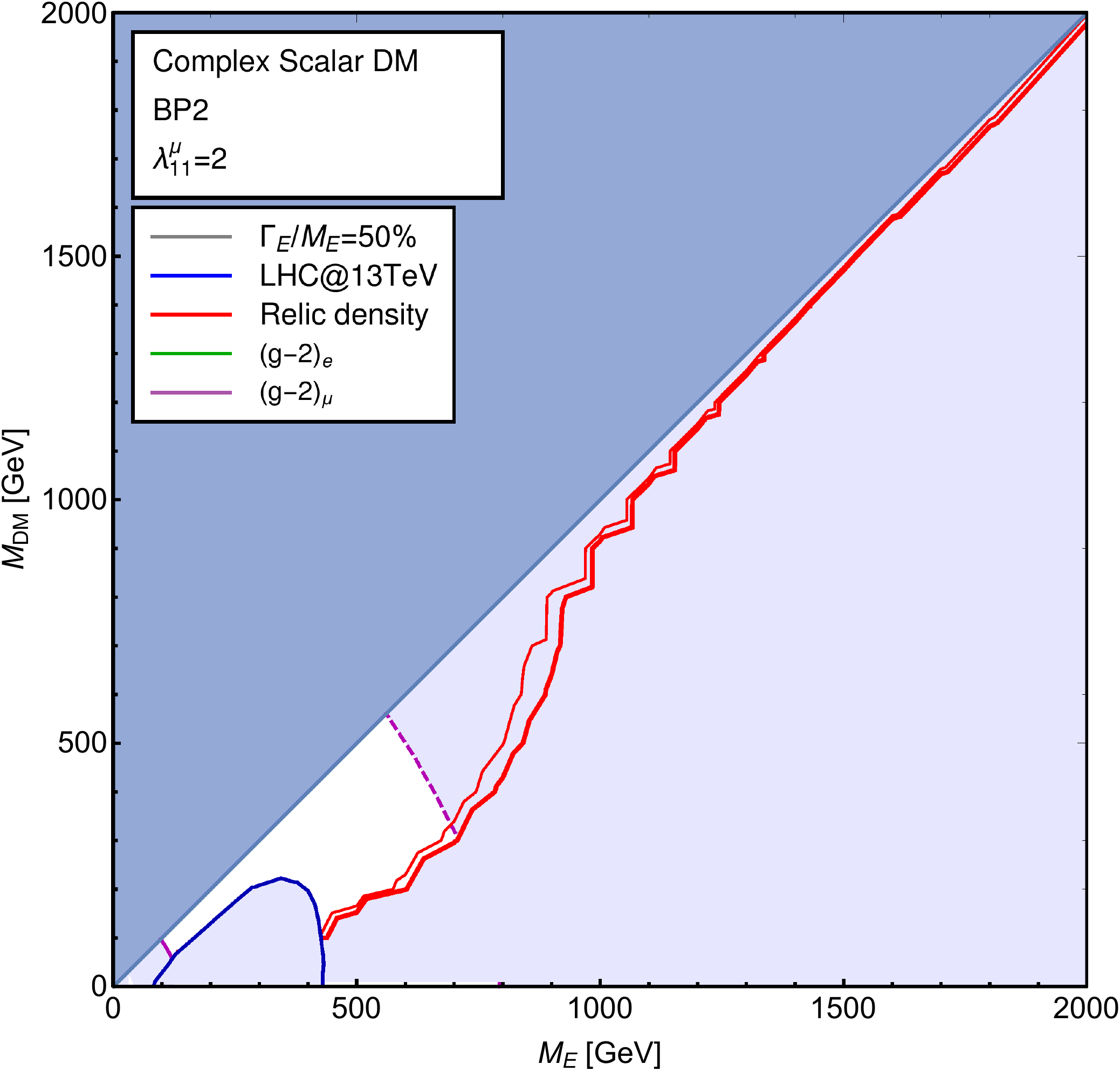}
\end{minipage}
\begin{minipage}{.32\textwidth}
\includegraphics[width=\textwidth]{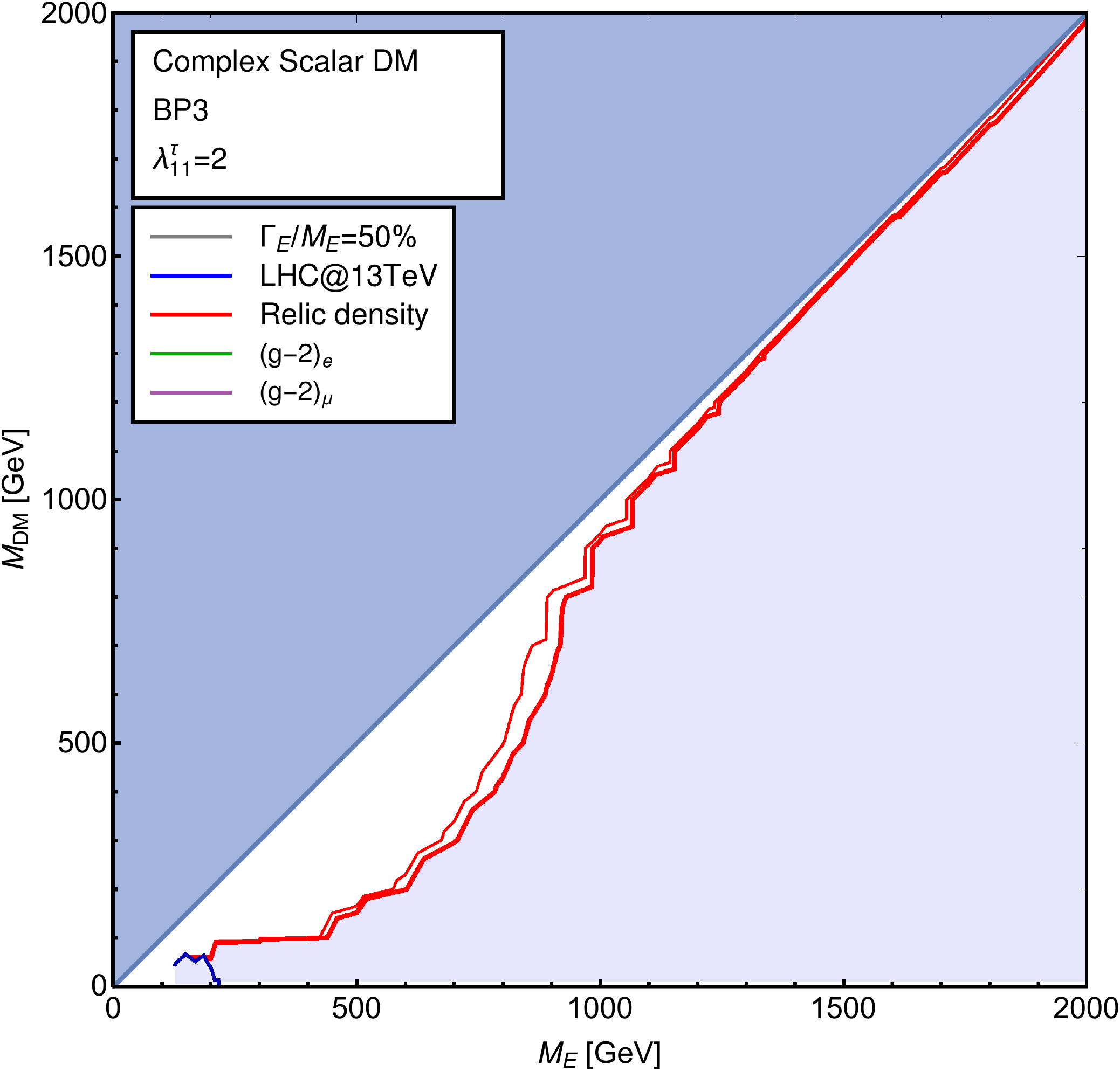}
\end{minipage}\\
\begin{minipage}{.32\textwidth}
$$\text{excluded}$$
\end{minipage}
\begin{minipage}{.32\textwidth}
$$\text{excluded}$$
\end{minipage}
\begin{minipage}{.32\textwidth}
\includegraphics[width=\textwidth]{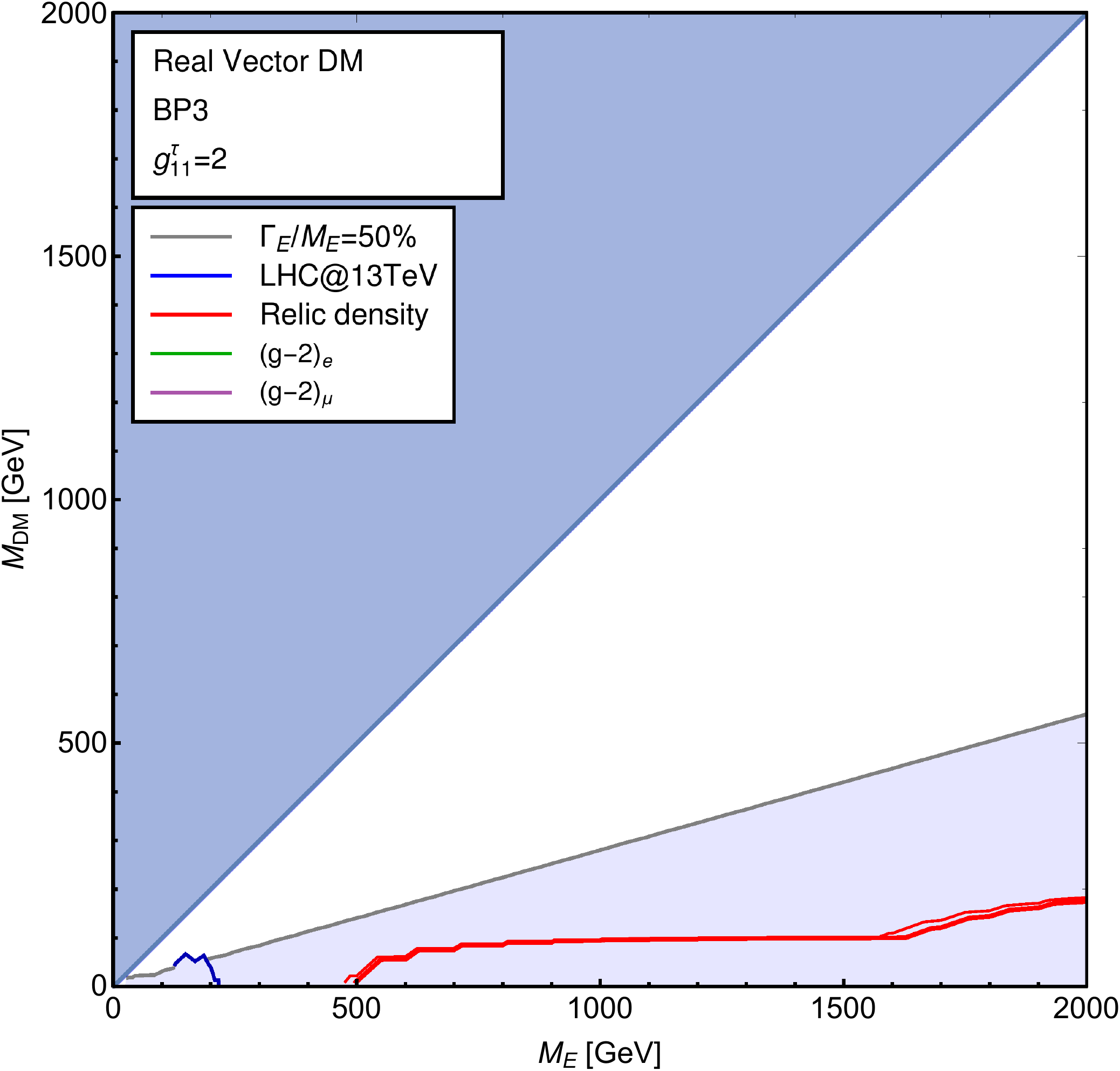}
\end{minipage}\\
\begin{minipage}{.32\textwidth}
$$\text{excluded}$$
\end{minipage}
\begin{minipage}{.32\textwidth}
$$\text{excluded}$$
\end{minipage}
\begin{minipage}{.32\textwidth}
\includegraphics[width=\textwidth]{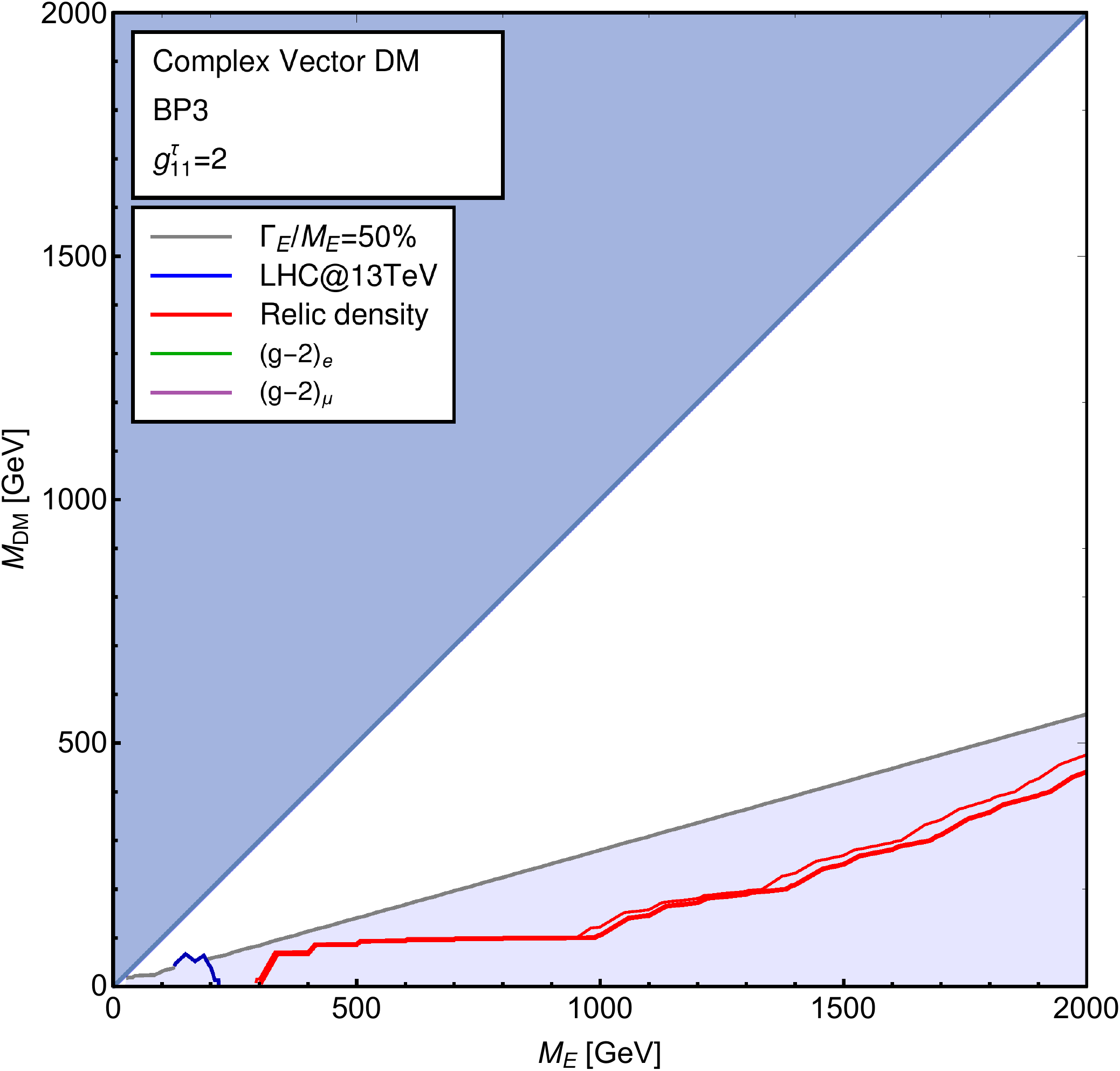}
\end{minipage}
\caption{\label{fig:combined123coup2} Excluded region (in light blue) for a fixed value of the VLL coupling $\lambda_{11}^f=g_{11}^f=2$ for BP1, 2 and 3.}
\end{figure}

\begin{figure}[ht!]
\centering
\begin{minipage}{.32\textwidth}
\includegraphics[width=\textwidth]{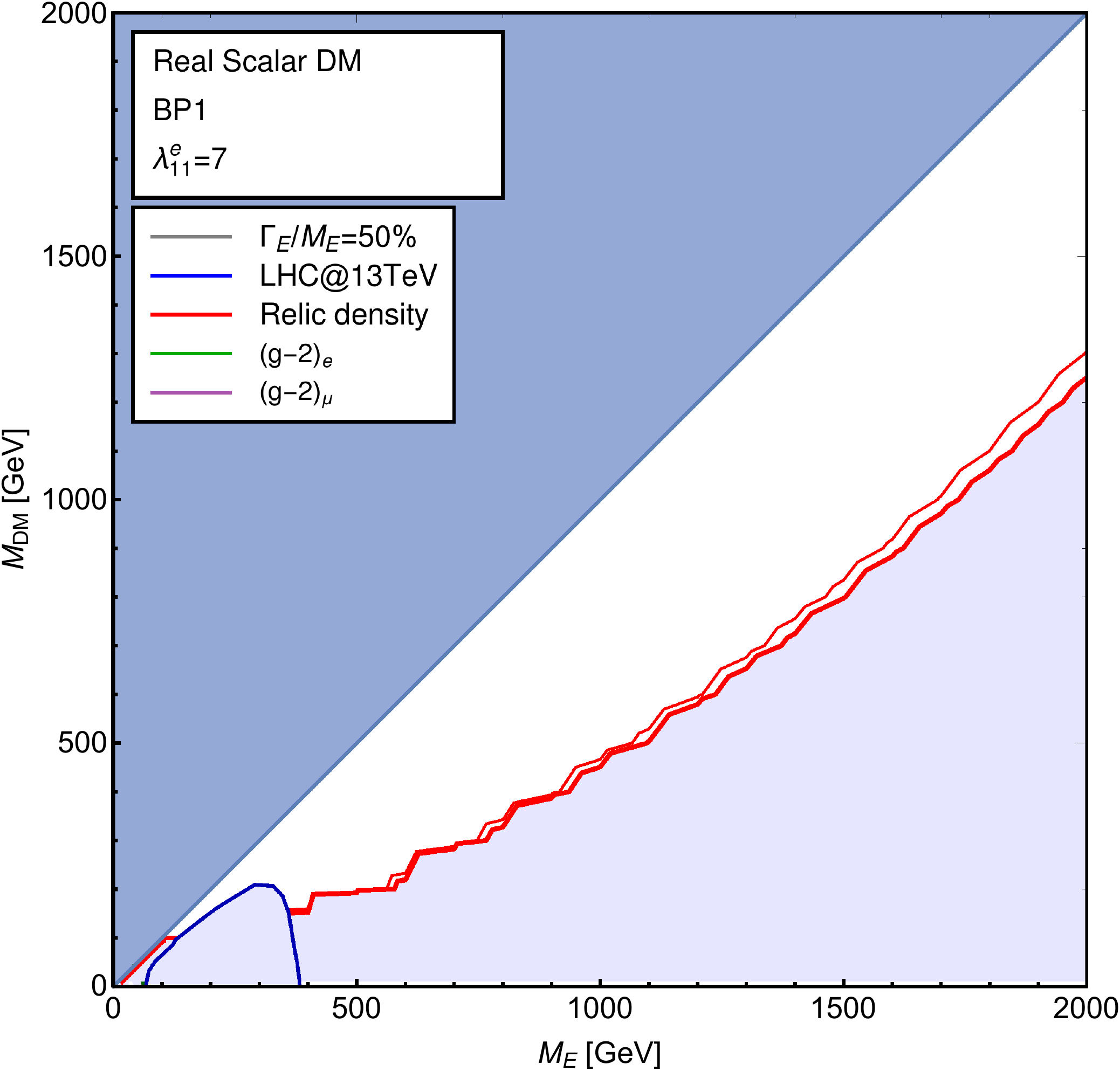}
\end{minipage}
\begin{minipage}{.32\textwidth}
\includegraphics[width=\textwidth]{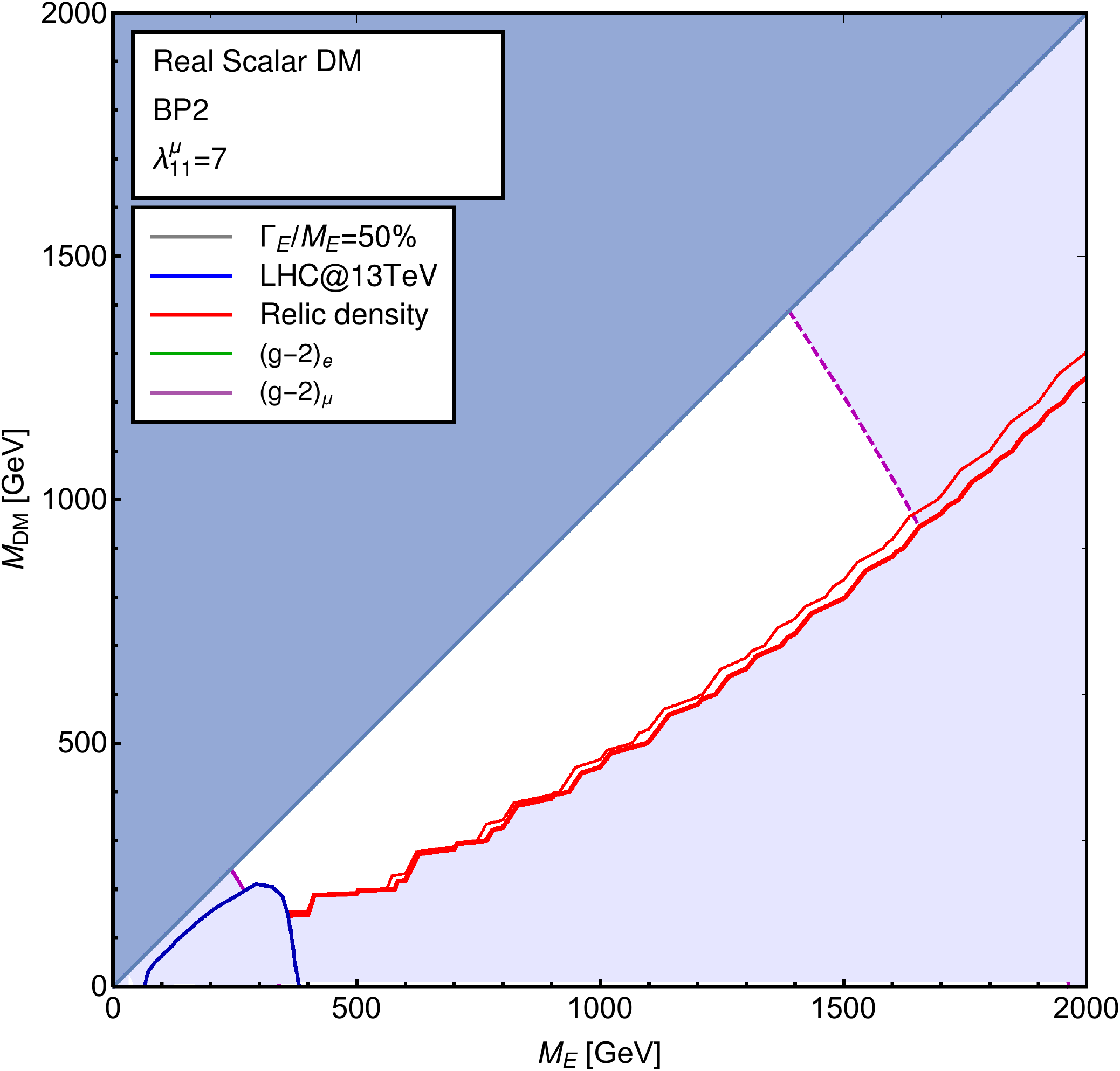}
\end{minipage}
\begin{minipage}{.32\textwidth}
\includegraphics[width=\textwidth]{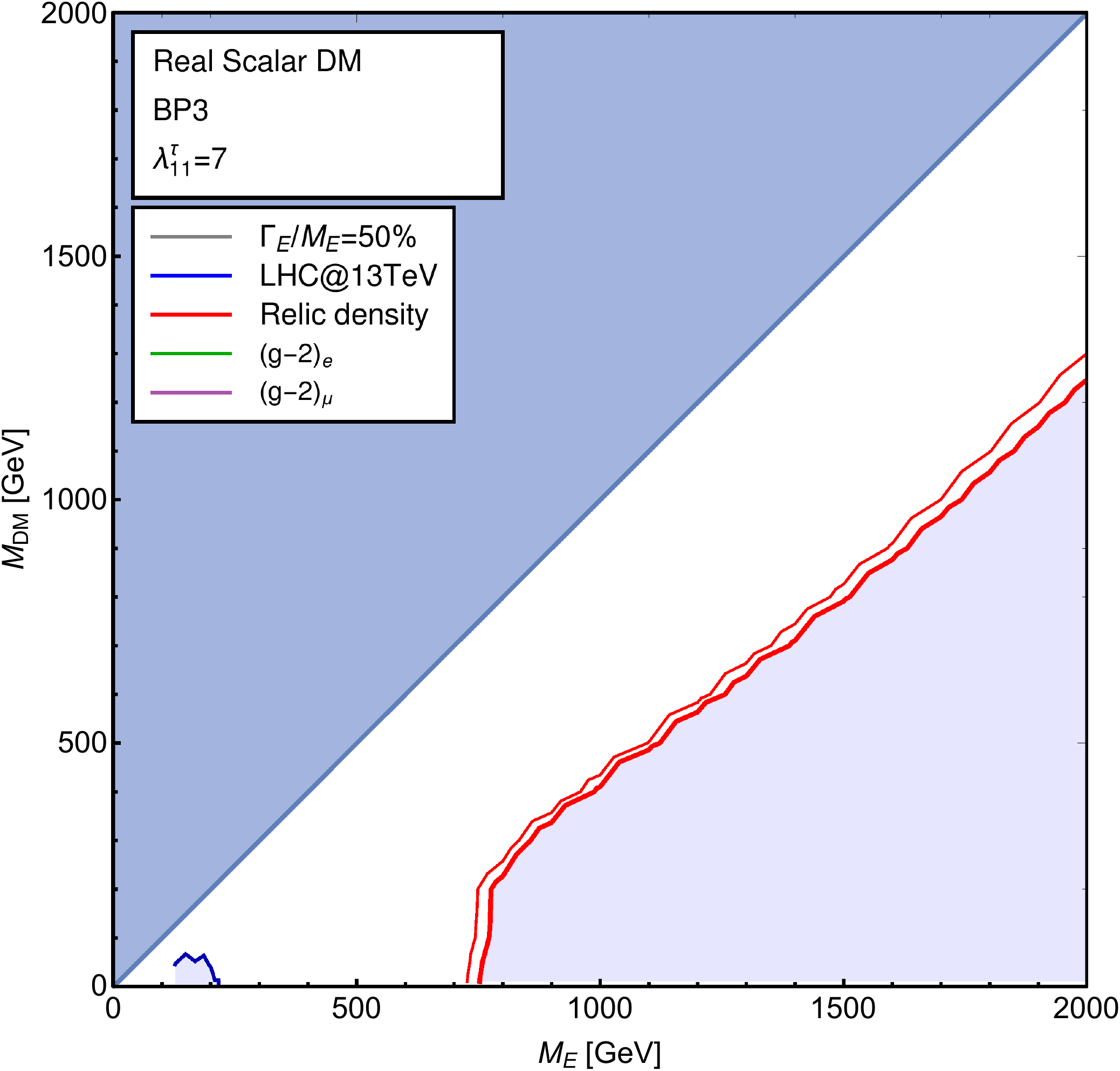}
\end{minipage}\\
\begin{minipage}{.32\textwidth}
\includegraphics[width=\textwidth]{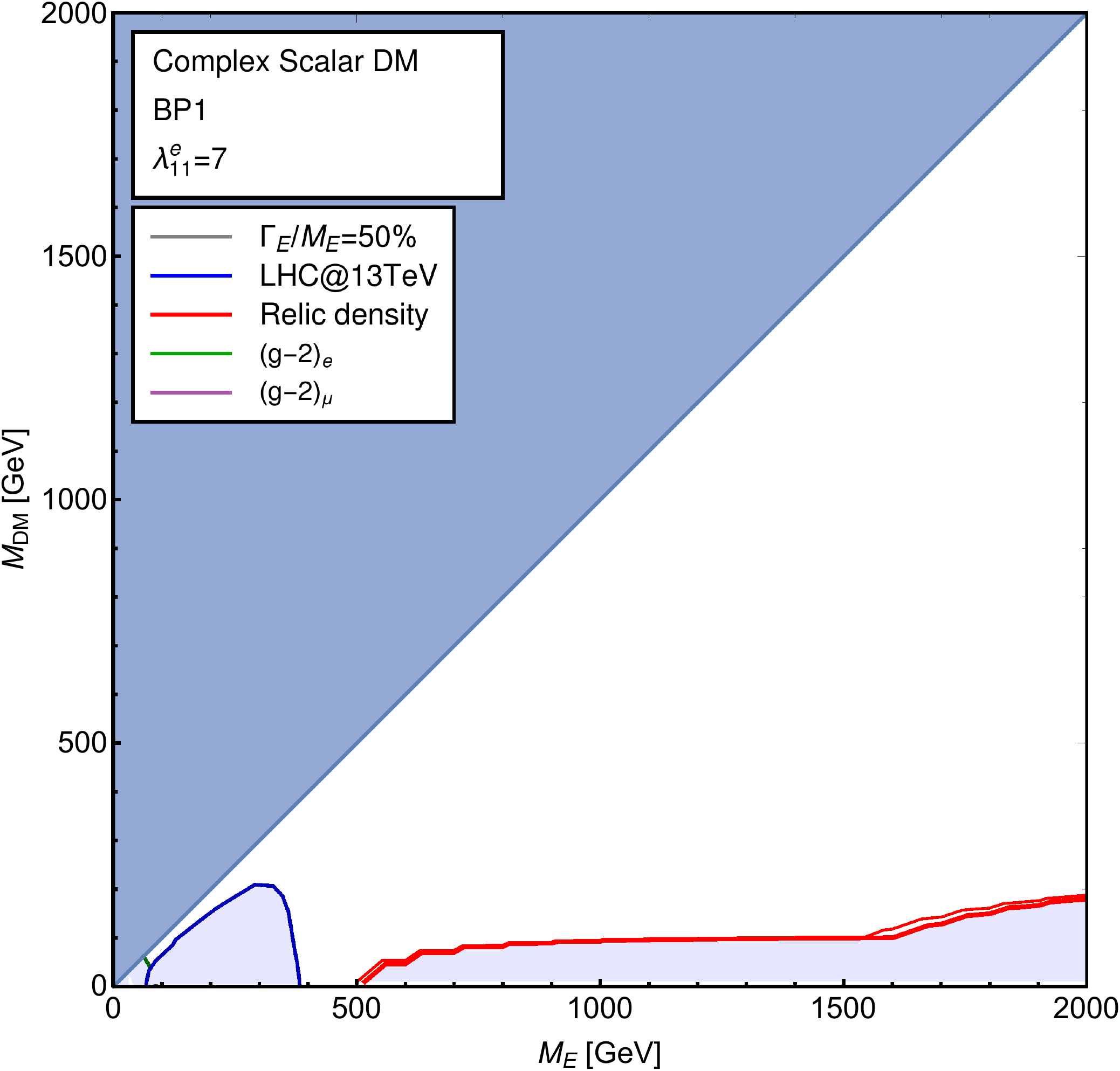}
\end{minipage}
\begin{minipage}{.32\textwidth}
\includegraphics[width=\textwidth]{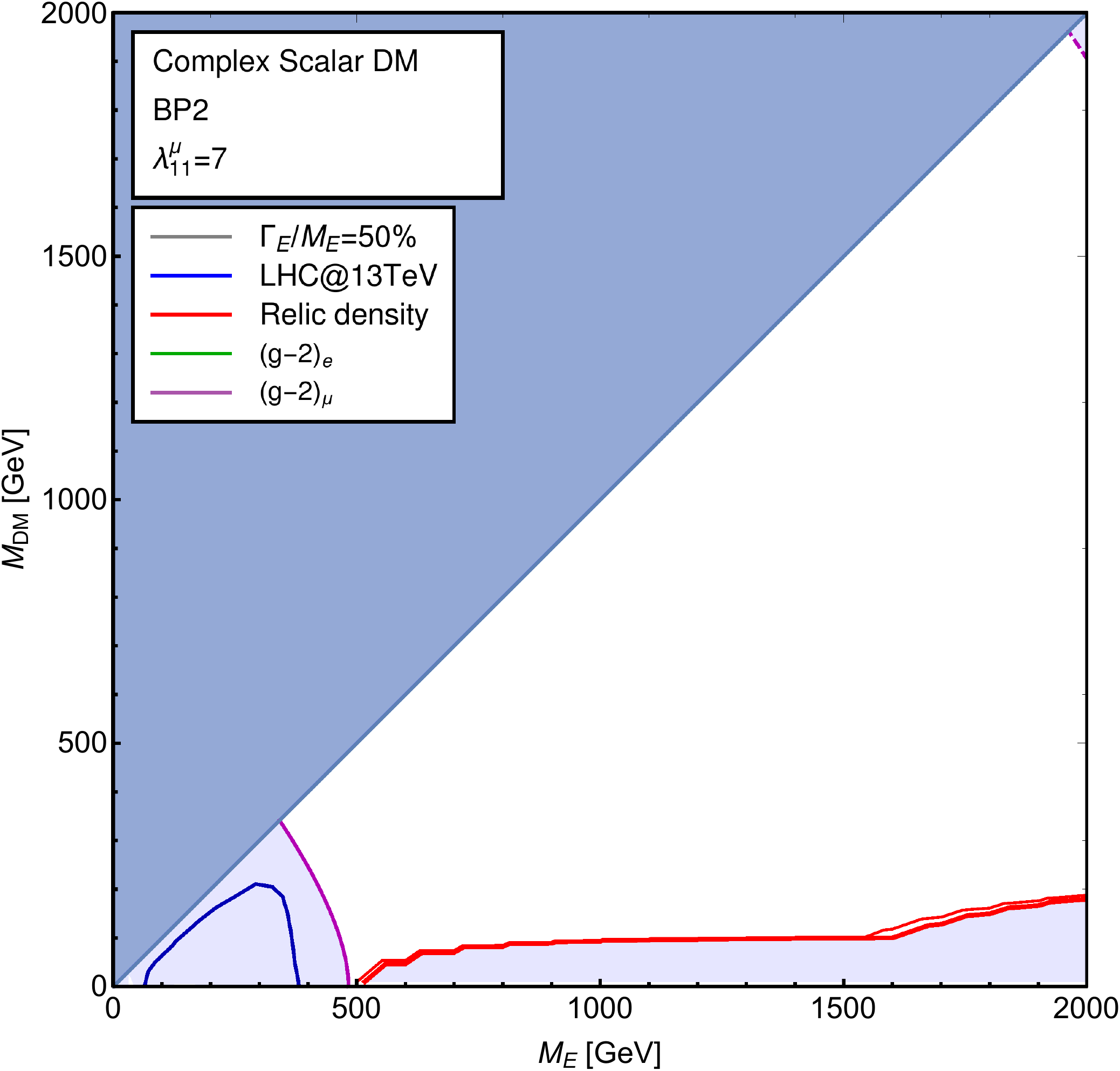}
\end{minipage}
\begin{minipage}{.32\textwidth}
\includegraphics[width=\textwidth]{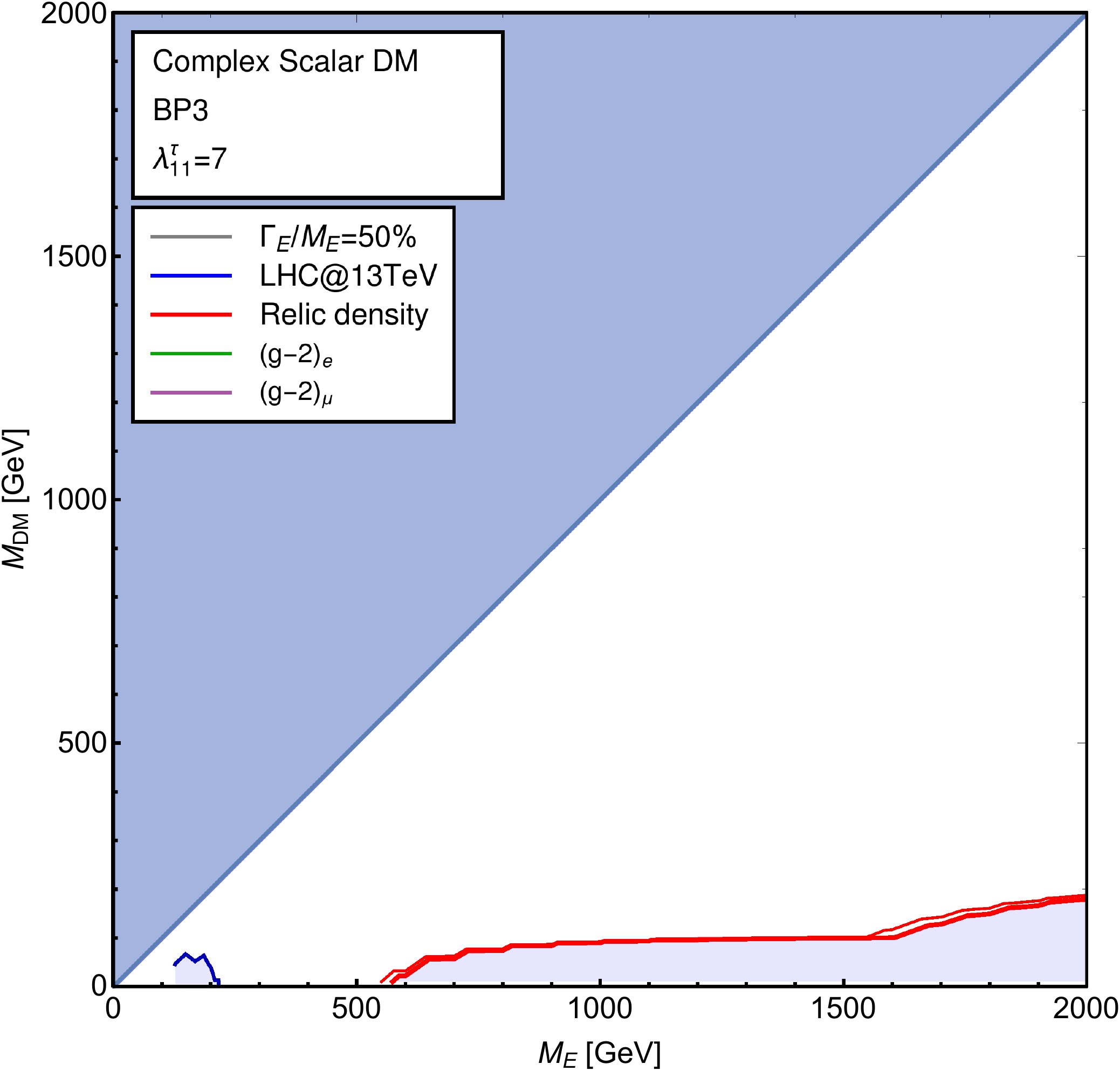}
\end{minipage}\\
\begin{minipage}{.32\textwidth}
$$\text{excluded}$$
\end{minipage}
\begin{minipage}{.32\textwidth}
$$\text{excluded}$$
\end{minipage}
\begin{minipage}{.32\textwidth}
\includegraphics[width=\textwidth]{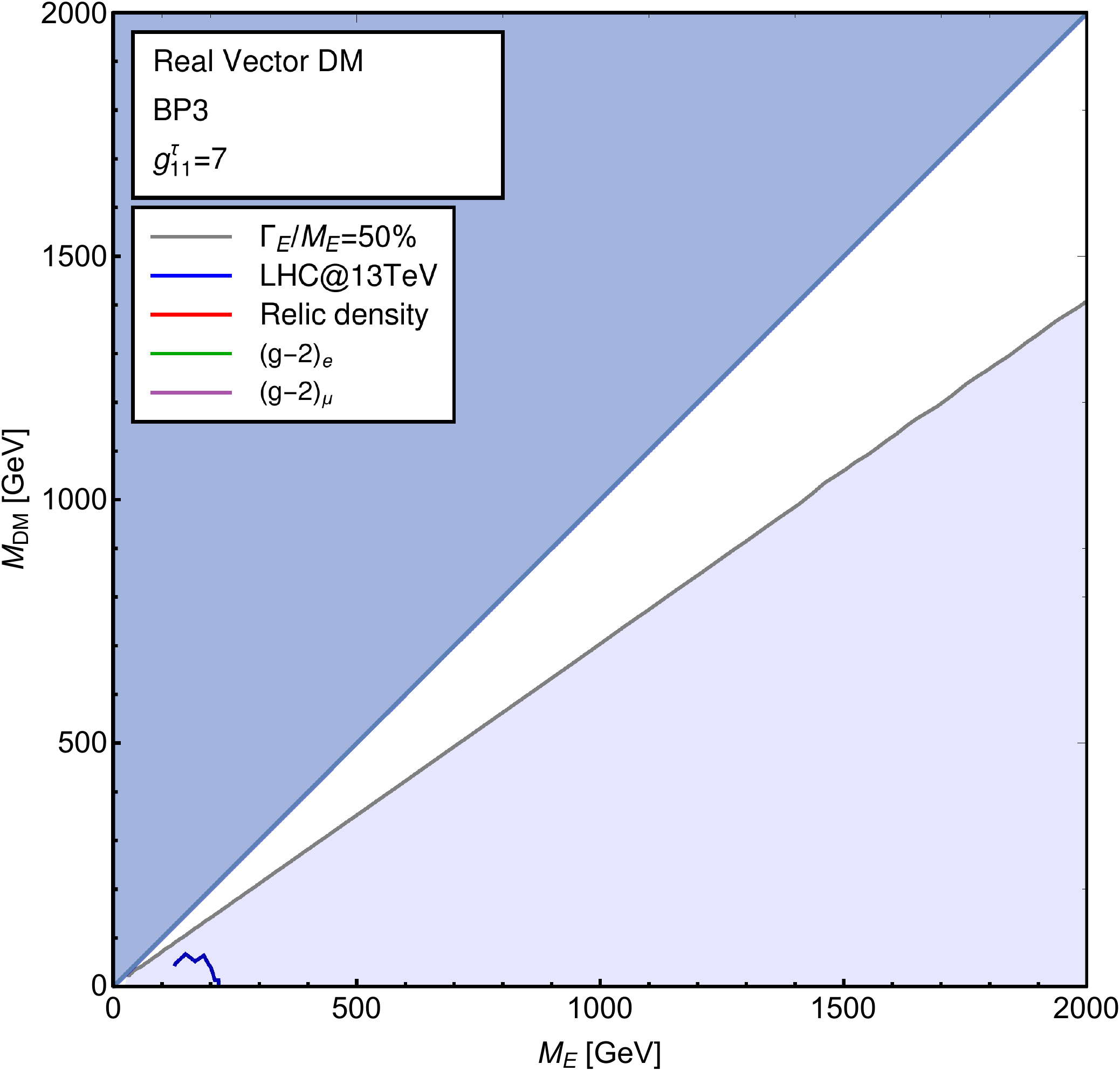}
\end{minipage}\\
\begin{minipage}{.32\textwidth}
$$\text{excluded}$$
\end{minipage}
\begin{minipage}{.32\textwidth}
$$\text{excluded}$$
\end{minipage}
\begin{minipage}{.32\textwidth}
\includegraphics[width=\textwidth]{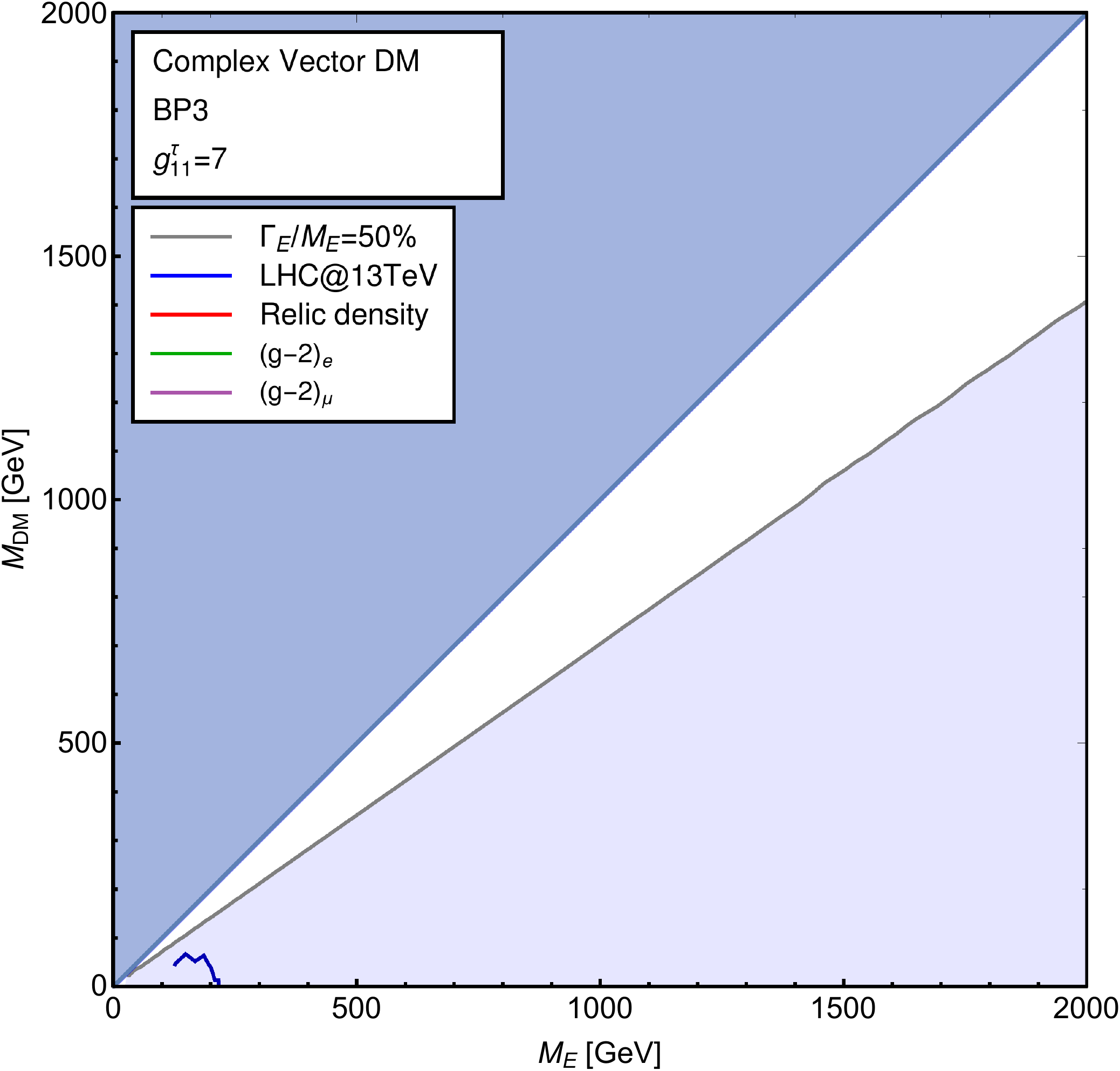}
\end{minipage}
\caption{\label{fig:combined123coup7} Excluded region (in light blue) for a fixed value of the VLL coupling $\lambda_{11}^f=g_{11}^f=7$ for BP1, 2 and 3.}
\end{figure}

Comparisons of constraints for specific values of the coupling are shown in Figs.~\ref{fig:combined123coup2} and \ref{fig:combined123coup7} for small and large values of the coupling (within the allowed range). The excluded region is shown in light blue. 

\begin{figure}[ht!]
\centering
\includegraphics[width=.32\textwidth]{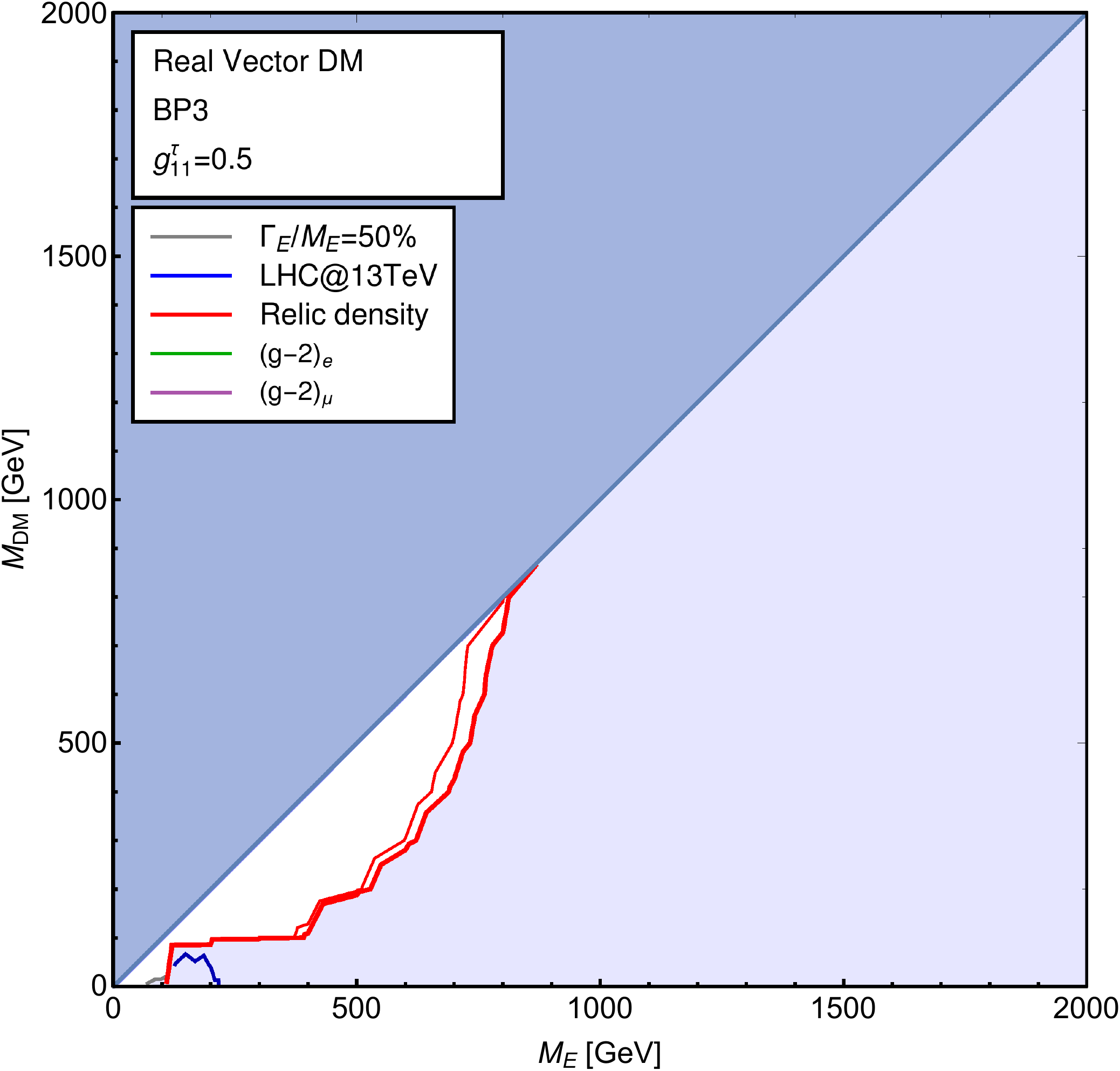}
\includegraphics[width=.32\textwidth]{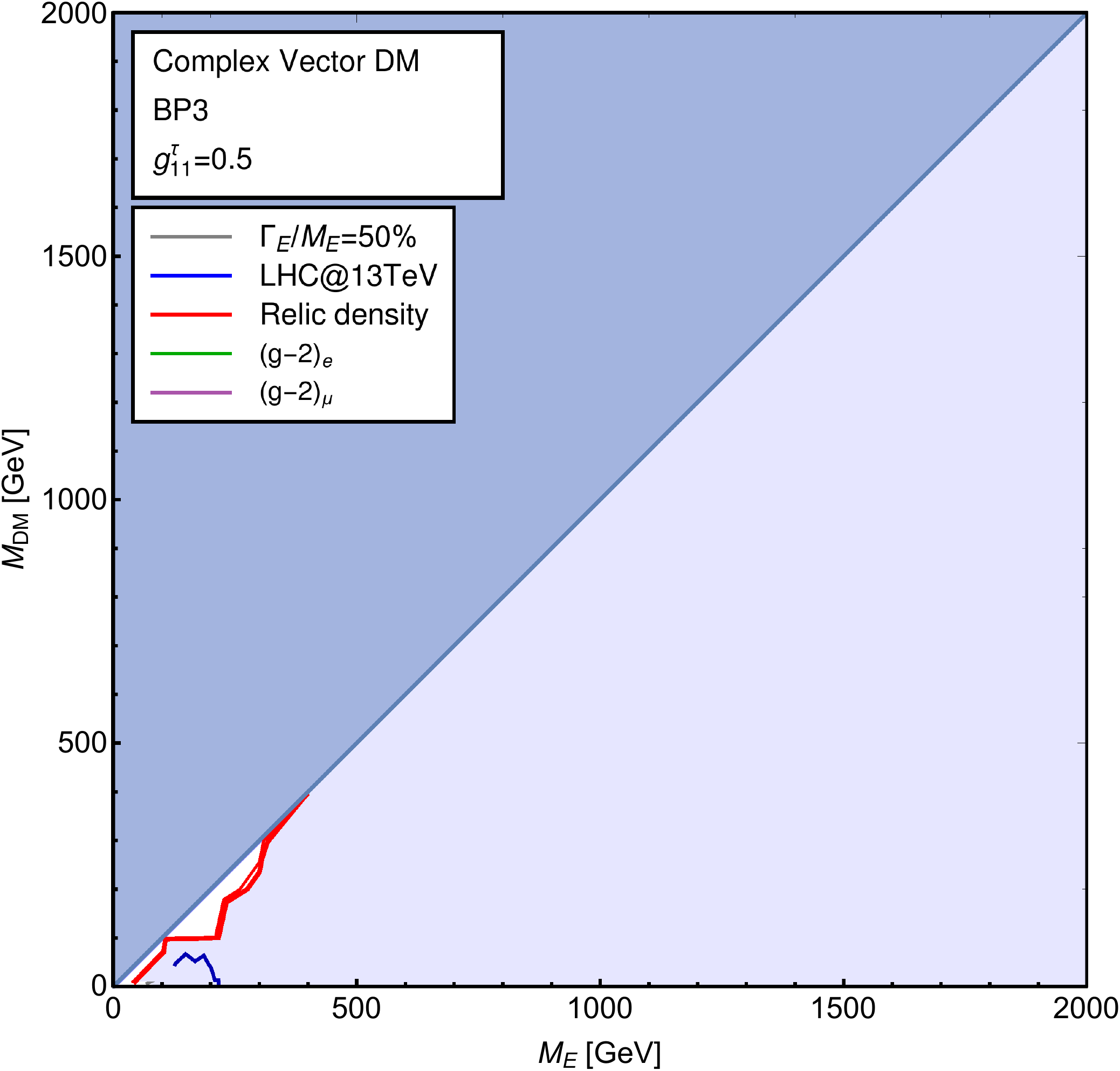}
\caption{\label{fig:combined3coup0.5} Excluded region (in light blue) for a fixed value of the VLL coupling with vector DM $g_{11}^f=0.5$ for BP3.}
\end{figure}

Focusing on the BPs we have considered and in the context of a minimal extension of the SM with just a DM bosonic candidate and a vector-like fermion carrying leptonic charge, the main conclusions of our analysis can be summarised as follows:
\begin{itemize}
\item For \textbf{\textit{BP1}} a vectorial DM is excluded by the complementarity between $(g-2)_e$, which requires a coupling $g_{11}^e \lesssim 0.02$, and relic density, which requires a coupling $g_{11}^e \gtrsim 0.3$. Therefore, if a signal of bosonic and leptophilic DM interacting with the SM electron or muon is observed, it has to be interpreted in terms of a scalar DM. In such scenario, for the smallest values of the interaction coupling VLL DM-lepton compatible with relic density, only the region with small mass gap between VLL and DM is allowed, while a larger region of parameter space becomes available as the coupling increases.
\item For \textbf{\textit{BP2}} a vectorial DM is excluded by $(g-2)_\mu$ alone, for which the parameter $\delta a_\mu$ is positive and not compatible with zero within 3$\sigma$, while the contribution of the VLL DM loop is negative. Therefore, analogously to BP1, a future signal of bosonic and leptophilic DM has to be interpreted in terms of a scalar DM. The relic density constraint has a similar qualitative behaviour as for BP1. However, in this case, the $(g-2)_\mu$ bound constrains the allowed region of parameter space into a band which becomes larger and encompasses larger VLL and DM masses as the coupling increases.
\item For \textbf{\textit{BP3}}, the absence of a $(g-2)_\tau$ constraint allows the possibility to have both scalar and vector DM scenarios. However, depending on the value of the coupling, a larger phase space can be available for either scalar or vector DM. For values of the coupling which allow a tiny strip in the degenerate VLL DM region for the real scalar DM scenario (Fig.~\ref{fig:combined123coup2}), the complex scalar and vector DM scenarios have a larger allowed space, and the vector DM case allows combinations with relatively light DM candidates. However, due to the stronger sensitivity of the VLL to the coupling with vector DM candidates, the width of the VLL can acquire very large values, above 50\% of its mass. The same happens of course for values of couplings which open a larger allowed region of parameter space for scalar DM (Fig.~\ref{fig:combined123coup7}); however, for such values of couplings, the region of allowed parameter space for vector DM shrinks towards the degenerate region. Finally, scenarios allowed only in the vector DM case can be possible, corresponding to small values of the couplings and represented in Fig.~\ref{fig:combined3coup0.5}. In this case, the allowed region for vector DM is for almost degenerate masses. In the degenerate mass regions, therefore, the only possibility to distinguish a scalar from vector DM $\tau$-philic DM scenario is to look at the kinematical properties of a future signal.
\item For \textbf{\textit{BP4}}, flavour changing interactions open the possibility to impose LFV constraints, which require small couplings in almost all the parameter space. In particular, for this BP the strongest constraint is given by $\mu\to e \gamma$, which requires couplings smaller than $\mathcal{O}(10^{-1})$ for both scalar and vector DM. Such strong constraint is in tension with the relic density measurement, which requires a large coupling, $\mathcal{O}(1)$ for scalar DM and $\mathcal{O}(10{-1})$ for vector DM. Therefore a scenario with universal leptophilic couplings is completely excluded. 
\end{itemize}

In summary, in minimal extensions of the SM, a vector DM scenario is only allowed for scenarios where the heavy new lepton interacts with the SM $\tau$ lepton. If a signal with vector DM is seen in observables involving lighter leptons, only non-minimal scenarios can be invoked for its interpretation, where possible cancellations of contributions from different topologies may relax some of the above constraints.

\section{Conclusions}

In the attempt to extract information on the properties of DM, and/or a new mediator to its creation, we have put forward here a  scenario where the former is a boson, of spin 0 or 1, and the latter  a  vector-like fermion, of spin 1/2, carrying lepton number, each being an odd eigenstate of a discrete \Z symmetry which distinguishes SM particles from these two new states, thereby  providing the means to render such a DM candidate stable and the mediator to decay exclusively to it. Hence, such a construct is rather minimal and designed to be sensitive to not only  DM data themselves but also to others where potential anomalies have been seen, like the anomalous magnetic moments of leptons and LFV processes. Hence, it is {\it per se} an attractive DM scenario while it can also be perceived as a simplified version of a more fundamental theory.
Within such a framework, we have been able to prove the potential of several data  in distinguishing between the two DM spin hypotheses and/or identifying the chiral structure of the mediator 
over regions of parameter space compliant with a variety of experimental constraints, from flavour to collider samples, from cosmological to laboratory probes. The sensitivity emerges not only indirectly in  response to the relic density experiments, but also directly in the characteristics of a potential signal detectable at both high energy colliders and low energy experiments. 

For a start, as the DM mediator is charged, we have assessed the impact of its presence in 125 GeV Higgs data collected at the LHC, most notably in di-photon final states, as a heavy charged lepton would enter the $h\to\gamma\gamma$ transition at loop-level. (In fact, a companion heavy neutral lepton would also affect such data, by altering the rate of the Higgs boson  leptonic signatures, just like its charged counterpart.) We have found that, due to the non-decoupling property of new chiral families, the constraints on these states
are quite stringent, allowing only extremely light objects, roughly lighter than 2 GeV. We have therefore moved on to see whether such states can be allowed by direct searches.   
Specifically, we have continued by sketching the parameter space of the model surviving constraints emerging from the measurement of $l^+_il^-_j + \MET$ final states produced at LHC from $pp$ annihilations. 
Herein, we have verified that the shape of the excluded regions depends minimally on the DM nature, whether vector or scalar, whether real or complex, in the NWA. For larger couplings, generating sizeable VLL width of order $\mathcal{O}(10\%)$ of its mass, the contours are only slightly deformed, leaving unchanged the qualitative behaviour of the results.
At any rate, we have been able to  broadly identify the regions in the $(M_E , M_{\rm DM})$ parameter space where such a difference would be manifest in future data to be collected at the LHC. A somewhat orthogonal pattern appears from the study of relic density data, wherein the sensitivity contours are now significantly dependent upon the assumption made on whether the DM is scalar or vector (but not  whether it is real or complex).
Finally, from the study of current limits from the LFV process $\mu\to e\gamma$ and the anomalous magnetic moment of both electron and muon, we have discovered that the dependence of the exclusion contours on the $(M_E , M_{\rm DM})$ plane are remarkably different depending upon whether the DM is scalar or vector, though the established trend of the differences is not the same in both sets of observables. All this, therefore, points to the fact that, if an excess is observed in the future in one of these channels, and, if the masses of the new states ($M_E$ and
$M_{\rm DM}$) are determined via independent measurements,  unequivocal determination of the DM spin will be possible in such LFV observables. This could well be achieved through the use of LHC data. (Interestingly, the recently reported excess in the measurement of the cosmic ray electron and positron reported by the DAMPE collaboration~\cite{Ambrosi:2017wek} at an energy of about 1.4 TeV could be originated from scalar DM annihilating to an $e^+e^-$ through the exchange of a $t$-channel VLL and interpreted within BP1 of the simplified model studied in this paper.)

In short, our study paves the way towards a programme of charactering the nature of both (bosonic) DM and a new (leptonic) mediator with upcoming data. We have reached this conclusion based on numerical analyses adopting sophisticated numerical tools for both the theoretical predictions of the underlying BSM scenario 
and the up-to-date constraints imposed upon it by current experimental data. Crucially, in doing so, we have allowed for finite width effects of the mediator, an aspect which can impinge greatly on the results obtained, in both LHC and relic density data, and which is normally overlooked in routine studies.

\paragraph{Acknowledgements}
AD is partially supported by the ``Institut Universitaire de France'', the Labex-LIO (Lyon Institute of 
Origins) under grant ANR-10-LABX-66 and FRAMA (FR3127, F\'ed\'eration de Recherche ``Andr\'e Marie Amp\`ere").
SM is supported in part by the NExT Institute, the  grant H2020-MSCA-RISE-2014  no. 645722 (NonMinimalHiggs) and the STFC Consolidated Grant  grant number ST/L000296/1.

\bibliography{XF}
\bibliographystyle{JHEP}

\end{document}